\documentclass[twocolumn]{aastex62}

\newcommand{\kms}{\mbox{km s$^{-1}$}}

\newcommand{\mhi}{$M_{\rm HI}$}%

\newcommand{\xone}{\mbox{$x_{1}$}}
\newcommand{\xtwo}{\mbox{$x_{2}$}}

\received{}
\revised{}
\accepted{}
\submitjournal{ApJ}

\shorttitle{Molecular gas and Star Formation Properties in Early Stages Mergers}
\shortauthors{Espada et al.}

\begin{document}

\title{Molecular gas and Star Formation Properties in Early Stage Mergers:  SMA CO(2-1) Observations of the LIRGs NGC\,3110 and NGC\,232}

\correspondingauthor{Daniel Espada}
\email{daniel.espada@nao.ac.jp}

\author{Daniel Espada} 
\affiliation{National Astronomical Observatory of Japan, 2-21-1 Osawa, Mitaka, Tokyo 181-8588, Japan}
\affiliation{The Graduate University for Advanced Studies (SOKENDAI), 2-21-1 Osawa, Mitaka, Tokyo, 181-0015, Japan}

\author{Sergio Martin}
\affiliation{Joint ALMA Observatory, Alonso de C{\'o}rdova, 3107, Vitacura, Santiago 763-0355, Chile}
\affiliation{European Southern Observatory, Alonso de C{\'o}rdova 3107, Vitacura, Santiago 763-0355, Chile}

\author{Simon Verley}
\affiliation{Departamento de F{\'i}sica Te{\'o}rica y del Cosmos, Universidad de Granada, 18010 Granada, Spain}
\affiliation{Instituto Universitario Carlos I de F{\'i}sica Te{\'o}rica y Computacional, Facultad de Ciencias, 18071 Granada, Spain}

\author{Alex R. Pettitt}
\affiliation{Department of Physics, Faculty of Science, Hokkaido University, Kita 10 Nishi 8 Kita-ku, Sapporo 060-0810, Japan}

\author{Satoki Matsushita}
\affiliation{Academia Sinica Institute of Astronomy and Astrophysics,
  11F of Astro-Math Bldg, AS/NTU, No.1, Sec. 4, Roosevelt Rd,
  Taipei 10617, Taiwan, Republic of China}

\author{Maria Argudo-Fern{\'a}ndez}
\affiliation{Centro de Astronom\'ia, Universidad de Antofagasta, Avenida Angamos 601, Antofagasta 1270300, Chile}
\affiliation{Chinese Academy of Sciences South America Center for Astronomy, China-Chile Joint Center for Astronomy, Camino El Observatorio, 1515, Las Condes, Santiago, Chile}

\author{Zara Randriamanakoto}
\affiliation{Department of Astronomy, University of Cape Town, Private Bag X3, Rondebosch 7701, South Africa}

\author{Pei-Ying Hsieh}
\affiliation{Academia Sinica Institute of Astronomy and Astrophysics, P.O. Box 23-141, Taipei 10617, Taiwan, Republic of China}

\author{Toshiki Saito}
\affiliation{National Astronomical Observatory of Japan, 2-21-1 Osawa, Mitaka, Tokyo 181-8588, Japan}
\affiliation{Max-Planck-Institut f{\"u}r Astronomie, K{\"o}nigstuhl 17, D-69117 Heidelberg, Germany}

\author{Rie E. Miura} 
\affiliation{National Astronomical Observatory of Japan, 2-21-1 Osawa, Mitaka, Tokyo 181-8588, Japan}

\author{Yuka Kawana}
\affiliation{Japan Women's University, 2-8-1 Mejirodai, Bunkyo, Tokyo, 112-8681, Japan}
\affiliation{Tokyo Metropolitan Mihara High School, 1-33-1, Omorihigashi, Ota-ku, Tokyo, Japan}

\author{Jose Sabater}
\affiliation{Institute for Astronomy, University of Edinburgh, EH9 3, HJ Edinburgh, UK}

\author{Lourdes Verdes-Montenegro}
\affiliation{Instituto de Astrof{\'i}sica de Andaluc{\'i}a (CSIC), Apdo. 3004, 18008, Granada, Spain}

\author{Paul T. P. Ho}
\affiliation{Academia Sinica Institute of Astronomy and Astrophysics, P.O. Box 23-141, Taipei 10617, Taiwan, Republic of China}
\affiliation{East Asian Observatory, 660 North Aohoku Place, University Park, Hilo, HI 96720, USA}

\author{Ryohei Kawabe}
\affiliation{National Astronomical Observatory of Japan, 2-21-1 Osawa, Mitaka, Tokyo 181-8588, Japan}
\affiliation{The Graduate University for Advanced Studies (SOKENDAI), 2-21-1 Osawa, Mitaka, Tokyo, 181-0015, Japan}

\author{Daisuke Iono}
\affiliation{National Astronomical Observatory of Japan, 2-21-1 Osawa, Mitaka, Tokyo 181-8588, Japan}
\affiliation{The Graduate University for Advanced Studies (SOKENDAI), 2-21-1 Osawa, Mitaka, Tokyo, 181-0015, Japan}

\begin{abstract}
Mergers of galaxies are an important mode for galaxy evolution because they serve as an efficient trigger of powerful starbursts. However, observational studies of the molecular gas properties during their early stages are scarce. 
We present interferometric CO(2--1) maps of two luminous infrared galaxies (LIRGs), NGC\,3110 and NGC\,232, obtained with the Submillimeter Array (SMA) with $\sim$ 1\,kpc resolution.
While NGC\,3110 is a spiral galaxy interacting with a minor (14:1 stellar mass) companion,
NGC\,232 is interacting with a similarly sized object. 
We find that such interactions have likely induced in these galaxies enhancements in the molecular gas content and central concentrations, 
 partly at the expense of atomic gas.
The obtained molecular gas surface densities in their circumnuclear regions are $\Sigma_{\rm mol}~\gtrsim10^{2.5}$\,M$_\sun$\,pc$^{-2}$, higher than in non-interacting objects by an order of magnitude. Gas depletion times  of ~0.5--1\,Gyr are found for the different regions, lying in between non-interacting disk galaxies and the starburst sequence. { In the case of NGC\,3110, the spiral arms show { on average 0.5\,dex shorter depletion times} than in the circumnuclear regions if we assume a similar { H$_{\rm 2}$-CO} conversion factor.
 We show that even in the early stages of the interaction with a minor companion, a starburst is formed along the circumnuclear region and spiral arms, where a large population of SSCs is found ($\sim$350), and at the same time a large central gas concentration is building up which might be the fuel for an active galactic nucleus. { The main morphological properties of the NGC\,3110 system are reproduced by our numerical simulations and allow us to estimate that the current epoch of the interaction is at $\sim$ 150\,Myrs after closest approach.}}
\end{abstract}


\section{Introduction}
\label{introduction}

Interactions between galaxies are among the most important mechanisms modifying the properties of galaxies through their cosmological lifetime \citep[e.g.][]{1992ARAA..30..705B,2005Natur.435..629S}, especially because the merger rate is seen to be higher in the early universe \citep[e.g.][]{2010ApJ...709.1067B,2015A&A...576A..53L}.
Several mechanisms are at play to drive gas from external regions towards the nuclei of galaxies and produce starbursts (SBs) and active galactic nuclei (AGNs). 
The dissipative nature of the gas leads to loss of angular momentum in shocks, resulting in subsequent inflows which then trigger SBs within the inner few kiloparsecs of the galaxy, as well as feeding the AGNs  \citep{1996ApJ...471..115B}.  Observationally, interactions are seen to play a main role in increasing Star Formation (SF) and concentrating the gas towards the center \citep[e.g.][]{1987ApJ...320...49B,2007AJ....133..791S,2013MNRAS.430..638S}.
Probably the most extreme objects are Ultra/Luminous Infrared Galaxies (U/LIRGs), where IR luminosities are above 10$^{11}$ L$_\sun$, and represent one of the most dramatic phases in galaxy evolution \citep[e.g.][]{1996ARA&A..34..749S}. 

The transformation of disk galaxies in the merging process has been widely studied numerically \citep{1972ApJ...178..623T,1992ApJ...400..153M,1996ApJ...464..641M,1992ARAA..30..705B,1996Natur.379..613M,2005Natur.435..629S,2008MNRAS.384..386C}.
Simulations including the stellar and gaseous components of merging systems have covered a wide range of the parameter space, but usually  numerical studies presenting the star formation history have focused on major mergers of similarly sized disk galaxies (mass ratios up to 3:1) \citep[e.g.][]{2013MNRAS.430.1901H} rather than galaxies accreting smaller objects.
The galaxy mass ratio (together with closest approach distance and relative velocities) is indeed found to be a major parameter characterizing the resulting merger driven starburst. The induced SF in the primary galaxy for  large mass ratio mergers $>$10:1 is found to be just marginally larger \citep{2008MNRAS.384..386C}.

Observationally it is found that there are enhancements of the Star Formation Rate (SFR) especially for galaxy pairs at small projected separations of $<$ 30-40\,kpc \citep{2008AJ....135.1877E}, but can be seen up to 150\,kpc \citep{2012MNRAS.426..549S,2013MNRAS.433L..59P}. These enhancements are larger for galaxies of approximately equal mass, but such a SFR enhancement can { also be seen for galaxy pairs whose masses vary by up to a factor of $\sim$10} \citep{2008AJ....135.1877E}, although it has been argued that this is true mostly in the least massive galaxy of the pair \citep{2007AJ....134..527W}.  As for the molecular gas content, from which stars form, \citet{1999ApJ...512L..99G} found that it decreases as merging advances. \citet{1999ApJ...512L..99G} also argued that the starburst is probably not due to the formation of more molecular clouds from atomic gas, but enhanced star formation in preexisting molecular clouds.
On the other hand, \citet{2017ApJ...844...96Y} recently found that the molecular gas masses in the central regions of galaxies are relatively constant from early to late merger stages, { which they interpreted as molecular gas inflow replenishing} the consumed gas by SF. Observations with better resolution are needed in order to probe the detailed properties of molecular gas and onset of SF in these systems.

High angular resolution CO observational studies at different stages of the merging process and for different mass ratios are essential to understand the molecular gas response to the interaction, which have consequences on the subsequent SF activities. A large number of high resolution observational studies of intermediate 
to late-stage mergers 
and merger remnants  have been performed \citep[][]{1998ApJ...507..615D,1999AJ....117.2632B,2005ApJS..158....1I,2008ApJS..178..189W,2009ApJ...692.1432G,2009ApJ...695.1537I,2009AJ....137.3581I,2012ApJ...753...46S,2013PASJ...65L...7I,2013AJ....146...47I,2014ApJS..214....1U,2014ApJ...787...48X,2015ApJ...803...60S,2016PASJ...68...20S,2017ApJ...834....6S,2017ApJ...840....8S,2017ApJ...835..174S}. Although intermediate and late stage major mergers garner a significant fraction of all the observational studies, early stage phases when separations are still relatively large ($>$ 40\,kpc) have rarely been part of these studies. 
Therefore, our understanding from an observational point of view of the early stages of the interaction is relatively poor, and { it} is essential to investigate what is the impact of smaller objects, as this kind of events are expected to be more frequent. Although these kind of systems can not be easily found locally given the short time scales (a few 10$^8$\,Myr) involved close to the first approach, it is of key importance to complete the sequence of snapshots at every merger stage in order to show how the interstellar medium and SF evolve along the merger process sequence.

In the early stages of a merger, simulations show that galaxies approach each other but are still distinct and the morphology starts to be just slightly distorted. It is established that in this phase the generation of disturbed two arm barred spirals and tidal bridges/tails occurs \citep{1972ApJ...178..623T}. A direct consequence in this phase of this kind of interactions are galactic spiral features \citep{2006ARep...50..785T,2010MNRAS.403..625D}, warps \citep{2009ApJ...703.2068D,2014ApJ...789...90K}, bars \citep{2014ApJ...790L..33L,2018MNRAS.474.5645P}, (stellar and gaseous) bridges which may connect the galaxies, and tidal tails extending well beyond the main body of the galaxies \citep{2010ApJ...725..353D}.
Although the tidal interaction forms these kind of spiral arms and probably relatively long-lived bars, the gas response varies very rapidly and flows efficiently toward the central regions \citep{2004ApJ...616..199I}.

In this paper we present molecular gas maps at $\sim$1\,kpc scale resolution as traced by CO(2--1) for two cases of early-stage tidal interactions between galaxies at relatively large distances ($\gtrsim$ 40\,kpc) and with different mass ratios: NGC\,3110, interacting with a minor companion of 14:1 its mass, and NGC\,232, which is interacting mostly with NGC\,235, an object of comparable mass. 
Our aim is to shed  light on the effect that a different galaxy mass ratio has on the molecular gas properties of the primary galaxies and the onset of the starburst in the early stages of the interaction. 
This paper is organized as follows. First we describe the properties of the two galaxies subject to study in this paper in \S~\ref{intro}. We introduce our Submillimeter Array (SMA) observations and data reduction in \S~\ref{observationReduction}. In \S~\ref{result} we focus on the identification of the different molecular components and study their major physical properties. We compare the gas content and SF properties, and derive the spatially resolved SF law in \S~\ref{discuscaling} with the help of H$\alpha$ and 24\,$\mu m$ maps. 
Finally,  in \S~\ref{discussion} we compare { the observational maps with our own numerical simulations as well as with others in the literature}. Our main conclusions are summarized in \S~\ref{conclusion}.  We adopt a cosmology of $\Omega_\Lambda$ = 0.73, $\Omega_M$ = 0.27, and $H_0$ = 73\,\kms\,Mpc$^{-1}$.

\section{The LIRGs NGC\,3110 and NGC\,232 }
\label{intro}

\begin{figure*}[tbh]
\begin{center}
\includegraphics[width=8cm]{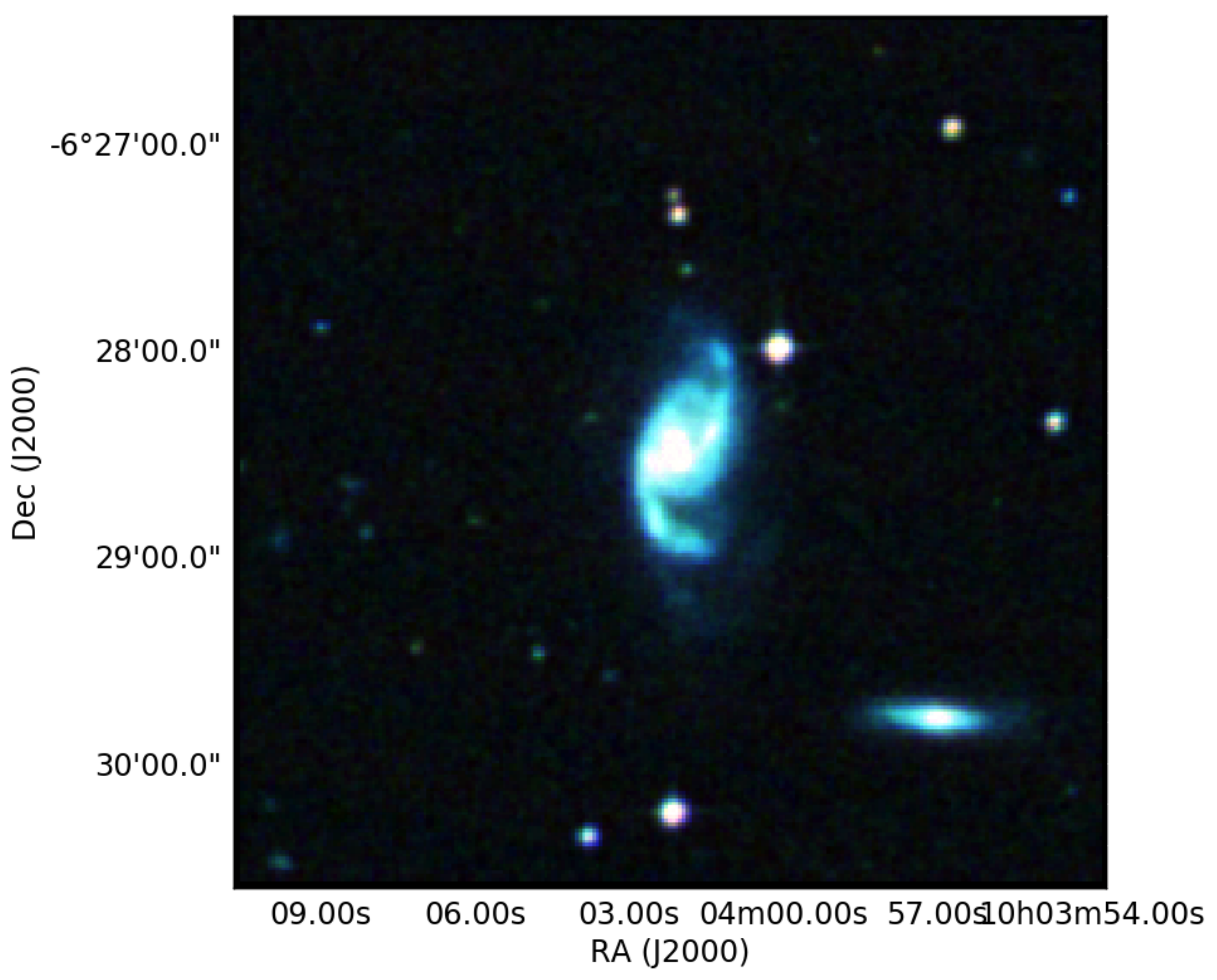}
\includegraphics[width=8cm]{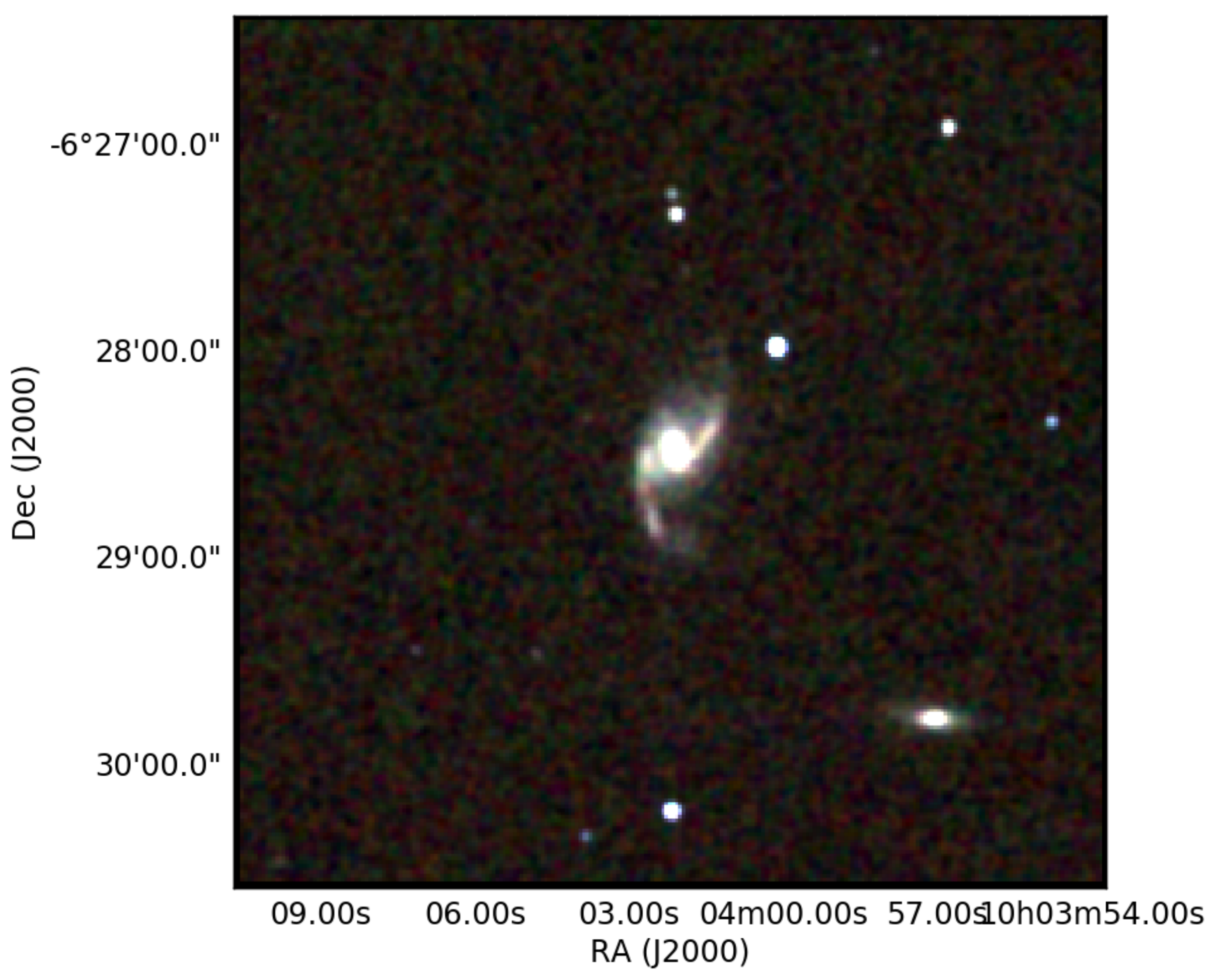}
\includegraphics[width=8cm]{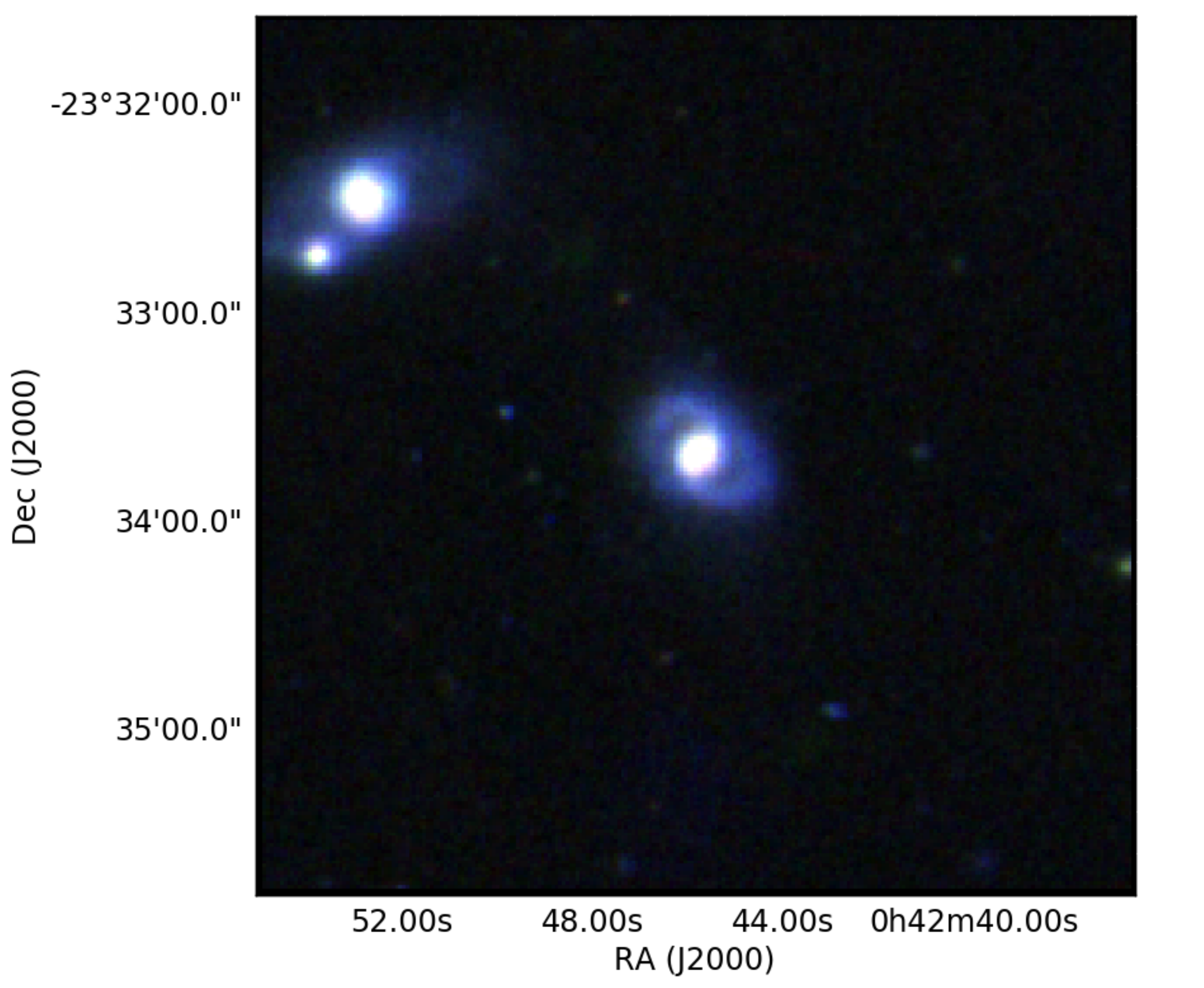}
\includegraphics[width=8cm]{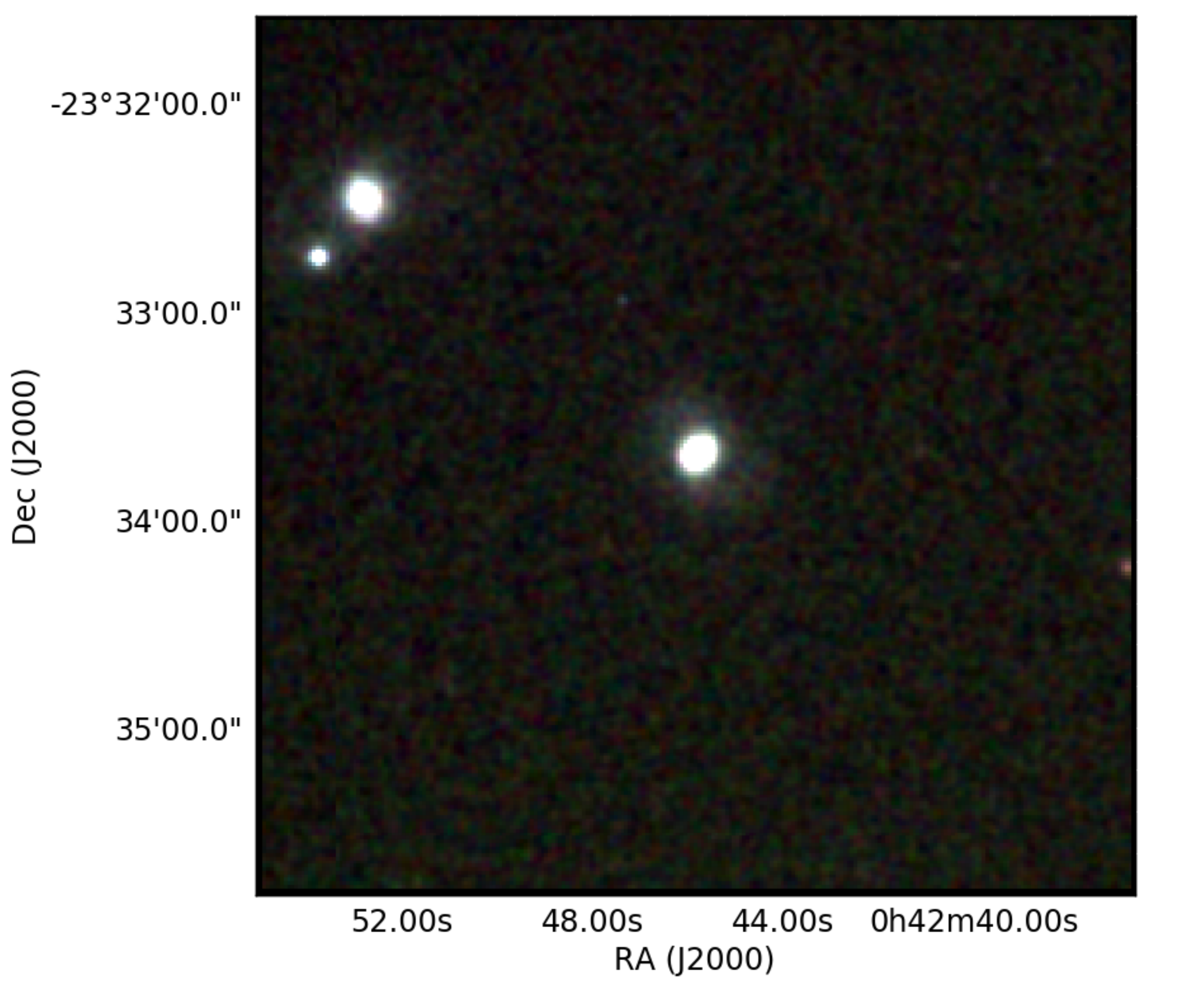}

\end{center}
\caption{DSS2 blue, IR, red (left) and 2MASS JHK (right)  composite RGB images of the NGC\,3110 (top) and NGC\,232/5 (bottom) pairs. NGC\,3110 and NGC\,232 are the galaxies in the center of the fields. The size of the maps is 4\arcmin. 
\label{fig1}}
\end{figure*}

We first summarize the main properties of NGC\,3110 and NGC\,232, including their morphological properties, environment, activity and gas properties. The main parameters characterizing these two galaxies are listed in Table~\ref{tbl-1}. Fig.~\ref{fig1} shows optical DSS2 and NIR 2MASS composite images, and Fig.~\ref{fig2} the H$\alpha$ maps. 

These two objects are part of the SMA B0DEGA project (SMA Below 0 DEgree GAlaxies, \citealt{2010gama.conf...97E}), where the CO(2--1) line was observed for about a hundred galaxies with the Submillimeter Array (SMA). Because they are located in the southern sky, most of the galaxies in the sample did not have previous interferometric CO observations published in the literature. 
The sample is composed of infrared (IR) bright galaxies, selected with the criteria: 2.58 $\times$ $S_{\rm 60\mu m}$ + $S_{\rm 100\mu m}$  $>$ 31.5~Jy  ($S_{\rm 60\mu m}$ and $S_{\rm 100\mu m}$ are the IRAS  60\,$\rm \mu m$ and 100\,$\rm \mu m$ fluxes), recession velocities $V <$ 7000~\kms , and located in the southern sky up to  $\delta$ = --45$\arcdeg$. The galaxies in the sample are mostly of spiral type and members of interacting systems such as groups and pairs of galaxies, but we excluded intermediate and late stage major mergers.
 The data can be accessed via the CfA SMA database\footnote{https://www.cfa.harvard.edu/cgi-bin/sma/smaarch.pl}. 
 
NGC\,3110 and NGC\,232 are the most IR luminous galaxies in the B0DEGA sample, and among the most IR luminous objects that can be found located nearby which are not intermediate/late stage major mergers. 
Both galaxies  were classified as mildly interacting (i.e. separate nuclei, weak tidal features), as opposed to objects with strong tidal features, mergers and merger remnants \citep{2002ApJS..143...47D}. 
These two local LIRGs are also part of the Great Observatories All-Sky LIRG Survey (GOALS, \citealt{2009PASP..121..559A}), which is a flux-limited sample ($S_{\rm 60\mu m} >$ 5.24\,Jy) composed of a total of 203 galaxies with $L_{\rm IR}$ $>$ 10$^{11}$ L$_\odot$ based on the IRAS Revised Bright Galaxy Survey (RBGS) presented in \citet{2003AJ....126.1607S}.

\subsection{NGC\,3110}
\label{intro3110}

\begin{figure*}[tbh]
\begin{center}
\includegraphics[width=8cm]{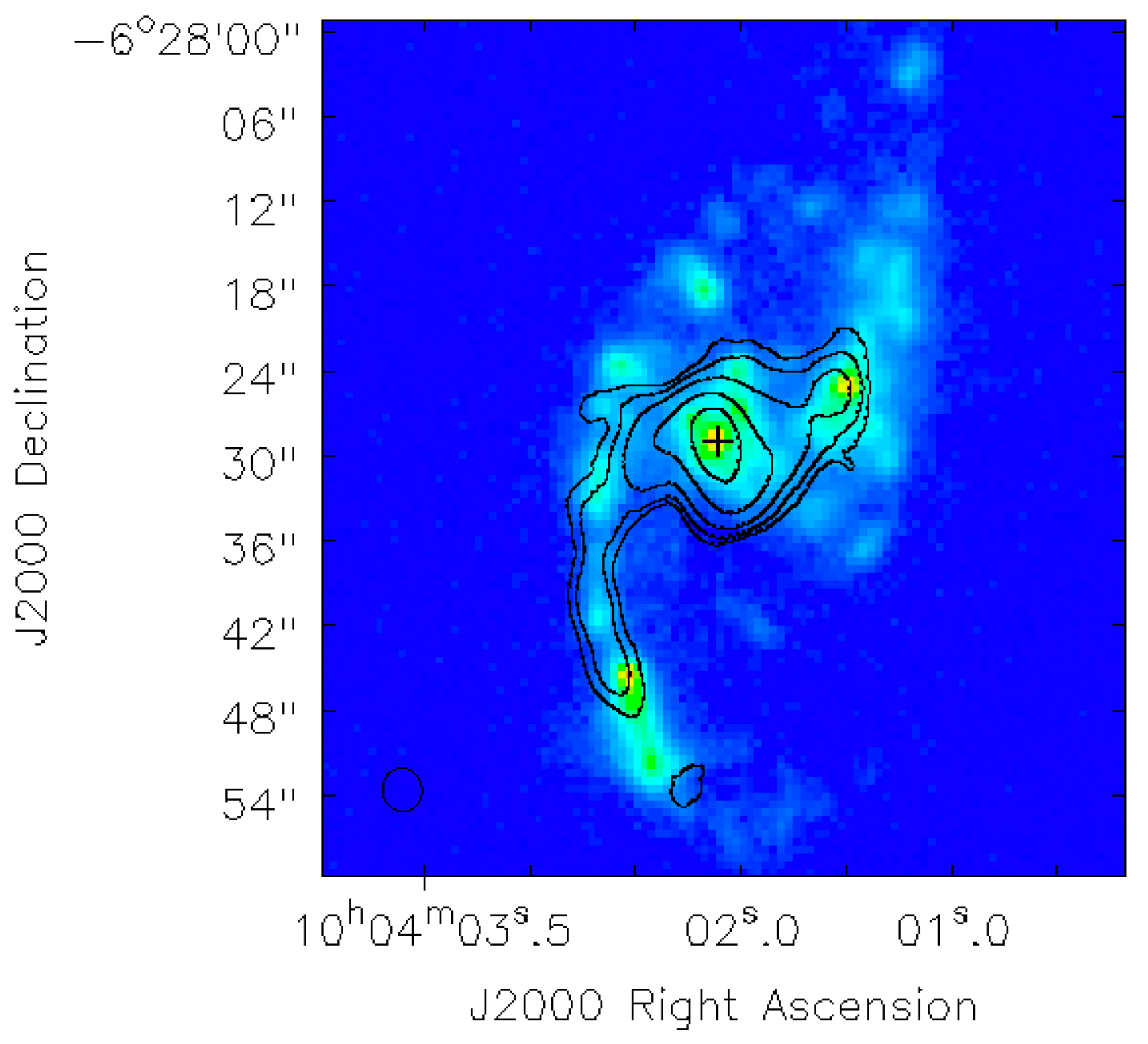}
\includegraphics[width=8.2cm]{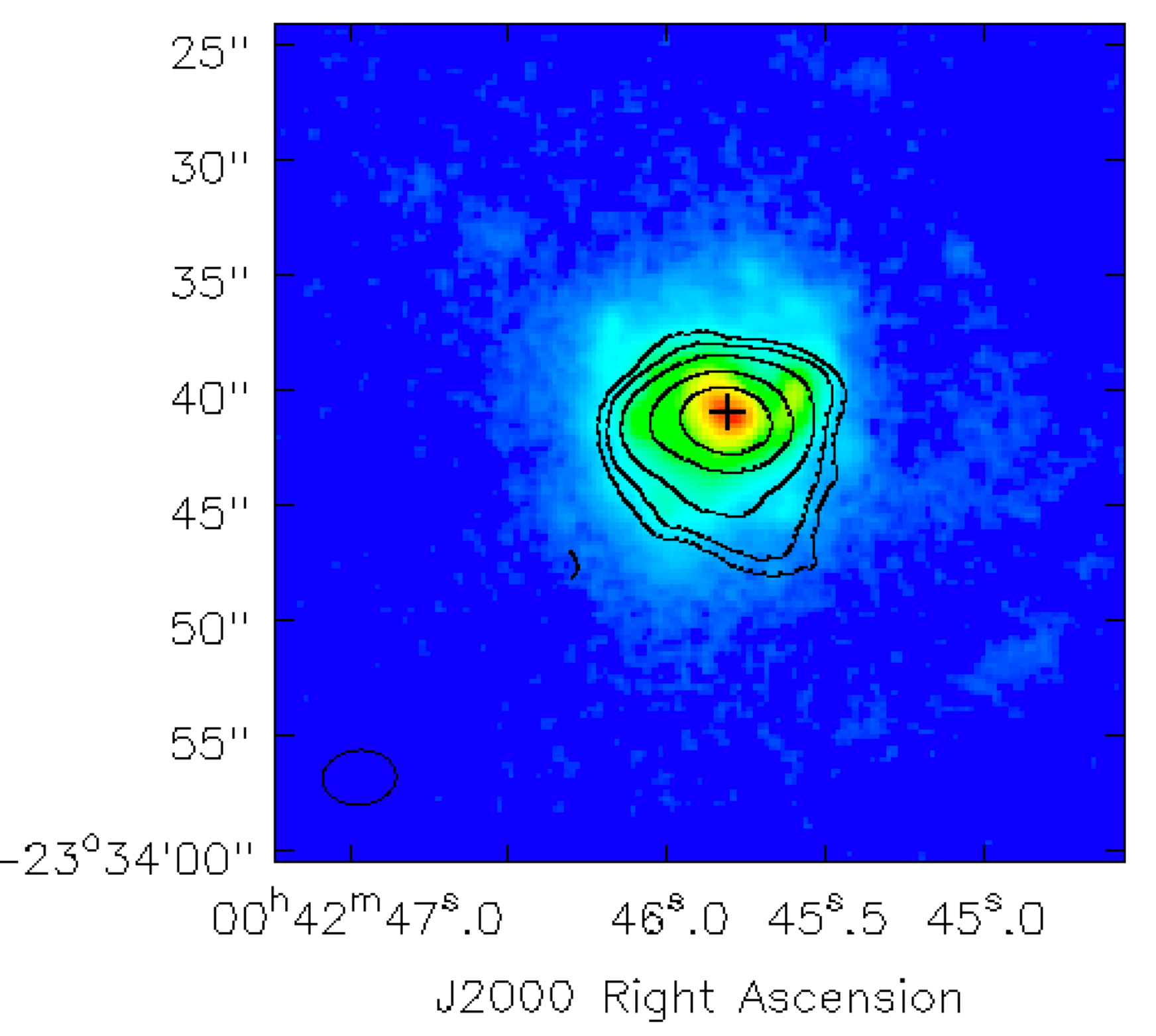}
\end{center}
\caption{H$\alpha$ images of NGC\,3110 \citep{2004AJ....127..736H} and NGC\,232 \citep{2006ApJ...643..173S} with the SMA CO(2--1) contours overlaid. The field of view are 60\arcsec\ $\times$ 57\arcsec\ and 36\arcsec $\times$ 36\arcsec , respectively. The color scale ranges from 0 to 6.51 $10^{-15}$\,erg\,s$^{-1}$\,cm$^{-2}$ and from 0 to 3.85 $10^{-15}$\,erg\,s$^{-1}$\,cm$^{-2}$. The CO(2--1) contours (black) are as in Fig.\,\ref{fig7} and \ref{fig8}, { i.e. at 3, 5, 10, 25, and 50$\sigma$, where $\sigma$ = 1.4\,Jy\,beam$^{-1}$\,\kms\ for NGC\,3110 and $\sigma$ = 1.7\,Jy\,beam$^{-1}$\kms\ for NGC\,232}.  The HPBW of the synthesized beams of the SMA CO(2--1) observations are indicated at the bottom left of the panels. The cross sign shows the center of the galaxies.
\label{fig2}}
\end{figure*}

NGC\,3110 is usually classified as a barred Sb galaxy (Table\,1). It is located at a distance of $D$ = 75.2\,Mpc (1\arcsec\ corresponds to 352\,pc), and it is interacting with a minor companion located to the SW, which likely contributed to the formation of its two asymmetric spiral arms and a central bar-like feature as seen in Fig.\,\ref{fig1}. The optical diameter is 1\farcm86, or 39.3\,kpc. The structure that is usually classified as a bar feature has a length $\sim$20\arcsec\ (7\,kpc) as seen in H$\alpha$ emission \citep{2004AJ....127..736H}. However, a more careful inspection of high resolution images reveals that this feature does not resemble the distribution of a standard bar. 
This galaxy has an overall high SFR of about 20--30\,M$_\odot$\,yr$^{-1}$,  as obtained from IR and H$\alpha$ data \citep{2002ApJ...568...88Y,2002ApJS..143...47D}, mostly concentrated towards its center. 
The SF is fueled by a { large mass of molecular gas}, estimated to be $M_{\rm mol}$ $\simeq$ 2 $\times$ 10$^{10}$ M$_\odot$ (Table\,1).
There is also abundant SF in the two arms of the galaxy based on H${\alpha}$ maps (Fig.\,\ref{fig2}), with several bright \ion{H}{2} knots distributed along them \citep{2004AJ....127..736H,2002ApJS..143...47D} and in fact the arms contribute to up to 35\% of the FIR flux \citep{2000ApJS..131..413Z}.
These spiral arms wind up to the center within the inner 9\arcsec\  (3.2\,kpc) as seen in Br$\gamma$, H$_2$ S(1--0), and \ion{Fe}{2} in VLT/SINFONI data, and there is also an elongated structure along the N-S direction \citep{2015A&A...578A..48C}. Hubble Space Telescope (HST) images also show some filaments orienting towards its nuclear region \citep{1998ApJS..117...25M}.
The nuclear activity was classified as \ion{H}{2} \citep{1995ApJS...98..171V,2003ApJ...583..670C}, and so far there is no evidence of AGN in its nucleus.

Following the visual morphological scheme by \citet{2016ApJ...825..128L} for the GOALS sample, from objects in the early stages of interaction to merger remnants, NGC\,3110 was classified as an early stage major merger (stage 1, or M1, following their nomenclature), where an early stage major merger is defined as a galaxy pair with a velocity difference between the two galaxies of $\Delta V < 250$\,\kms\ and separations of less than 75\,kpc, which have no prominent tidal features, and appear to be on their initial approach. This classification also implies that the mass ratio between the two interacting galaxies are comparable ($<$4:1), although the companion of NGC\,3110 is likely smaller than this. 
With their morphological classification scheme it would probably fit better into the minor merger case. 
The companion galaxy is MCG-01-26-013 (PGC029184), which is located at just 1\farcm8 from NGC\,3110, or 38\,kpc (Fig.~\ref{fig1}), and with a (projected) velocity difference of 235\,\kms . Its optical diameter is 0\farcm98, about half of that of the main galaxy.
It is seen nearly edge-on and it is classified as S0$^+$pec or S0-a \citep[][Hyperleda]{1991rc3..book.....D}. Its (extinction corrected) blue-band luminosity is more than one order of magnitude smaller than that of NGC\,3110, 7.3 $\times$ 10$^{9}$\,L$_\sun$ versus 10$^{11}$\,L$_\sun$ (Table~\ref{tbl-1}), so the mass difference is estimated to be 14:1. 
There exists H${\alpha}$ emission towards the nucleus of the companion as well, and it is more centrally concentrated than that in NGC\,3110 \citep{2002ApJS..143...47D}.
From Very Large Array (VLA) \ion{H}{1} observations, it is seen that a local \ion{H}{1} minimum is found at the centre of { NGC\,3110}, and although the distribution is asymmetric with its peak towards the northern arm, no tidal tail is seen in these maps to a sensitivity of $\sim$10$^{20}$~atoms~cm$^{-2}$ \citep{2002MNRAS.329..747T}.

\subsection{NGC\,232}
\label{intro232}

At a distance of $D$ = 90.5\, Mpc (1\arcsec\ corresponds to 420\,pc), NGC\,232 is a barred Sa galaxy seen with an inclination of $\sim$47\arcdeg (Table\,\ref{tbl-1}), and with two arms which seem to form a ring-like structure in optical images. The size of the disk is 0\farcm97, or 24\,kpc.  In the optical image (Fig.~\ref{fig1}) one can discern that the two spiral arms (to the N and S) are asymmetric and emerge from the barred central component, which is elongated along the SE to NW direction. 
A high SFR of about 30\,M$_\odot$\,yr$^{-1}$ can be derived from its FIR flux \citep[e.g.][]{2006ApJ...643..173S}. 
The H$\alpha$ maps in \citet{2006ApJS..164...52S} show a bright compact nuclear region of about 10\arcsec\ in diameter elongated along the E--W direction (P.A. $\simeq$ 100\arcdeg), 
as well as extended diffuse emission with a somewhat different P.A., which is the direction along the bar-like structure seen in optical and NIR images at P.A. $\simeq$ 140\arcdeg (Fig.\,\ref{fig1}), as well as in radio emission \citep{2006ApJS..164...52S}.
The brightest component in the H$\alpha$ map along the E--W direction is also detected in Pa$\alpha$ \citep{2015ApJS..217....1T}.
The molecular gas mass is  
estimated to be $M_{\rm mol}$ $\simeq$ 2 $\times$ 10$^{10}$\,M$_\odot$ (Table\,\ref{tbl-1}).
NGC~232 was classified to host a nuclear starburst (\ion{H}{2}) \citep{1995ApJS...98..171V,2002ApJ...564..650C}, although it has also been classified as a low ionization nuclear emission-line region (LINER) by \citet{2000AJ....120...47C}. A very extended ($\sim$ 3\,kpc) and highly collimated linear structure extending from the nucleus  has been recently found using VLT/MUSE \citep{2017ApJ...850L..17L}, which indicates that this is in fact an AGN. This is the second longest optical emission-line jet reported so far. 

NGC\,232 is classified as a major merger in the stage 2, where galaxy pairs show obvious tidal bridges and tails \citep{2016ApJ...825..128L}. Although the companion galaxy to the NE, NGC\,235 (ESO 474-G016, Fig.\,\ref{fig1}), might be an evolved merger by itself because it has a double nucleus, the separation to NGC\,232 and the lack of disturbed features may indicate that from the point of view of NGC\,232 it is in an earlier stage of interaction as well.  
NGC\,235 is a peculiar S0 which is separated by 2\arcmin\ (or 50\,kpc) and with a (projected) velocity difference of $\sim$120\,\kms\ with respect to NGC\,232. Both NGC\,232 and 235 are members of a southern compact group of four objects (SCG\,10, or SCG~0040-2350,  \citealt{1994AJ....107.1235P,2000AJ....120...47C,2002AJ....124.2471I}), although we note that two of them are probably merging and are both associated with NGC\,235 itself and the fourth is an edge-on spiral galaxy (NGC\,230, $V$ = 6759\,\kms) to the SW separated by about 5\farcm9 ($\sim$ 148\,kpc) from NGC\,232 and with no signs of interaction. On the other hand there is a signature of interacting debris between NGC~232 and NGC~235 from the H$\alpha$ knot that is lying between the two galaxies \citep{1994AJ....107...99R,2002ApJS..143...47D}.  
These two objects are of comparable size.
Assuming the same distance, the blue-band optical luminosities for NGC~232 and NGC~235 are $L_{\rm B}$ = 2.8 $\times$ 10$^{10}$\,L$_\odot$ and 3.5 $\times$ 10$^{10}$\,L$_\odot$. The optical diameters of NGC~232 and NGC\,235 { are comparable, with that of NGC\,235 being slightly} larger (24.4\,kpc vs 34.8\,kpc).

\section{Observations and Data Reduction}
\label{observationReduction}

We present in this paper CO(2--1) observations using the Submillimeter Array (SMA, \citealt{2004ApJ...616L...1H}) towards NGC\,3110 and NGC\,232. 
NGC\,3110 was observed for 162\,min (on source) with 7 antennas in a compact configuration, with maximum projected baselines of 53.795\,k$\lambda$, on 2008 Oct 11, and 
NGC\,232 was observed for 89\,min (on source) with 8 antennas in a slightly more extended configuration with maximum projected baselines of 90.950\,k$\lambda$, on 2008 May 8 and 21.

 Table~\ref{tbl-2} summarizes the set-up of the interferometric observations targeting the CO(2--1) line ($\nu_{\rm rest}$ = 230.538\,GHz), calibrators used, as well as the resulting synthesized beams and rms noise levels.
The digital correlator was configured with 3072 channels (2\,GHz bandwidth), resulting in a velocity resolution of about 1\,\kms. Although the velocity resolution of our CO(2--1) observations is 1\,\kms, we have binned the data cubes to 20\,\kms\ to have better signal to noise ratio (S/N) { while still Nyquist sampling the CO(2--1) line}. Enough line-free channels are present to subtract the continuum emission. 						
The field of view is characterized by a Half Power Beam Width (HPBW) of the primary beam at 230\,GHz of the SMA 6-m antenna of $52\arcsec$ (18.3\,kpc for NGC\,3110, and 21.8\,kpc for NGC\,232), which is sufficient to cover in a single pointing the whole extent of the two galaxies. 

The data were reduced using the SMA-adapted MIR-IDL package\footnote{MIR is a software package to reduce SMA data based on the package originally developed by Nick Scoville at Caltech. See https://www.cfa.harvard.edu/$\sim$cqi/mircook.html.}.
The absolute flux scale for the data was determined by observing one planet or satellite, and quasars were used for the bandpass and the time variable gain corrections (phase and amplitude). By comparing with flux measurements of the phase calibrators close to the time of the observations, we estimate that the absolute flux uncertainty is of the order of 20\%.

The imaging of the CO(2--1) line was conducted in MIRIAD \citep{1995ASPC...77..433S}. The { continuum was fit with a constant offset} from
all the line-free channels with the task \verb!UVLIN! and subtracted from the line. The data were \verb!INVERT!ed using \verb!ROBUST! = 0.5. 
For NGC\,3110 the obtained synthesized beam is $3\farcs1 \times 2\farcs$8 (or $1.1 \times 1.0$\,kpc) with a major axis of P.A.\ = 10$^{\rm o}$.7 (East of north).  The  rms noise level is 14\,mJy\,beam$^{-1}$  for a 20~\kms\ channel. 
As for NGC\,232, the resulting synthesized beam is $3\farcs2 \times 2\farcs$4 (or $1.3 \times 1.0$\,kpc) with a major axis of P.A.\ = --86$^{\rm o}$.2.  
The rms noise level is found to be 15\,mJy\,beam$^{-1}$  for a 20~\kms\ channel. 

{ The dirty images were then deconvolved with the task \verb!CLEAN!. First we carried out the deconvolution without any supporting region for each channel. We created a mask with regions where emission was apparent, which were selected large enough (at least $\sim$10\arcsec\ diameter) not to miss any possible extended emission. The regions were refined manually in several iterations and the goodness of the selection was assessed by minimizing the noise level in line free regions of the channels, as well as reducing the contribution by sidelobes. The stopping criteria was a clean cutoff limit, 14\,mJy\,beam$^{-1}$ for NGC\,3110 and 15\,mJy\,beam$^{-1}$ for NGC\,232. The cumulative fluxes as a function of number of clean components were seen to converge in all channels. We considered other less restrictive cutoff limits but they were seen not to converge in channels with more complex distribution. The number of clean components ranged from $\sim$200 (in the channels corresponding to the edges of the profiles) to $\sim$800 in the case of NGC\,3110, and $\sim$ 200 -- 400 for NGC\,232. The final residual channel images looked like noise in all channels.}

\section{Results}
\label{result}

We present in this section the results obtained from the SMA CO(2--1) data, including spectra, channel and moment maps, as well as position-velocity (P--V) diagrams.

\subsection{CO(2--1) Profiles and Recovered Flux}
\label{subsect:co2-1emissionline}

\begin{figure*}
\centering
\includegraphics[width=8.4cm]{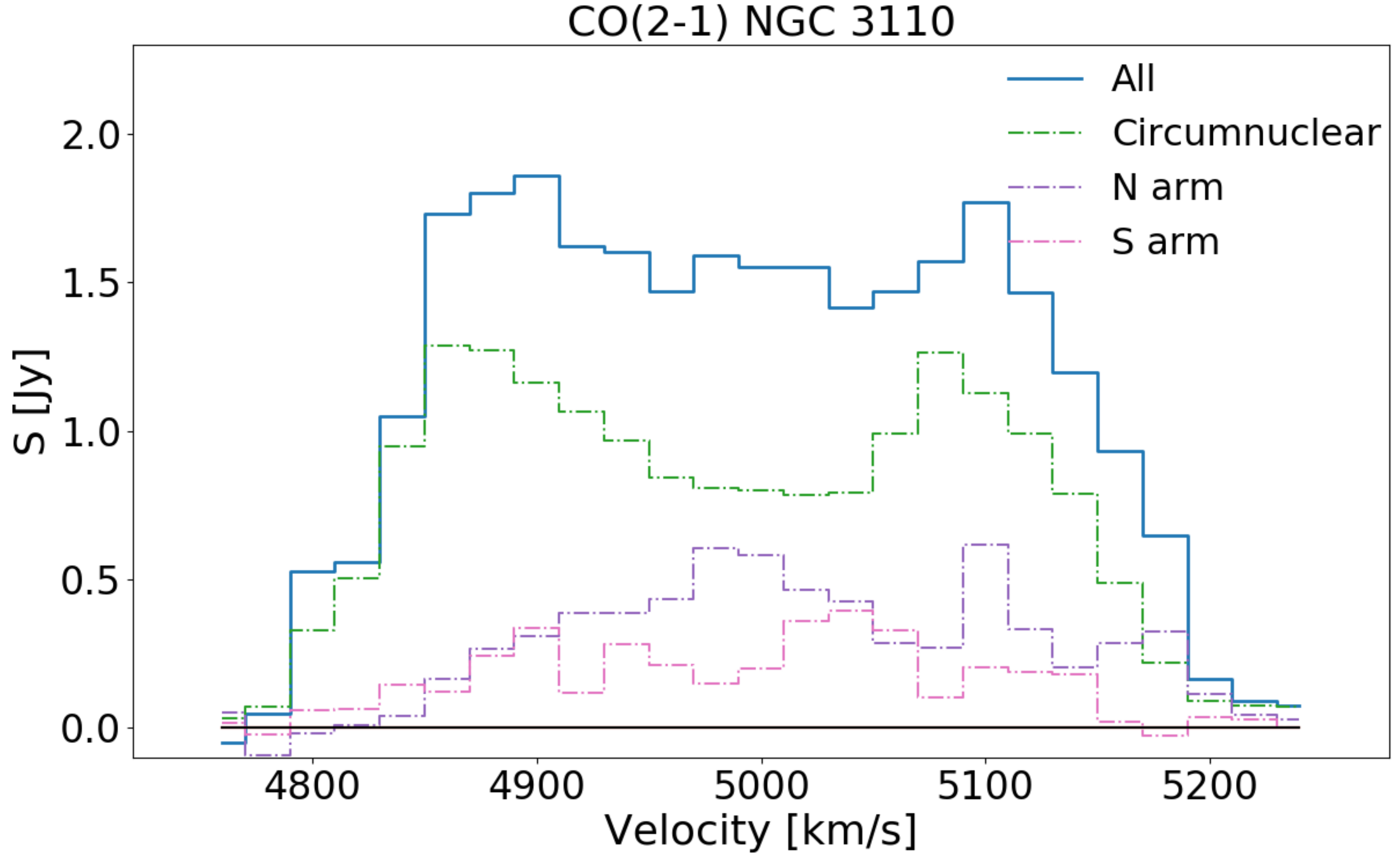}
\includegraphics[width=8.4cm]{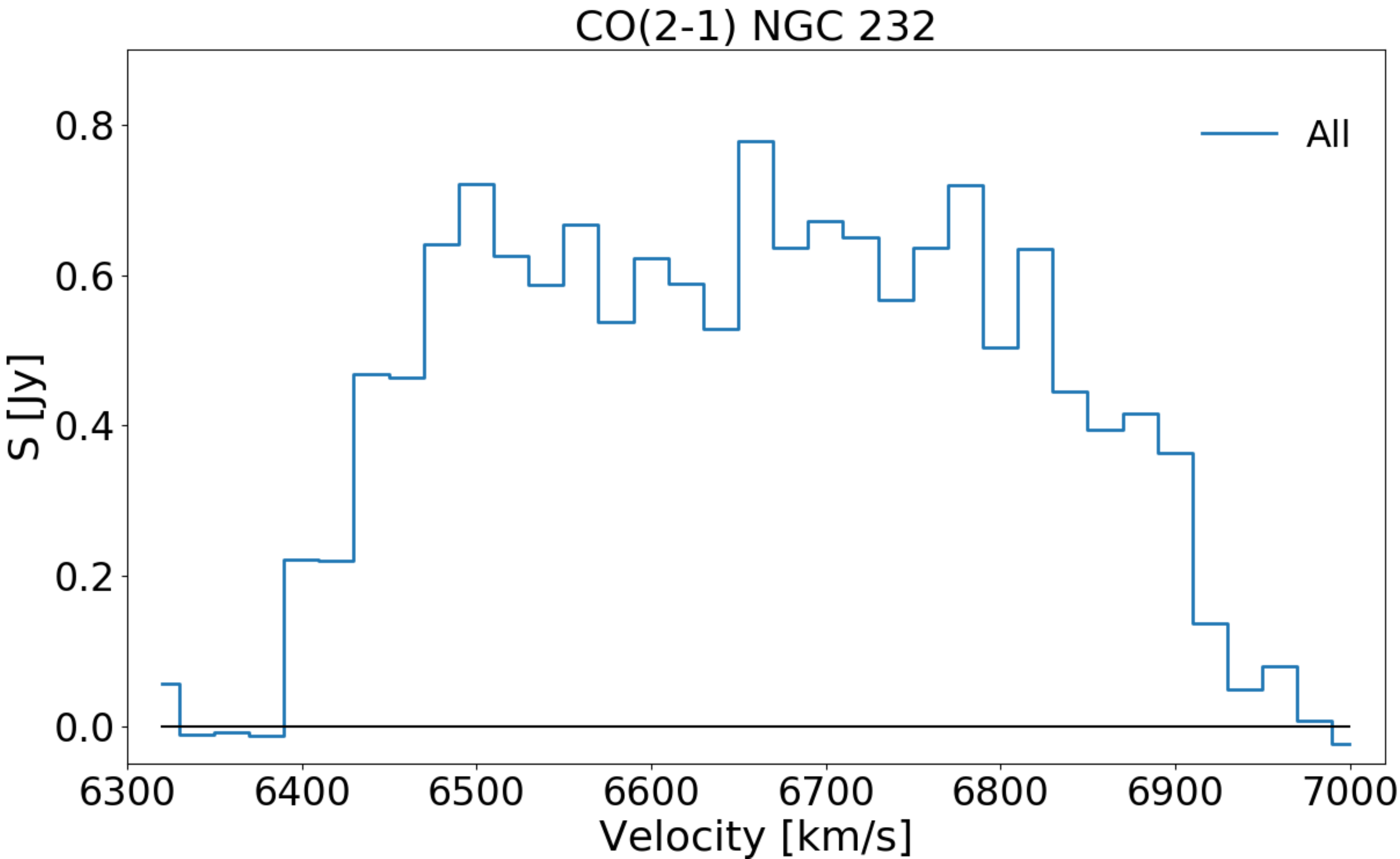}\\
\vspace{0.5cm}
\includegraphics[width=8.4cm]{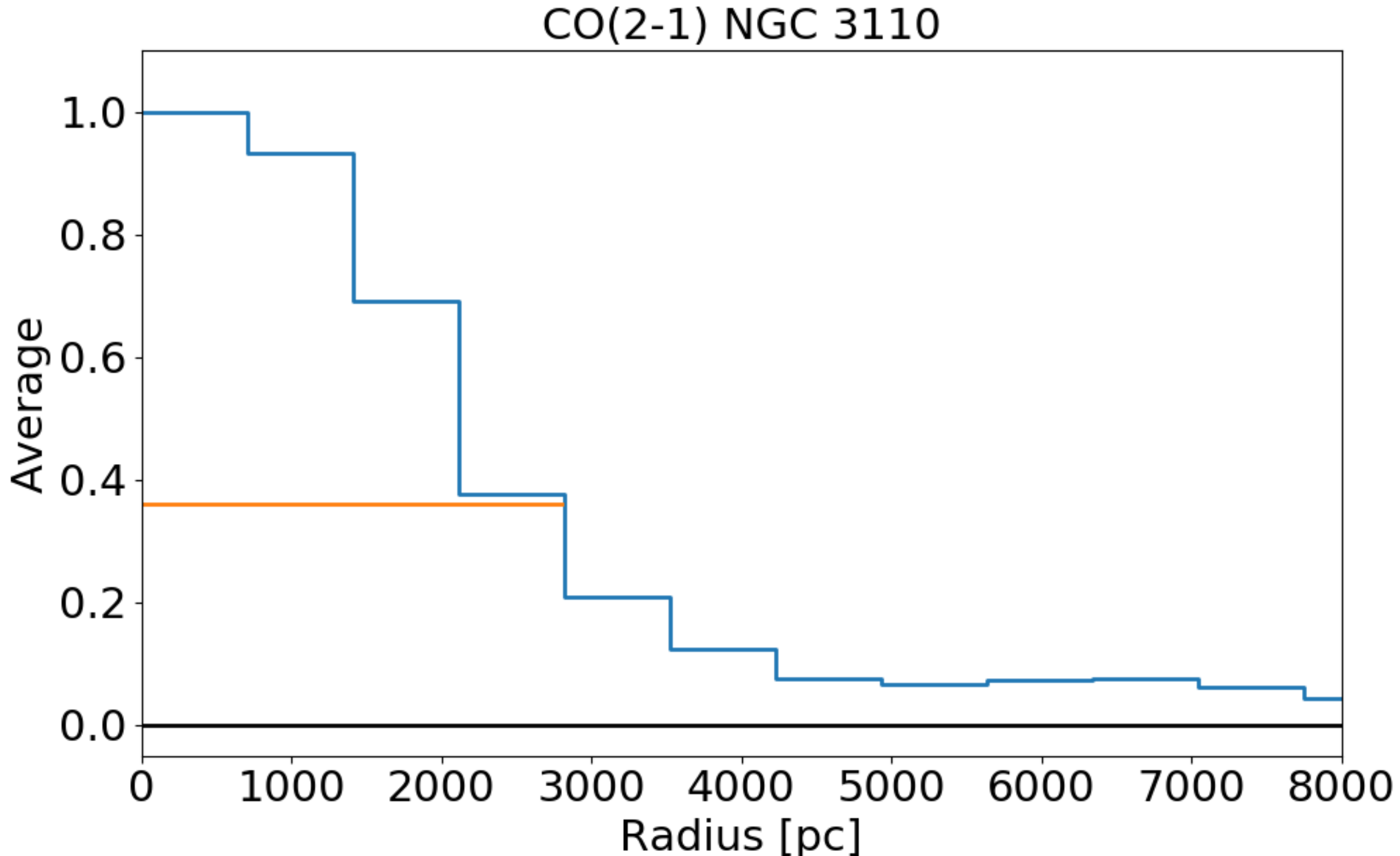}
\includegraphics[width=8.4cm]{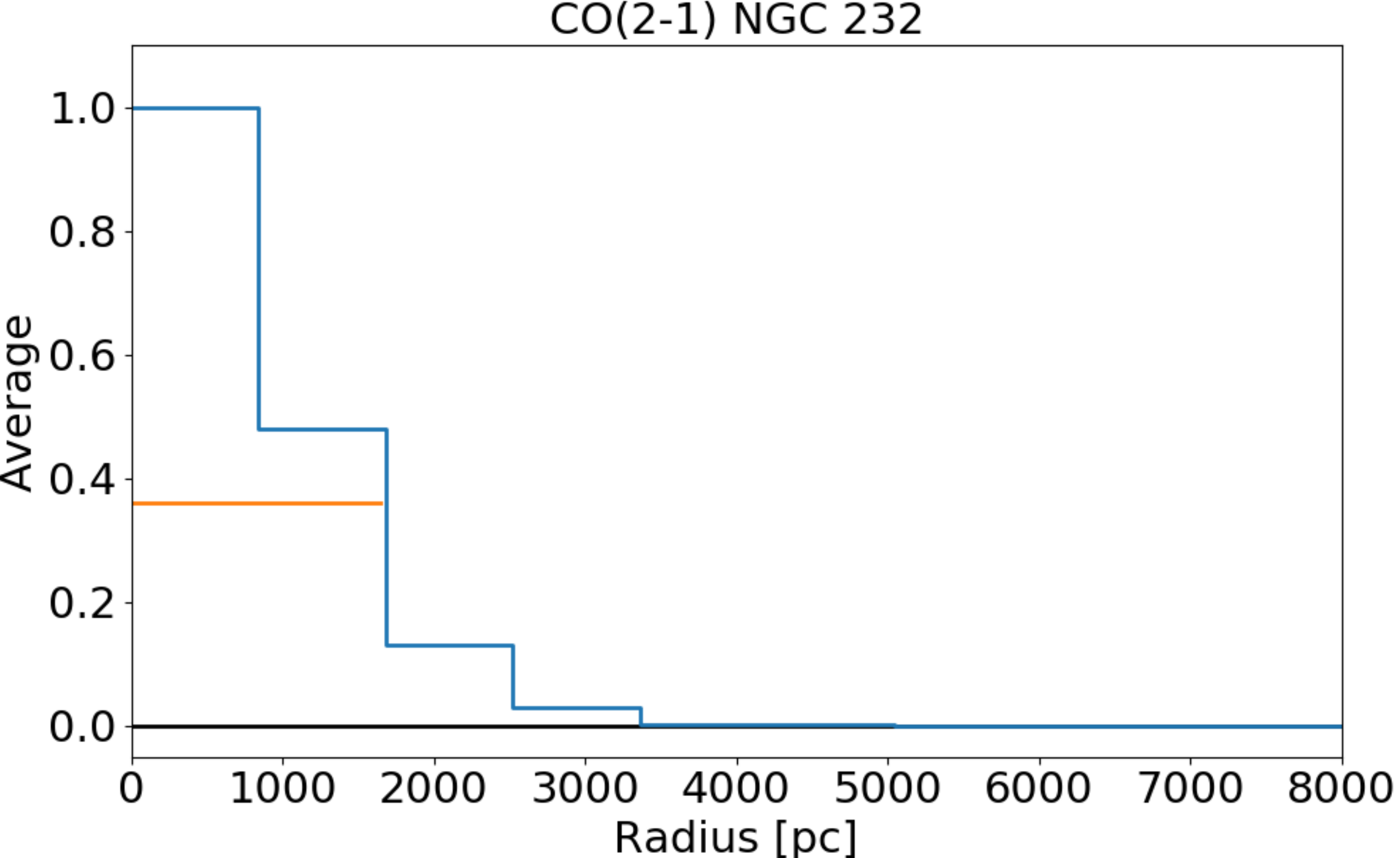}
\vspace{0.1cm}
\caption{ { Top)} CO(2--1) spectra for NGC\,3110 and NGC\,232. The y-axis shows fluxes in units of Jy and the x-axis the radio LSRK velocities in bins of 20\,\kms , as in the channel maps (Fig.~\ref{fig5} and \ref{fig6} for each galaxy). For NGC\,3110 (top left panel), dashed-dotted lines correspond to the { circumnuclear} region (green), the northern (purple) and southern (pink) arms. { Bottom)} The azimuthally averaged CO(2--1) normalized radial profiles for NGC\,3110 (left panel) and NGC\,232 (right).  The x-axis shows the radius from the center of the galaxies up to 8\,kpc and is expressed in pc.
The horizontal (red) line shows the width at 1/e of the central peak.
\label{fig4}}
\end{figure*}

In Fig.~\ref{fig4} we present the CO(2--1) spectra of both galaxies integrated over the area with detected emission within the HPBW of 52\arcsec\ at 230\,GHz. Note that { throughout this paper radial velocities are expressed in the radio definition and with respect to the Local Standard of Rest (LSRK frame)}. 
The CO(2--1) profiles of both objects are relatively flat. In the case of NGC\,3110, we can slightly discern a double peak, at 4900\,\kms\ and 5120\,\kms , the approaching side being brighter than the opposite one. 
Emission is detected from 4780\,\kms\ to 5240\,\kms\ in NGC\,3110, and from 6370\,\kms\ to 6970\,\kms\ in NGC\,232. The velocity widths are then $\Delta V$ = 460 $\pm$ 10\,\kms\ and 600 $\pm$ 10\,\,\kms , and if corrected by inclination (65\arcdeg\ and 47\arcdeg , Table\,\ref{tbl-1}),  $\Delta V_{\rm corr}$ = 508\,\kms\ and 820\,\kms , respectively.
The systemic velocities obtained from the average of the profile edges at a 20\% level of the peak are 4990 $\pm$ 10\,\kms\ and 6650 $\pm$ 10\,\kms\ for NGC\,3110 and NGC\,232, respectively. The systemic values are consistent with those in the literature if we correct the offsets due to the different velocity frame conventions.

To quantify how much flux is recovered by our interferometric observations, we compare the flux 
with previously obtained single-dish CO(2--1) data. 
The total SMA CO(2--1) flux measurements for NGC\,3110 and NGC\,232 corrected by the primary beam response are 679 and 295\,Jy\,\kms , respectively. 
Unfortunately no CO(2--1) single-dish data could be found in the literature for NGC\,3110. We use for this comparison the CO(1--0) single-dish integrated flux of 395\,Jy\,\kms\ obtained from the NRAO 12\,m spectrum presented in \citet{1991ApJ...370..158S}, with a 55\arcsec\ HPBW, which almost matches the SMA field of view, and we assume a Jy-to-K conversion factor of 30.4\,Jy/K. 
We then convert to the CO(2--1) integrated flux by assuming that the integrated intensity ratio between CO(2--1) and CO(1--0) (in Tmb scale) is $R_{2-1/1-0}$ $\sim$ 0.8. This is reasonable because the line intensity ratio R$_{3-2/1-0}$ within the inner 14\farcs5 is R$_{3-2/1-0}$ =  0.78 \citep{2003ApJ...588..771Y}. Assuming $R_{2-1/1-0}$ = 0.8, the CO(2--1) total flux yields 1264\,Jy\,\kms , and although with large uncertainties due to the various assumptions, we estimate that the recovered flux in our experiment is $\sim$ 54\%. 
The comparison for NGC\,232 is more straightforward because a direct CO(2--1) flux measurement exists in the literature. The CO(2--1) integrated intensity (in $T_{\rm mb}$ scale) obtained using the SEST telescope is 17.8\,K\,\kms\  with a beam of 23\arcsec\ \citep{1996A&AS..118...47C,2007A&A...462..575A}. Given the compact CO(2--1) distribution seen in the SMA maps, as well as in the H$\alpha$ maps, it is reasonable to think that most of the emission will arise well within the SEST beam. The total flux is then 409\,Jy\,\kms\  assuming a 23\,Jy/K conversion factor. We thus estimate that $\sim$ 70\% of the flux is recovered.

{ We also tested the existence of missing flux in the SMA data by tapering visibilities to increase the brightness sensitivities. We found that for NGC\,3110 a taper resulting in angular resolutions  of 5\arcsec\ (and 7\arcsec ) increase the total integrated flux by 7\,\%. Flux loss was found to be up to $\sim$20\,\% in some channels $\pm$ 100\,\kms\ of the systemic velocity where there is more complex distribution. This effect was not noticeable at channels corresponding to the profile edges as a result of the more compact distribution there.  As for NGC\,232, no clear difference in total integrated flux was found for tapers at 5\arcsec\ and 7\arcsec .}

\subsection{CO(2--1) Interferometric Maps}
\label{subsec:cointmaps}

\begin{figure*}
\begin{center}
\includegraphics[width=16cm]{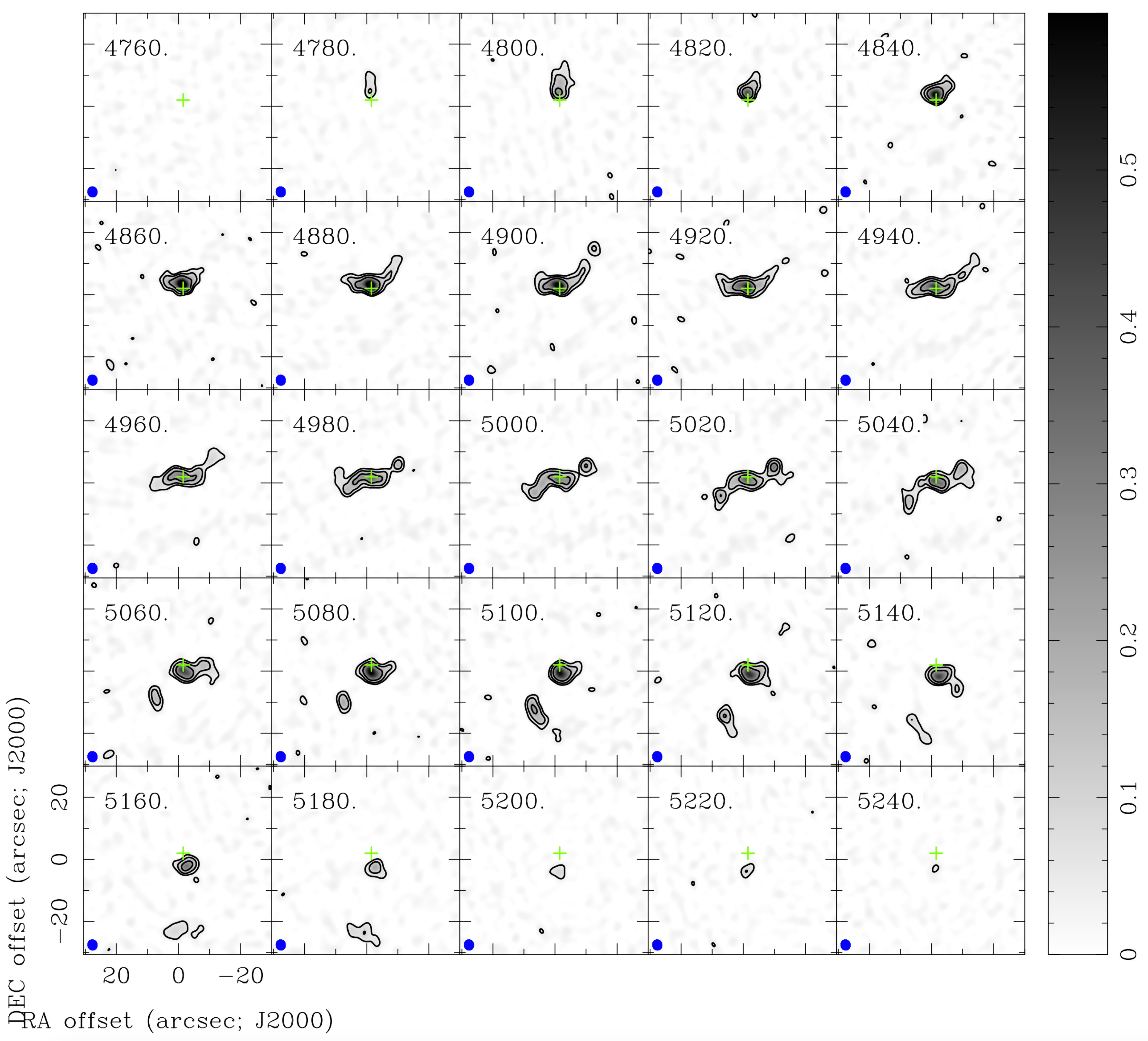}
\end{center}
\caption{CO(2--1) channel maps of NGC\,3110 in the velocity range  $4760 - 5240$~\kms\ (radio LSRK) in 20~km~s$^{-1}$ bins. The size of the maps is 52\arcsec. The velocities are shown in the upper right corner of each panel and the synthesized beam is shown in the bottom left corner.  The rms noise of an individual channel is $\sigma$ = 0.014~Jy~beam$^{-1}$.  The contour levels are at 3, 7, and 14 $\sigma$. The gray scale ranges from 0 to 0.6\,Jy/beam. The cross sign shows the center of the galaxy.
\label{fig5}}
\end{figure*}

\begin{figure*}
\begin{center}
\includegraphics[width=16cm]{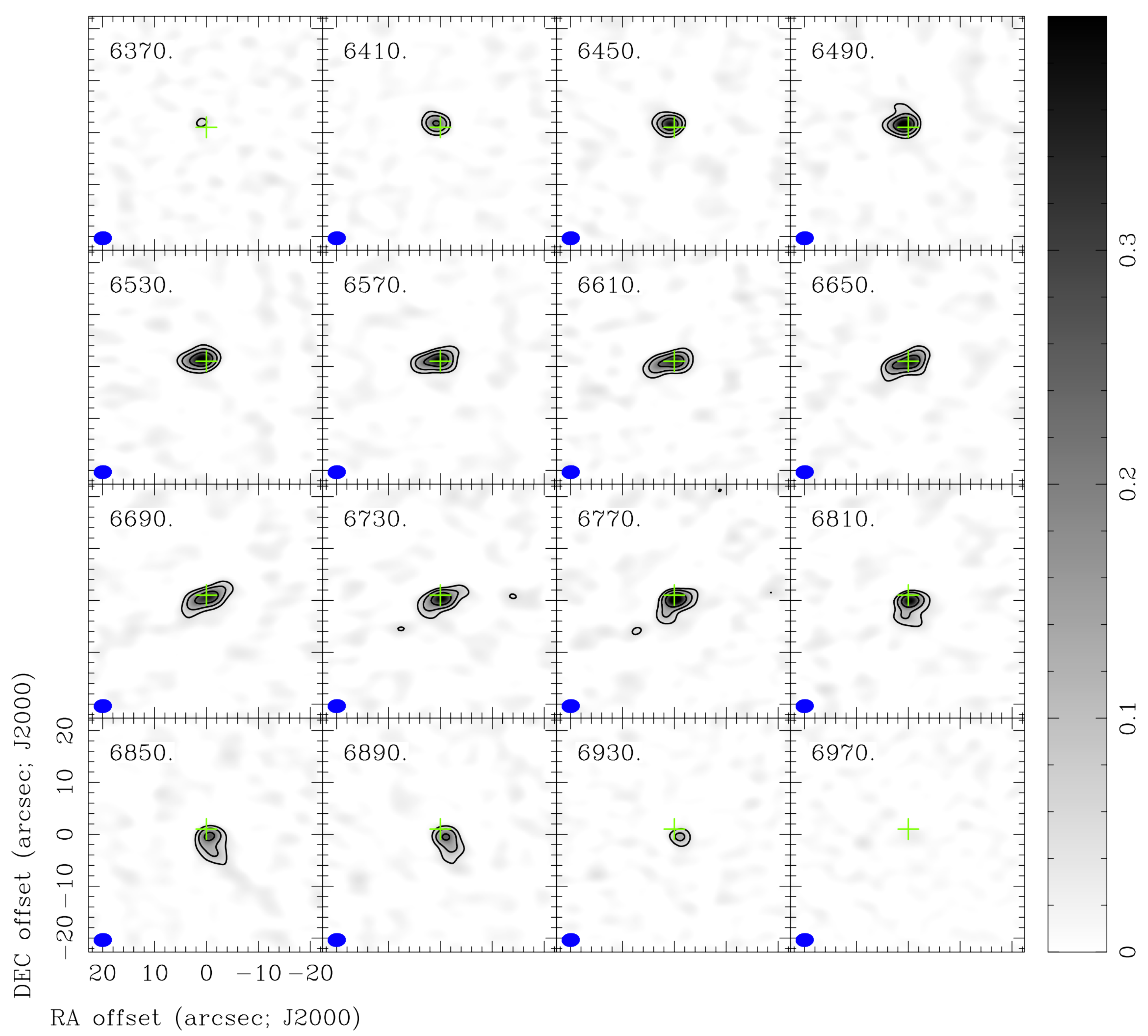}
\end{center}
\caption{CO(2--1) channel maps of NGC\,232 in the velocity range  $6370 - 6970$~\kms\ (radio LSRK) in 40~km~s$^{-1}$ bins. The size of the maps is 45\arcsec. The velocities are shown in the upper right corner of each panel and the synthesized beam is shown in the bottom left corner.  The rms noise of an individual channel is $\sigma$ = 0.011~Jy~beam$^{-1}$.  The contour levels are at 3, 7, and 14 $\sigma$. The gray scale ranges from 0 to 0.4\,Jy/beam. The cross sign shows the center of the galaxy.
\label{fig6}}
\end{figure*}

Fig.~\ref{fig5} shows the CO(2--1) channel maps of NGC\,3110. 
The CO(2--1) channel maps show blue-shifted (approaching) emission on the N and red-shifted (receding) emission on the S. We distinguish in these maps an inner circumnuclear component within $\sim$10\arcsec\ in diameter (3.5\,kpc), which is the fastest rotating feature.  A more extended component is composed of two main spiral arms which wind up and are linked to the circumnuclear feature.
Fig.~\ref{fig6} shows the CO(2--1) channel maps of NGC\,232. 
The maps show a compact, although resolved, disk-like structure with blue-shifted emission on the NE and red-shifted on the SW. 

\begin{figure*}
\begin{center}
\includegraphics[width=8cm]{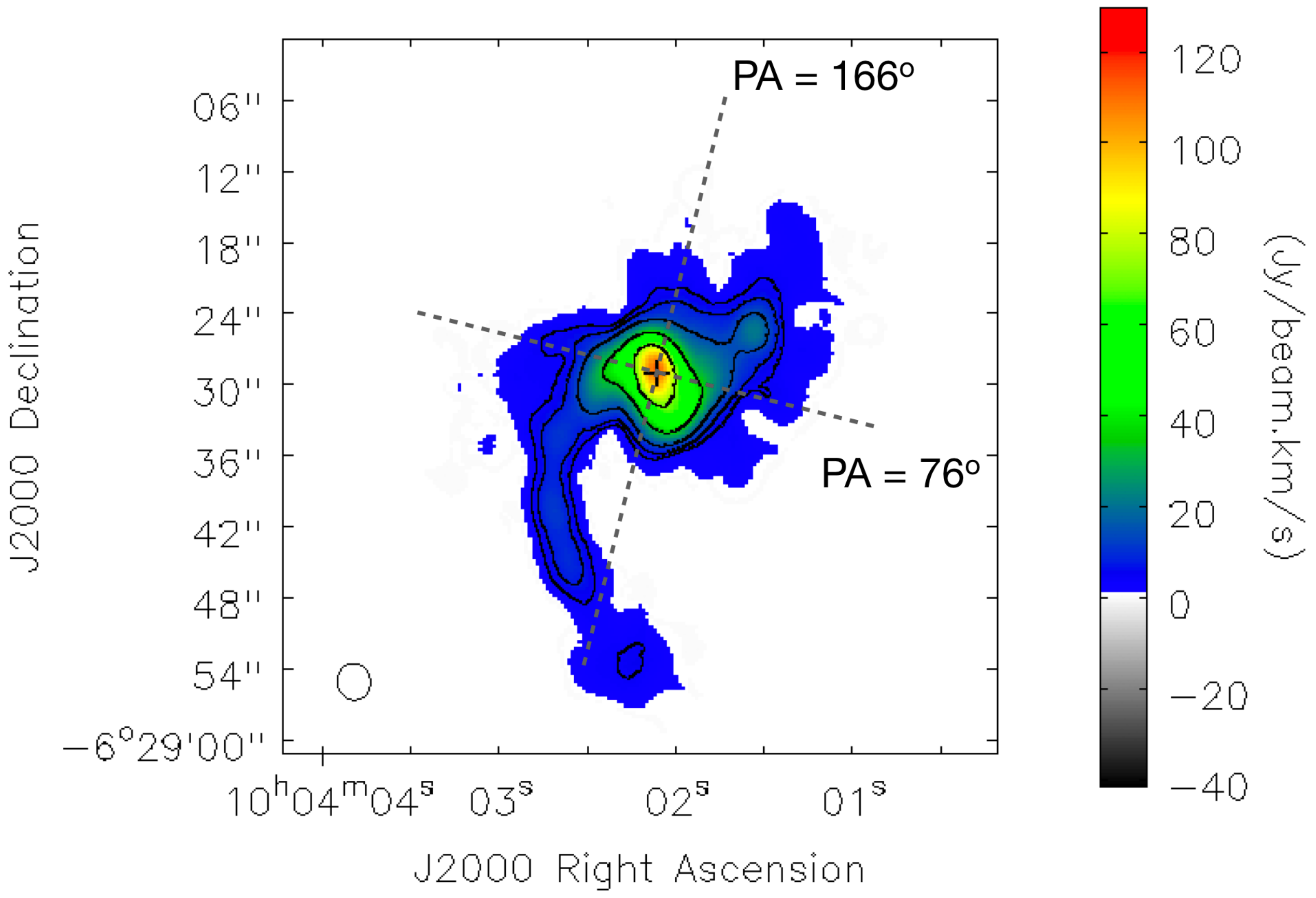}
\includegraphics[width=8.3cm]{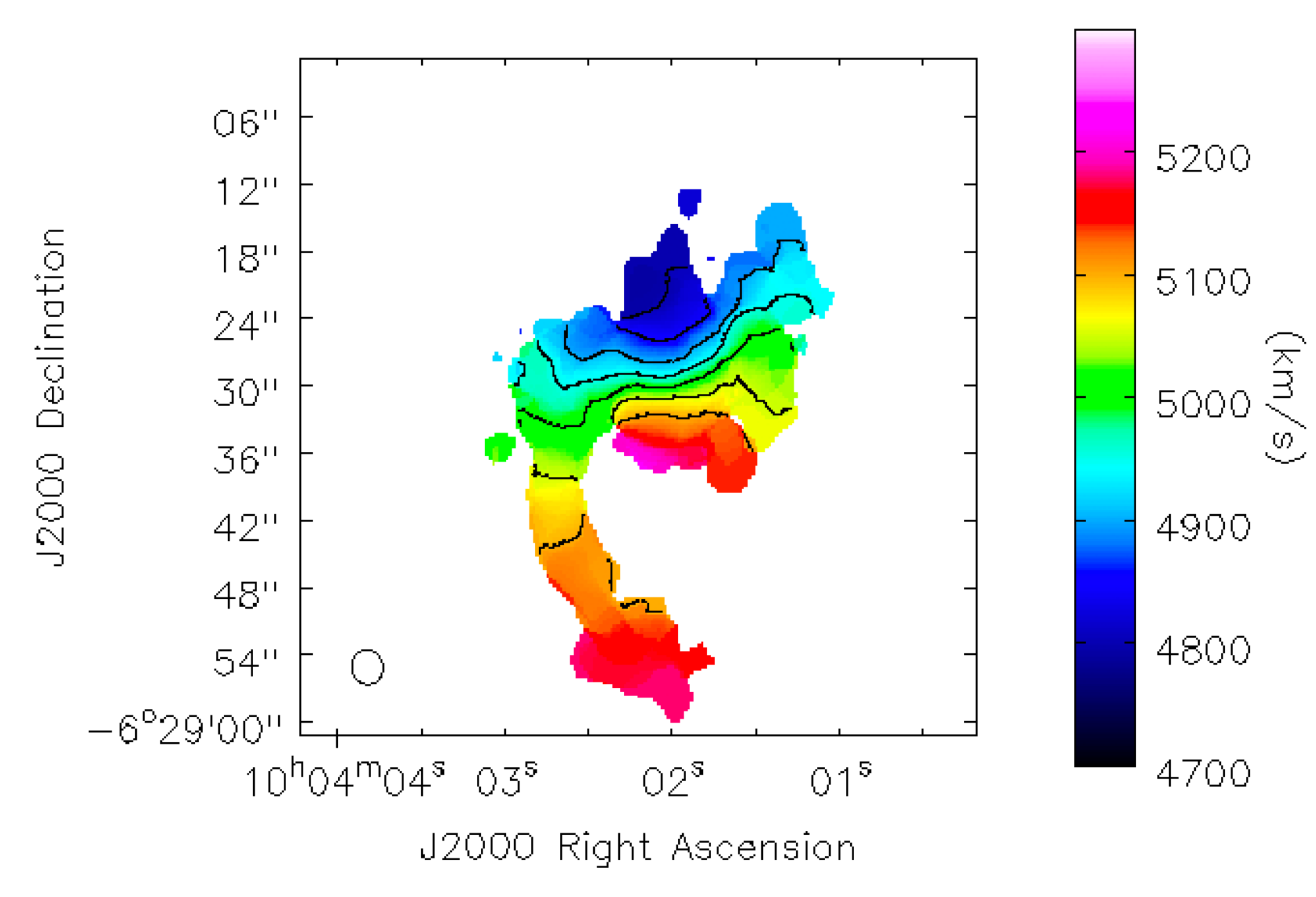}
\includegraphics[width=8.3cm]{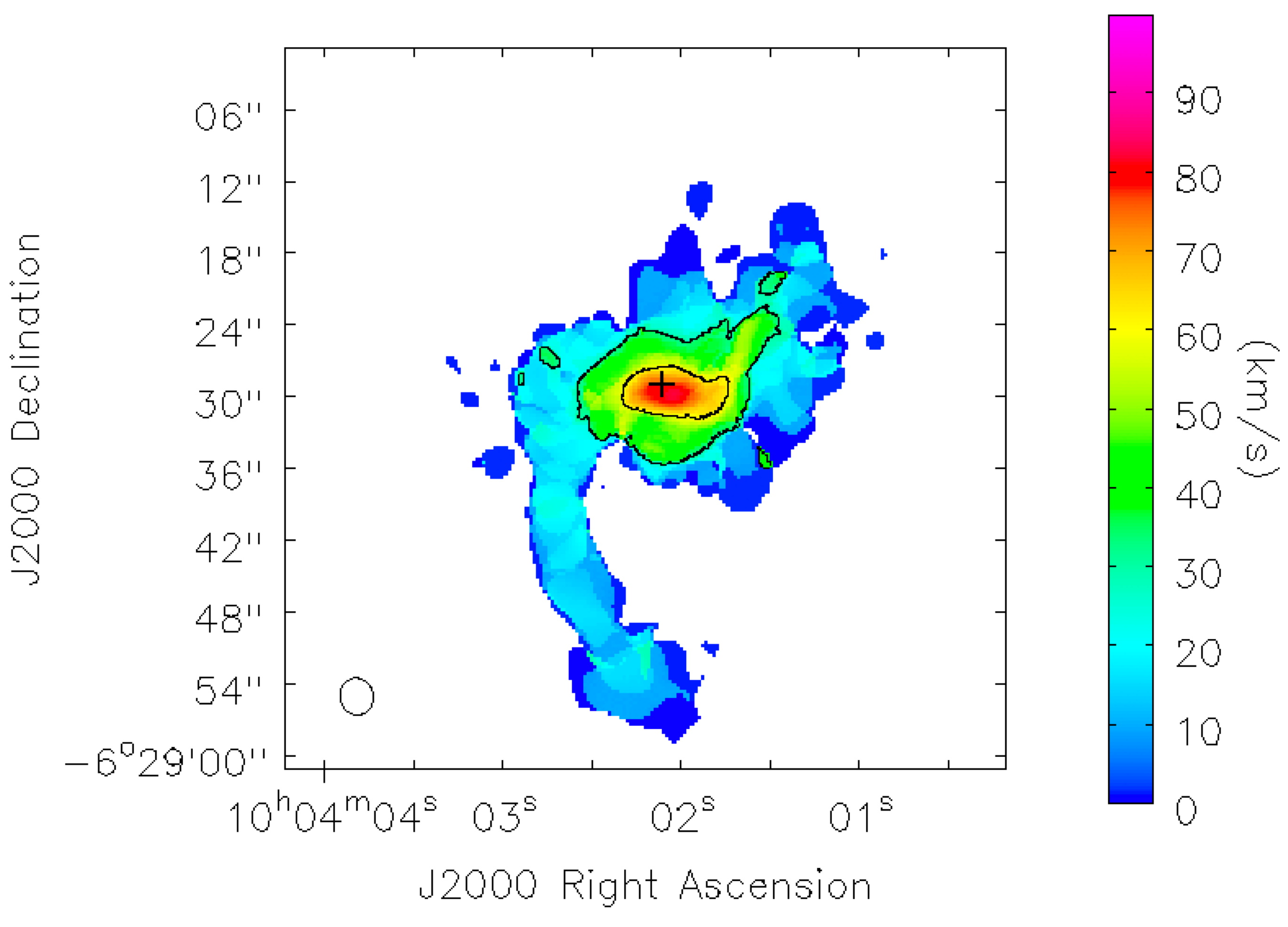}
\end{center}
\caption{ The integrated intensity map (upper left), the intensity-weighted velocity field (upper right), and the intensity-weighted velocity dispersion map (bottom) of NGC\,3110. The color scale is shown on the right side of each panel and the HPBW of the synthesized beam (3\farcs1 $\times$ 2\farcs8) is indicated as an ellipse at the bottom left.  The cross sign indicates the center of the galaxy.
Contour levels in the integrated intensity map are at 3, 5, 10, 25, and 50$\sigma$, where $\sigma$ = 1.4\,Jy\,beam$^{-1}$\,\kms . 
Contour levels in the velocity field are from 4750 to 5100\,\kms\ in bins of 50\,\kms. Contour levels in the velocity dispersion map range from 0 to 100\,\kms\ in 30\,\kms\ bins. 
The dashed lines in the moment 0 panel indicate the cuts for the two position-velocity diagrams in Fig.\,\ref{fig-pv}. 
\label{fig7}}
\end{figure*}

\begin{figure*}
\centering
\includegraphics[width=8cm]{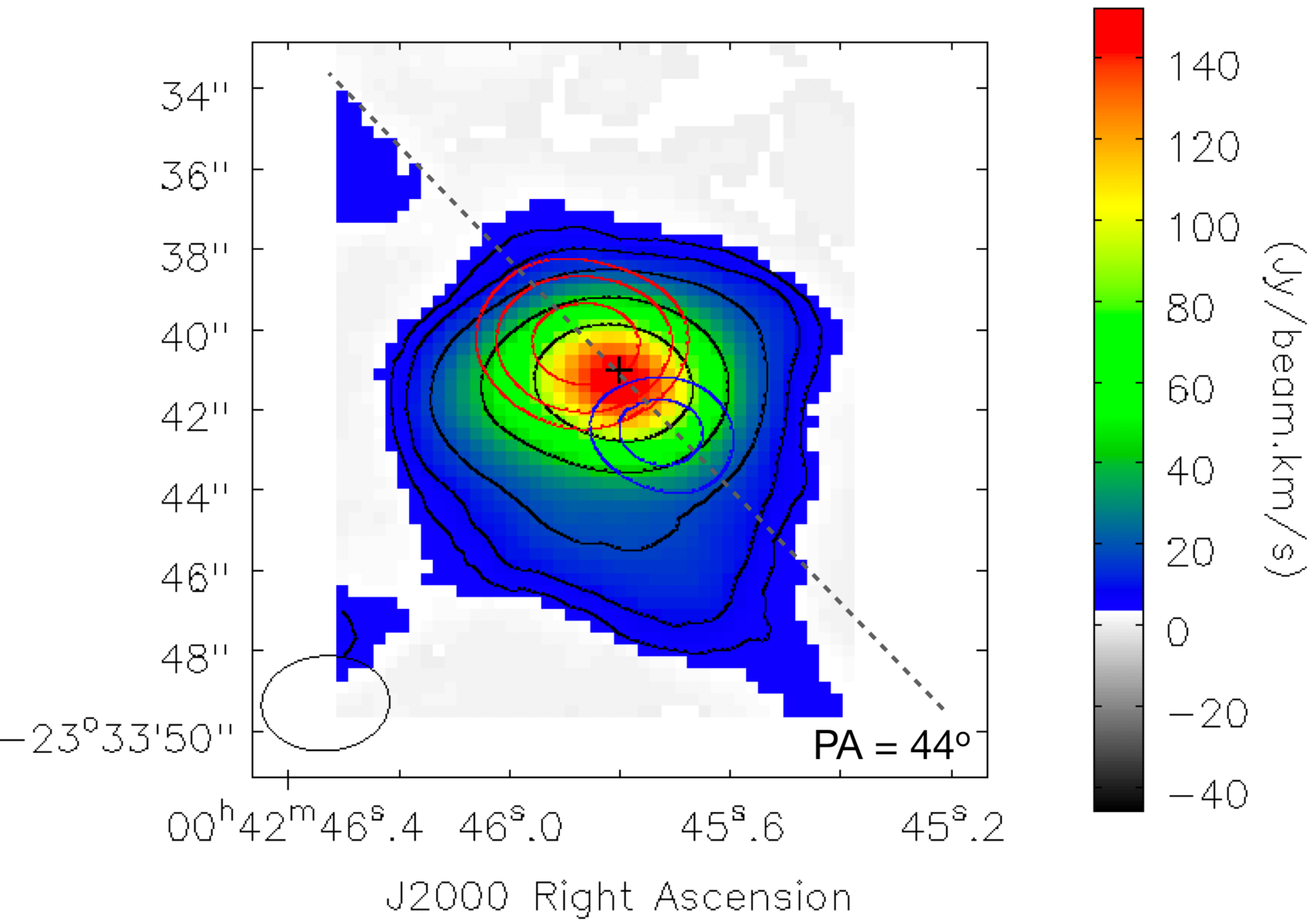}
\includegraphics[width=8.4cm]{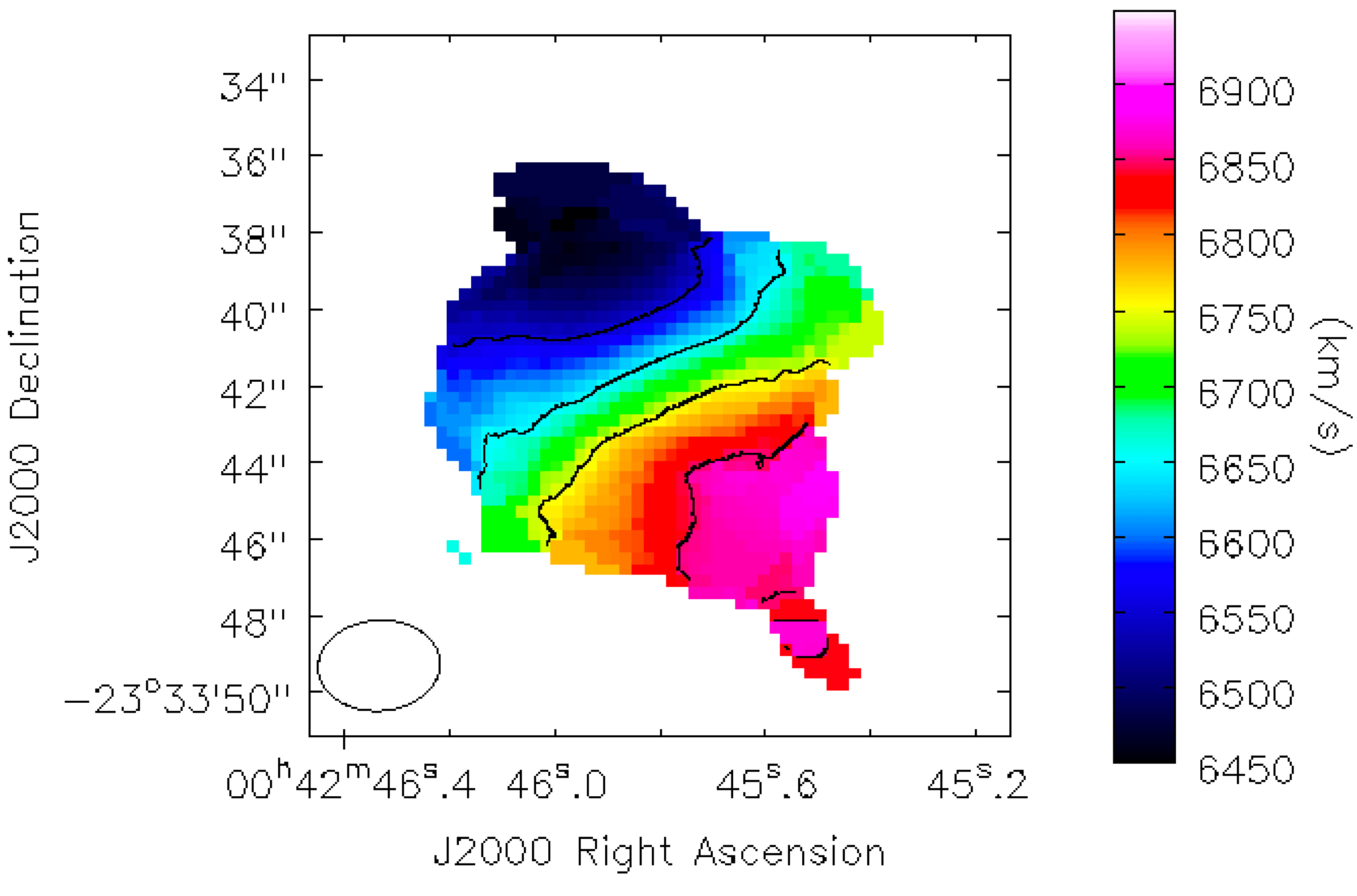}
\includegraphics[width=8.4cm]{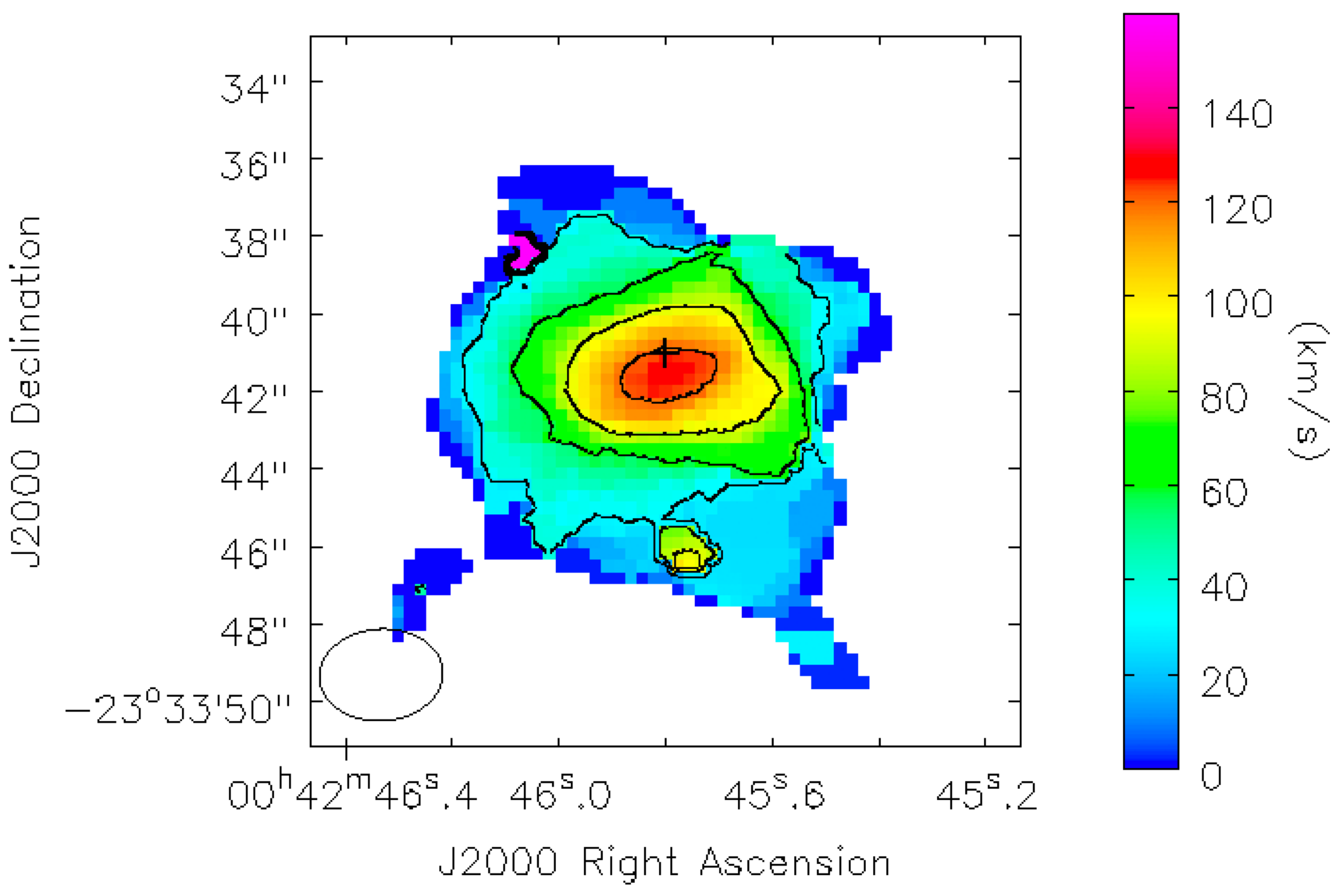}
\caption{ The integrated intensity map (upper left), the intensity-weighted velocity field (upper right), and the intensity-weighted velocity dispersion map (bottom) of NGC\,232. The color scale is shown on the right side of each panel and the HPBW of the synthesized beam (3\farcs2 $\times$ 2\farcs4) at the bottom left.  The cross sign indicates the center of the galaxy.
Contour levels in the integrated intensity map are at 3, 5, 10, 25, and 50$\sigma$, where $\sigma$ = 1.7\,Jy\,beam$^{-1}$\kms. Contours for the approaching (blue, $V$ $<$ 6460\,\kms) and receding (red, $V$ $>$ 6900\,\kms) wings of the profile are also shown. The contour levels for the approaching contours are 3, 5, and 10$\sigma$, and for the receding ones 3 and 5$\sigma$. 
Contour levels in the velocity field are 6550, 6650, 6750, and 6850\,\kms . Contour levels in the velocity dispersion map range from 0 to 120\,\kms\ in 30\,\kms\ intervals. The dashed line in the moment 0 panel  indicate the cut for the position-velocity diagram in Fig.\,\ref{fig-pv2}.
\label{fig8}}
\end{figure*}

The task \verb!IMMOMENT! in CASA \citep{2007ASPC..376..127M} was used to calculate the integrated-intensity maps, intensity-weighted velocity fields, and velocity dispersion distributions. { We applied masks to each channel as in the deconvolution process}. The latter two moment maps were clipped at 2 and 3 times the rms noise for NGC\,3110 and NGC\,232, respectively. Note that no primary beam correction was performed in the presented integrated-intensity map. The moment maps for NGC~3110 and NGC~232 are displayed in Figs.~\ref{fig7} and \ref{fig8}, respectively. In Fig.~\ref{fig2} we show the same CO(2--1) moment 0 contours over the H$\alpha$ maps.
There is a large concentration of molecular gas towards the inner few kiloparsecs of these two galaxies. The concentration width scale ($\sigma$)  obtained from a Gaussian fit on the azimuthally averaged intensity radial profile is found to be 1.4\,kpc and 0.8\,kpc, comparable to the resolution element, and at 1/e level with respect to the central peak the CO scale length is 2.8\,kpc and 1.7\,kpc, respectively. The radial profiles are shown in Fig.\,\ref{fig4} (bottom).

The CO(2--1) moment 0 map of NGC\,3110 in Fig.\,\ref{fig7} reveals the whole extent and morphology of the two asymmetric molecular spiral arms, one to the S and a relatively weaker and shorter one to the N, which wind up into the circumnuclear molecular gas component. The circumnuclear component is elongated and extends about 10\arcsec\ diameter in the NE to SW direction. The southern arm is on the receding side of the galaxy, emerging to the E of the central component and extending up to 26\arcsec\ (9.2\,kpc) from the nucleus. The northern molecular arm emerging from the W of the central component 
 is, on the other hand, much shorter, 18\arcsec\ (6.3\,kpc) from the nucleus, and its opening angle differs from that of the southern counterpart. The southern arm is particularly narrow and it is barely resolved in the perpendicular direction with the SMA data. The northern arm is more asymmetrically distributed and appears to be wider than the southern arm.

The CO emission of NGC\,3110 peaks very close to the center of the galaxy, but the kinematic center as seen from the CO(2--1) maps is offset by about 1\arcsec\ to the SE. This central component is characterized by the largest velocity gradient, as seen in the velocity dispersion map in Fig.\,\ref{fig7}, with typical values of 40\,\kms\ within the inner 5\arcsec\ (1.8\,kpc) (as opposed to the 10 -- 20\,\kms\ found along the spiral arms) and even $\geq$ 80\,\kms\ in the central 3\arcsec $\times$ 7\arcsec\ along the E-W direction. The circumnuclear molecular gas component in the inner few kiloparsecs has a P.A. $\simeq$ 10\arcdeg . This component is aligned with the brightest nuclear barred feature seen in optical/NIR emission (see \S\,\ref{intro3110}). The velocity width at zero intensity is the largest, $\Delta V$ = 460\,\kms , along this direction and the size of the circumnuclear component is about 10\arcsec\ $\times$ 3\arcsec\ (3.5 $\times$ 1.1\,kpc). Non-circular motions (and/or warp) within the circumnuclear component are clear as seen from the twisted morphology in the moment 1 map. The second brightest component in the CO(2--1) map is part of the northern arm, at a distance of 9\arcsec (3.2\,kpc) to the NW from the center. It has a relatively symmetric counterpart in the southern spiral arm as seen in the channels 5020-5040\,\kms\ in Fig.\,\ref{fig5}, although not as bright.

The CO(2--1) maps of NGC\,232 show that most of the emission is centrally concentrated within the inner 6\arcsec\  (2.5\,kpc), which we will refer to as the circumnuclear disk, although there is a more extended component of size $\sim$10\arcsec\ (4.2\,kpc), slightly elongated along the SW - NE direction. It is { well-centered on} the brightest nuclear component as seen in optical and NIR images. 
The P.A. of the kinematical axis of the inner portion is about 35$\arcdeg$ different to that of the most extended component, and is
almost perpendicular to the axis of the putative bar at P.A. = 140$\arcdeg$ (see \S\,\ref{intro232}). To our sensitivity limit we do not detect molecular gas external to the circumnuclear component in the form of spiral arms.
 An S-shape morphology is also seen in the velocity field (Fig.\,\ref{fig8}), which may suggest either non-circular motions or a warp. { This,} together with the nearly perpendicular barred-like feature seen in the optical/NIR images { indicates} that the circumnuclear disk may represent the family of x$_2$ orbits in a bar potential. 
 The moment 2 map (Fig.\,\ref{fig8}) shows large velocity dispersions within the circumnuclear disk up to 130\,\kms . 
 The CO(2--1) profile is very wide, 600\,\kms (or 820\,\kms\ if corrected by inclination) at full-width zero intensity, which is difficult to explain simply by non-circular motions. The emission corresponding to the wings of the spectrum ($V$ $<$ 6460\,\kms\ 
and $V$ $>$ 6900\,\kms ) 
are distributed along a P.A. = 44\arcdeg (see the high velocity component contour maps in the moment 0 panel of Fig\,\ref{fig8}) , and we speculate that it might be an outflowing component. Follow-up observations with higher angular resolutions are needed to confirm this.

\begin{figure*}
\begin{center}
\includegraphics[width=8.5cm]{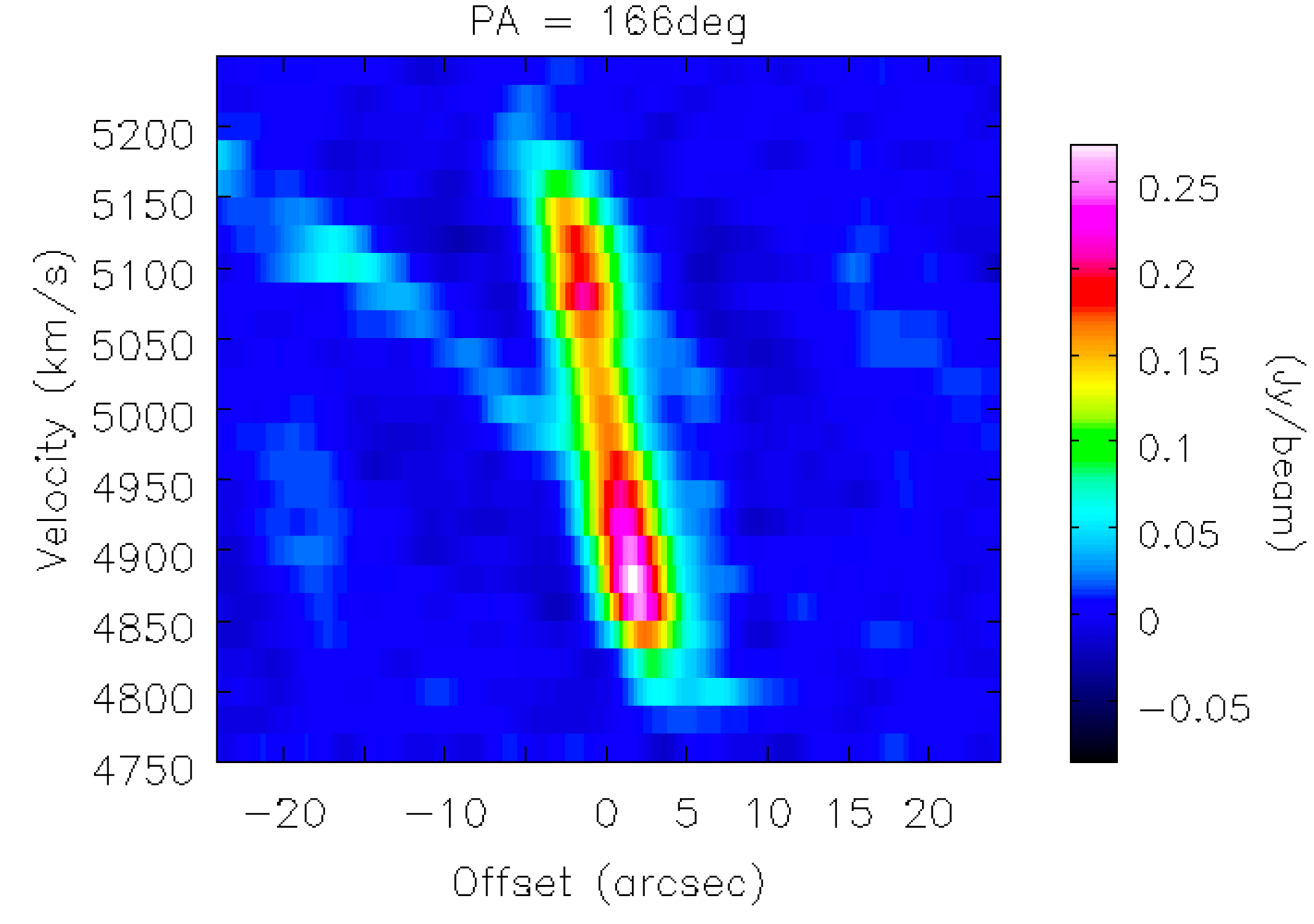}
\includegraphics[width=9cm]{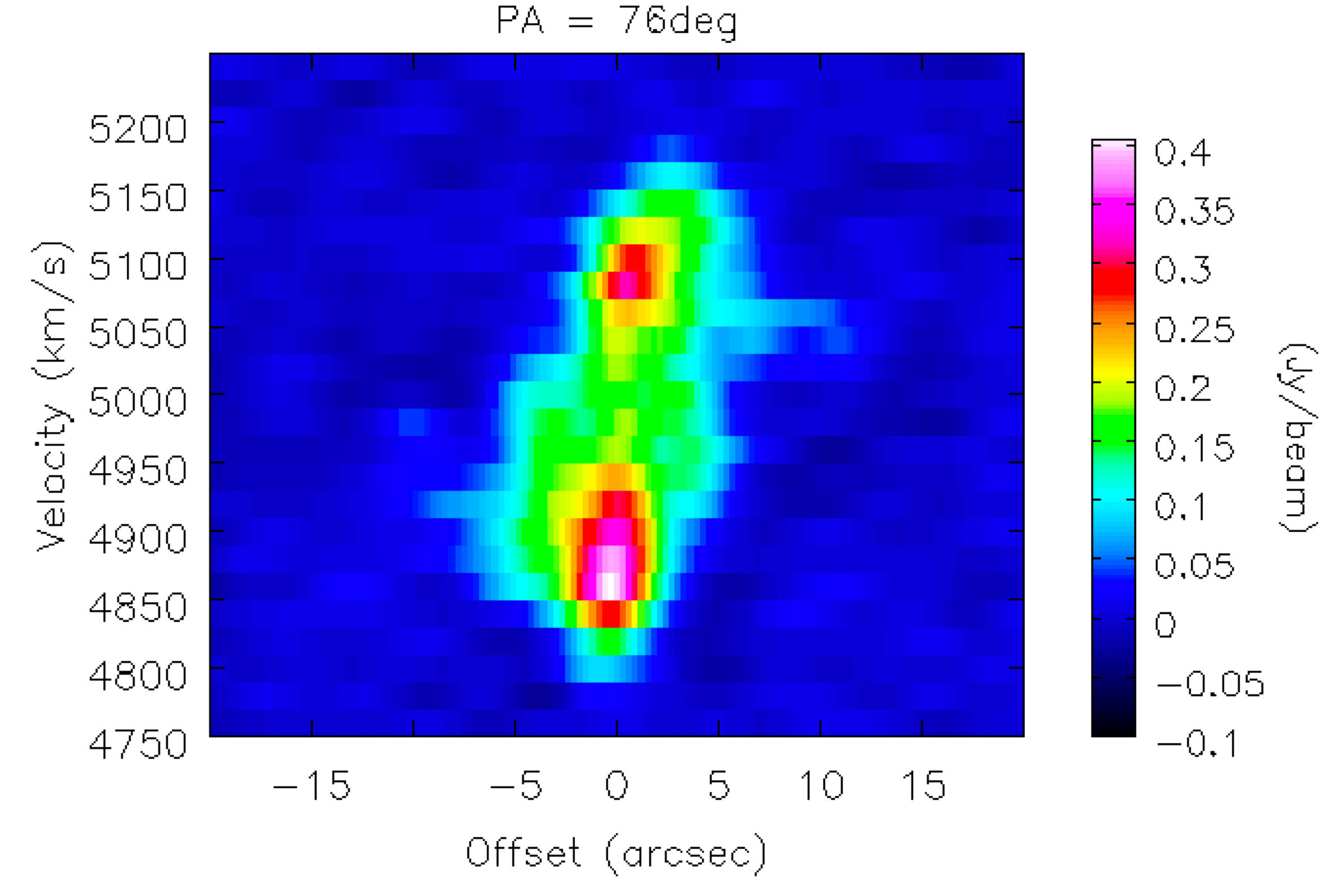}
\end{center}
\caption{ CO(2--1) Position - Velocity (P--V) diagrams of NGC\,3110: left) at PA = +166\arcdeg, length of 50\arcsec , and averaging width 16\arcsec,  right) at PA = 76\arcdeg, length of 40\arcsec , and averaging width of 7\arcsec . The color scales are shown on the right side of each panel. 
\label{fig-pv}}
\end{figure*}

\begin{figure*}
\begin{center}
\includegraphics[width=8.5cm]{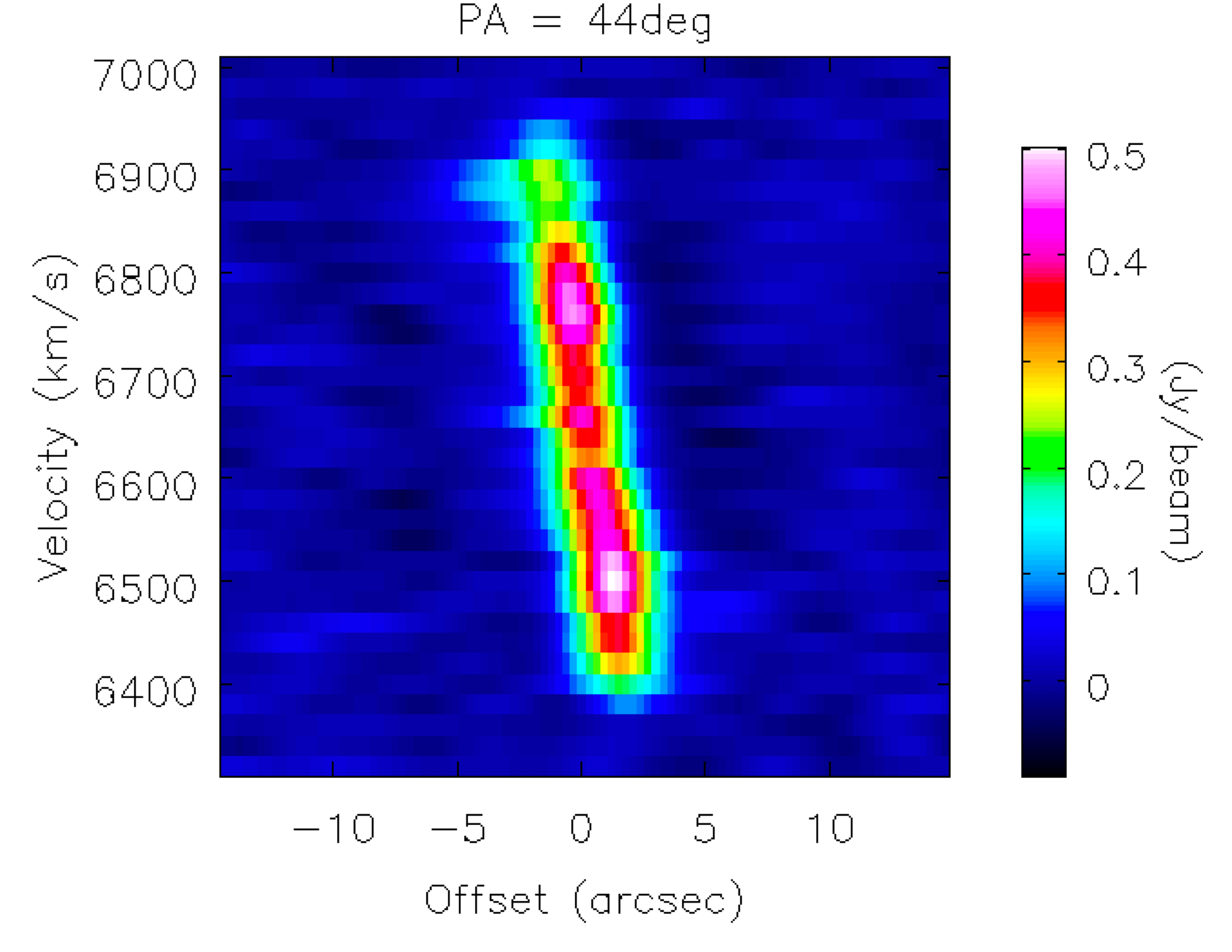}
\end{center}
\caption{  CO(2--1) Position - Velocity (P--V) diagram of NGC\,232 for a cut at PA = 44\arcdeg, length of 30\arcsec , and averaging width of 5\arcsec . The color scale is shown on the right side.
\label{fig-pv2}}
\end{figure*}

Finally, position-velocity (P--V) diagrams are shown in Fig.~\ref{fig-pv} for NGC\,3110 and in Fig.~\ref{fig-pv2} for NGC\,232.  Fig.~\ref{fig-pv} a) and b) show P--V cuts along P.A.\ = $166\arcdeg$ and perpendicular to it, $76\arcdeg$, both centered at NGC\,3110's nucleus. 
The slope in the central molecular component is $\Delta V/\Delta r$ $\simeq$ 0.15\,km~s$^{-1}$~pc$^{-1}$, as opposed to the slower rotation of the external CO components associated with the spiral arms. Note that one can see a distinct component with a smaller slope  on the left side of the P.A.\ = $166\arcdeg$ P--V diagram which corresponds to the southern arm.
  Fig.~\ref{fig-pv2} shows the P--V cut along P.A. = $44\arcdeg$ of NGC\,232.
  The slope is steeper in this case, $\Delta V/\Delta r$ $\simeq$ 0.4\,km~s$^{-1}$~pc$^{-1}$.
Note that beam smearing due to the relatively coarse spatial resolution likely makes the velocities of these curves lower limits. 

\subsection{Molecular Gas Mass}
\label{sub:moleculargasmass}

First we obtain the CO(2--1) luminosity, given by $L_{\rm CO(2-1)} = (c^2/2k_B)\,S_{\rm CO(2-1)}\,\nu_{\rm obs}^{-2}\,D_{\rm L}^2$, 
where $c$ is the light speed, 
$k_B$ is the Boltzmann constant, 
$S_{\rm CO(2-1)}$ is the integrated CO(2--1) line flux in Jy\,\kms, 
$\nu_{\rm obs}$ is the observed rest frequency in GHz, 230.538\,GHz, and 
$D_{\rm L}$ is the luminosity distance to the source in Mpc \citep[e.g.][]{2005ARA&A..43..677S}. 
It yields $L_{\rm CO(2-1)} = 611 \times S_{\rm CO(2-1)}\,D_{\rm L}^2$ [K\,\kms\,pc$^2$]. 
We use a conversion factor between the CO integrated intensity and H$_2$ column density of $X_{\rm CO}$
  $= N_{\rm H_2} /I _{\rm CO}$ = 2.0 $\times$ 10$^{20}$~cm$^{-2}$~(K~km~s$^{-1}$)$^{-1}$.
  This value is recommended for our Milky Way \citep[e.g.][]{2001ApJ...547..792D} and 'normal' (i.e. non-starbursting) galaxies \citep{2013ARA&A..51..207B}. 
 A caveat is that lower $X_{\rm CO}$ can be present in starburst systems \citep{2013ARA&A..51..207B,2003ApJ...588..771Y,2012ApJ...751...10P}, but the scatter found in these systems is large and it is not clear what the appropriate value might be for these two objects under study.
 We use for NGC\,3110 the integrated intensity CO(1--0) to CO(2--1) line ratio $R_{2-1/1-0}$ = 0.8 (see \S.~\ref{subsect:co2-1emissionline}), and for NGC\,232 $R_{2-1/1-0}$ = 0.46,  
as calculated from IRAM 30m CO(1--0) and SEST CO(2--1) observations \citep{1996A&AS..118...47C,2007A&A...462..575A}, where the calculated integrated intensities (in Tmb scale) were 38.2\,K\,\kms\ and 17.8\,K\,\kms, respectively. 

Finally we calculate the molecular gas masses as $M_{\rm mol} [{\rm M}_\odot] = 4.3 \times X_{\rm 2}~(R_{2-1/1-0})^{-1}   L_{\rm CO(2-1)} $ \citep{2013ARA&A..51..207B}, 
where the factor 1.36  for elements other than hydrogen \citep{2000asqu.book.....C} is taken into account, and $X_{\rm 2}$ is the $X_{\rm CO}$ factor normalized to 2 $\times$ $10^{20}$\,cm$^{-2}$\,(K\,\kms)$^{-1}$. 
Table~\ref{table2} shows the derived parameters of the different molecular gas components (i.e. circumnuclear region, NE and SW arms in NGC\,3110, and circumnuclear region in NGC\,232). The parameters include the velocity ranges, total CO(2--1) integrated fluxes for each galaxy and component, and the derived molecular gas masses. 
Both galaxies have molecular gas masses of about 1--2 $\times$ 10$^{10}$\,M$_\odot$, and these are consistent (if corrected by flux losses) with previously reported values.

\subsection{Molecular gas concentration and gas-to-dynamical mass ratios}
\label{discuconcentration}

Two mechanisms are usually invoked to explain how gas is driven to the circumnuclear regions of galaxies: interactions between galaxies and barred potentials.
First we calculate the molecular gas concentration (see Table\,\ref{tab4}) to see whether the two interacting objects { studied in this paper have comparable concentrations to those in literature samples of non-interacting or weakly interacting barred and unbarred galaxies}.

The molecular gas concentration is on the high side when compared to non-interacting and non-barred objects, although not as high as in some barred galaxies.
Within a radius of $\sim$ 1\,kpc (i.e. in the inner resolution element), the molecular gas surface densities (in the plane of the disk) are $\Sigma_{\rm mol}^{\rm 1kpc}$ = 306 and 455\,M$_\odot$\,pc$^{-2}$ for NGC\,3110 and NGC\,232 (see \S\,\ref{subsec:resolvedsflaw}). 
{ We note that within the inner regions, uncertainties due to flux loss are expected to be small (\S\,\ref{subsect:co2-1emissionline}).}
The concentration factor of molecular gas, defined as $f_{\rm con}$ = $\Sigma_{\rm mol}^{\rm 1kpc}$ / $\Sigma_{\rm mol}^{\rm disk}$, 
where $\Sigma_{\rm mol}^{\rm disk}$ is the disk surface density within the optical radius R$_{25}$ (isophotal radius at 25\,mag\,arcsec$^{-2}$ in B-band) { obtained from the total molecular gas content obtained with the single dish measurements (thus not affected by flux loss)}, is 38 for NGC\,3110 and 22 for NGC\,232. 
These values { for $f_{\rm con}$ are likely upper limits in case there is some flux loss in the 1\,kpc measurement}, and also because it is expected that the molecular gas distributions are not homogeneous within the inner 1\,kpc radius. 
 \citet{1999ApJ...525..691S} found that on average unbarred galaxies present a degree of concentration a factor of four times lower than barred systems. They measured $f_{\rm con}$ (defined with respect to a radius of 500\,pc) to be 100.2 $\pm$ 69.8 (the error is the standard deviation, for 10 objects) and 24.9 $\pm$ 18.5 (10 objects) for non-barred objects. We note that in the case that $X_{\rm CO}$ is smaller in the circumnuclear regions than in the outer disk, $f_{\rm con}$ will decrease. However, \citet{1999ApJ...525..691S} used a constant $X_{\rm CO}$ factor across galaxies and for different radii, so we can directly compare. The limitation is that we need to estimate the gas surface densities within a 500\,pc radius, $\Sigma_{\rm mol}^{\rm 500pc}$, which correspond to a size smaller than our resolution element. 
It is reasonable to expect that the CO(2--1) distributions of the two galaxies in this paper are centrally concentrated within 1\arcsec\ by comparing with the structures seen in other ISM components (see for example H$_2$ 1--0 S(1) and ionized emission for NGC\,3110 in \citealt{2015A&A...578A..48C}). 
We estimate that $\Sigma_{\rm mol}^{\rm 500pc}$, and therefore $f_{\rm con}$, will then increase by a factor of 50\%, which will result in concentrations closer to the typical value of barred galaxies, or at the high end of non-barred objects. It is thus likely that a large amount of gas is transported to the circumnuclear regions due to the interaction itself, and at least in the case of NGC\,3110, prior to the formation of a bar.
{ The $f_{\rm con}$ values estimated }here should be confirmed in the future with better angular resolution CO data. 

Next we estimated the gas-to-dynamical mass ratio within a radius of 1\,kpc ($f_{\rm dyn}$, see Table\,\ref{tab4}), which is closely related to the stability of the disk.
The dynamical masses within 1\,kpc are derived from equation $M_{\rm dyn}$ = $r$ ($V$ / sin($i$))$^2$ / $G$, where $r$ is radius in pc, $V$ the rotational velocity (230\,\kms\ for NGC\,3110 and 300\,\kms\ for NGC\,232, see \S\,\ref{subsec:cointmaps}), $i$ the inclination, and $G$ the gravitational constant. 
The gas-to-dynamical mass ratios we obtain, $f_{\rm dyn}$ = 0.06 for NGC\,3110 and 0.03 for  NGC\,232, are comparable to the values found in other galaxies,  which for a radius of 500\,pc, $f_{\rm dyn}$ = 0.01 -- 0.3 \citep[e.g.][]{1999ApJ...525..691S,2010AJ....139.2241M}.
{ Within the inner 1\,kpc uncertainties due to flux loss are expected to be small (\S\,\ref{subsect:co2-1emissionline}), so we do not expect that this issue will affect considerably the $f_{\rm dyn}$ measurements.}

\section{Properties of the interaction, conversion from \ion{H}{1} to H$_2$ and resulting Star Formation}
\label{discuscaling}

\subsection{Properties of the interacting systems}

From the relatively small separations and velocity differences, one can infer that it is likely that the two galaxies and their companions are probably gravitationally bound based on escape velocity arguments. 
However, the separations and velocity differences are just projected values along the line of sight so it is difficult to discard the fly-by hypothesis.
From the lack of strongly perturbed morphologies and tidal tails in the stellar, ionized, and (molecular) gas components, it is reasonable to expect that the interaction is in an early stage. This is backed by simulations \citep[e.g.][]{1996ApJ...464..641M,2004ApJ...616..199I}. { The \ion{H}{1} observations for NGC\,3110 reinforce this idea because of the lack of \ion{H}{1} tidal tails in the VLA \ion{H}{1} maps presented in \citet{2002MNRAS.329..747T}, at least to a sensitivity of about 10$^{20}$\,atoms\,cm$^{-2}$. { However, the VLA \ion{H}{1}  integrated flux is $S_{\rm HI, VLA}$ = 2.9 $\pm$ 0.9\,Jy\,\kms , and the Nancay single-dish measurement $S_{\rm HI, Nancay}$ =7.6\,Jy\,\kms\   \citep{1991A&A...245..393M}, which shows that the interferometric experiment is likely affected by by missing flux due to lack of zero spacings}. No high resolution \ion{H}{1} map exists for NGC\,232 to our knowledge.
\ion{H}{1} spectral asymmetries can also be a way to probe the level of interaction \citep{2011A&A...532A.117E}. The  \ion{H}{1} profile of NGC\,3110 in \citet{1991A&A...245..393M} is two peaked and relatively symmetric, but may reveal a small wing in the redshifted edge of the profile (which is to the North of the galaxy), and the  \ion{H}{1} line of NGC\,232 also shows a peculiar profile with three peaks, although the S/N is relatively poor. }

Next we quantify the tidal strength of the interactions. 
The local tidal force exerted by the companion galaxy (B) on the main galaxy (A) can be estimated from the following equation 
\begin{equation} 
Q_{\rm pair} = {\rm log} \left(\frac{M_{B}}{M_{A}} \left(\frac{D_{A}}{d_{AB}}\right)^3\right) \quad,
\end{equation} 
where $M_{X}$ is the stellar mass of $X$ = A or B, $D_{A}$ is the estimated diameter of galaxy A, and $d_{AB}$ is the projected physical distance between the two galaxies \citep{2007A&A...472..121V,2014A&A...564A..94A,2015A&A...578A.110A}. 
Since the luminosity is proportional to mass, we use the B-band luminosity ratio in this calculation. 
The tidal strengths are $Q_{\rm pair}$ = --1.1 and --0.85 for NGC\,3110 and NGC\,232, respectively.
Clearly the tidal strength for NGC\,232 is greater than that of NGC\,3110. Indeed, the interacting galaxies in the NGC\,232 pair are closer and the masses are comparable. 
These values of $Q_{\rm pair}$ for the two objects are, just by the contribution of their corresponding companions, much greater than those in galaxies considered to be mostly isolated \citep{2007A&A...472..121V,2013A&A...560A...9A}, and when compared to isolated pairs, we see that they are at the high end of the distribution, so these two interactions are generally stronger. The mean and standard deviation for a sample of isolated galaxies are $Q_{\rm isolated}$ = --5.19 and $\sigma_{Q, {\rm isolated}}$ = 0.84, and for isolated pairs $Q_{\rm pair}$ = --2.27 and $\sigma_{Q, {\rm pair}}$ = 1.24 \citep{2015A&A...578A.110A}.

\subsection{Molecular gas  excess and \ion{H}{1} deficiency}
\label{subsec:def}

Given the ongoing interaction, the galaxies in these two systems are not in equilibrium and far from evolving secularly.
NGC\,3110 and NGC\,232 have higher H$_2$-to-\ion{H}{1}  mass ratios (2.3 and 4.6, respectively) than the median value for isolated galaxies, $\sim$ 0.4 \citep{2011A&A...534A.102L}, for similar morphological types. Note that we multiplied by a factor of 1.36 because the contribution of He was not originally included in the values of \citet{2011A&A...534A.102L}. This large difference is due to the molecular gas content being larger and \ion{H}{1} content smaller for a given stellar mass. The molecular gas masses of these two interacting galaxies, $\sim$2 $\times$ 10$^{10}$\,M$_\odot$ (Table\,\ref{tbl-1}), are well above the median for Sa -- Sc isolated galaxies, $M_{\rm mol}$ $\simeq$ (0.6 -- 1) $\times$ 10$^{9}$\,M$_\odot$  \citep{2011A&A...534A.102L}. 
 Using the regression fit (ordinary least squares fit with y-axis as dependent variable, OLS(Y-X)) between  log($M_{\rm mol}$) and log($L_{\rm B}$) for Sb-Sc isolated galaxies in \citet[][Table\,7]{2011A&A...534A.102L}, we confirm that the molecular gas contents  are in fact larger and not due to different galaxy sizes.
The deficiency of molecular gas (or excess in this case) following the nomenclature in \citet{2011A&A...534A.102L}, $DEF_{mol}$ = log $M_{\rm mol,exp}$ -- log $M_{\rm mol,obs}$ (expected molecular gas mass, { also corrected by the contribution of He}, in log scale minus that observed), is --0.25 for NGC\,3110 and --0.65 for NGC\,232. 

We also calculate the expected $M_{\rm HI }$ following the regression fits between log($M_{\rm HI }$) and log($L_{\rm B}$) (also OLS(Y-X)) for isolated galaxies \citep[][Table\,6]{2017arXiv171003034J}, and compare it to the observed $M_{\rm HI}$ in Table\,\ref{tbl-1}.  { We follow similar conventions to obtain $M_{\rm HI }$ and $L_{\rm B}$ as in \citet{2017arXiv171003034J} (see also \citealt{2005A&A...436..443V})}. The \ion{H}{1}  content in these two objects { shows} some level of deficiency, defined as
$DEF_{HI}$ = log $M_{\rm HI,exp}$ -- log $M_{\rm HI,obs}$ (expected atomic gas mass in log scale minus that observed), and is quantified to be 0.39 for NGC\,3110 and 0.47 for NGC\,232.

Therefore the excess of molecular gas can be at least partially explained at the expense of \ion{H}{1}. This is further discussed in \S\,\ref{discussion}.

\subsection{Star Formation Activities \label{starformationactivities}}

The global SFRs in both NGC\,3110 and NGC\,232 are of the order of $\sim$ 20\,M$_\odot$\,yr$^{-1}$ (Table\,\ref{tbl-1}), larger than the typical values in non-interacting galaxies ($\sim$ 0.5\,M$_\odot$\,yr$^{-1}$,
 \citealt{2011A&A...534A.102L}) or even in galaxies hosting a bar, where in general large molecular gas concentrations can be found but the SFRs are typically of the order of  1\,M$_\odot$\,yr$^{-1}$ within the inner 1 kpc \citep{1999ApJ...525..691S}.
The derived global Star Formation Efficiencies (SFE = $SFR$ [M$_\odot$\,yr$^{-1}$] / $M_{mol}$ [M$_\odot$]) for both objects are 1 -- 2 $\times$ 10$^{-9}$ yr$^{-1}$.
This is closer to "normal" galaxies (10$^{-9}$\,yr$^{-1}$) than to the starburst sequence (10$^{-8}$\,yr$^{-1}$) \citep[e.g.][]{2010ApJ...714L.118D}. 

The H$\alpha$ images are shown in Fig.~\ref{fig2} for NGC\,3110 \citep{2004AJ....127..736H} and NGC\,232 \citep{2006ApJS..164...52S}. 
Note that their coordinate system presented some slight offsets of $<$ 2\arcsec\ and their astrometry was calibrated by comparing the images with continuum maps with similar resolution and with accurate astrometry. After this correction, in general we see agreement between the CO(2--1) and H$\alpha$ distributions in both objects. 

{
In NGC\,3110, the H$\alpha$ emission of the circumnuclear region agrees well in extent and morphology with the CO(2--1) emission. 
The second brightest H$\alpha$ component about 6\arcsec\ to the W of the nucleus is associated with another bright CO component as well. The third brightest H$\alpha$ component is located in the southern arm at about 18\arcsec\ from the nucleus, but CO(2-1) emission is much dimmer there.
The northern arm is substantially weaker in CO(2--1) emission than its southern counterpart, to the point that it is mostly undetected in our observations, even though \ion{H}{2} regions are of comparable brightness in both arms and the distribution is more spread. The southern arm's CO(2--1) and H$\alpha$ emissions overlap well, at least to our resolution element.
However, there are regions which are bright in H$\alpha$ emission but are not detected in the CO(2--1) observations. These regions are north of the southern arm ($\sim$6 -- 12\arcsec\ NE from the center), south of the northern arm ($\sim$6 -- 12\arcsec\ SW from the center), as well as some regions along both arms (and especially in the N arm).}

The H$\alpha$ emission morphology of NGC\,232 is quite different. One of the most remarkable differences is that the spiral arms that are clearly seen in optical and NIR images { do} not present signs of SF, nor molecular gas to our sensitivity limit. The main component that is seen in CO(2--1) and H$\alpha$ emissions is extended within a 10\arcsec\ (4.2\,kpc) region and with a somewhat chaotic morphology if compared to NGC\,3110. H$\alpha$ emission also seems to be more diffuse and not as compact as that of the CO(2--1) emission. H$\alpha$ emission and Pa$\alpha$ \citep{2015ApJS..217....1T} are centrally peaked and seem to trace two nuclear spirals or shock regions, which likely drive gas to the nucleus. That H$\alpha$ central and brightest component in the inner few arcsec correspond to the location of the wings in the CO(2--1) profile.
Also, there is some signature of molecular gas being more extended towards the SW from the center, but this is not clear in the H$\alpha$ maps.

\subsection{ Spatially Resolved Star Formation Laws}
\label{subsec:resolvedsflaw}
 
We present the spatially resolved SF laws using the SMA CO(2--1) data as a tracer of the molecular gas surface densities ($\Sigma_{\rm mol}$) and the H$\alpha$ and Spitzer 24\,$\mu$m maps for the SFR surface densities ($\Sigma_{\rm SFR}$). Since H$\alpha$ emission is affected by extinction, we estimate the amount of SF obscured by dust by using 24\,$\mu$m data obtained with the MIPS instrument (Multiband Imaging Photometer for Spitzer, \citealt{2004SPIE.5487...50R}) onboard of the Spitzer Space Telescope \citep{2004ApJS..154....1W}. We use for the three maps a common astrometric grid and align them so that there are no artificial offsets. The H$\alpha$ maps were compiled from \citet{2006ApJS..164...52S} for NGC\,232 and \citet{2004AJ....127..736H} for NGC\,3110  (\S\,\ref{starformationactivities}, see also Fig.~\ref{fig2}).
We degraded the resolution of the H$\alpha$ and CO(2--1) data to that of the Spitzer 24\,$\mu$m data by convolving them with a circular Gaussian beam. The FWHM of the MIPS point-spread function (PSF) at 24\,$\mu$m is 6\arcsec . The MIPS data are post-BCD (higher-level products) processed using the MIPS Data Analysis Tool \citep{2005PASP..117..503G}. We subtracted a background in both H$\alpha$ and 24 $\mu$m maps, which were quite small.

The molecular surface densities, $\Sigma_{\rm mol}$, were obtained using the same convention as already indicated in \S~\ref{sub:moleculargasmass} for $M_{\rm mol}$. The maps were corrected by the primary beam response.
They are in units of M$_\odot$~pc$^{-2}$, include the contribution from Helium, and have been corrected by inclination.
Our $\Sigma_{\rm mol}$ maps are sensitive to surface densities { of} $\Sigma_{\rm mol}$ $\simeq$ 10 \,M$\odot$~pc$^{-2}$.

Next we constructed maps of the SFR surface densities, $\Sigma_{\rm SFR}$. We use the combination of H$\alpha$ and 24\,$\mu$m calibration convention as in \citet{2010ApJ...714.1256C}.
We adopt the following prescription for the combination of the H$\alpha$ and 24 $\mu$m maps  to obtain the $\Sigma_{\rm SFR}$ \citep{2007ApJ...666..870C,2010ApJ...714.1256C}:
$\Sigma_{\rm SFR}$ [M$_\odot$ yr$^{-1}$ kpc$^{-2}$] = 5.45 $\times$ 10$^{-42}$ ( $S_{\rm H\alpha }$[erg s$^{-1}$ kpc$^{-2}$] + 0.031 $S_{24}$[erg s$^{-1}$ kpc$^{-2}$]  ) 
where $S_{24}$ and $S_{\rm H\alpha}$ denote the 24\,$\mu$m and H$\alpha$ luminosity surface densities, respectively. 
In this calibration of $\Sigma_{\rm SFR}$,  \citet{2007ApJ...666..870C} adopts the default initial mass function (IMF) in Starburst99 models \citep{1999ApJS..123....3L}, which is a Kroupa-type broken power law \citep{2001MNRAS.322..231K}. 
{ For comparison we also calculate  $\Sigma_{\rm SFR}$ only using 24\,$\mu$m data following $\Sigma_{\rm SFR}$ [M$_\odot$ yr$^{-1}$ kpc$^{-2}$] = 2.75 $\times$ 10$^{-43}$ $S_{24}$[erg s$^{-1}$ kpc$^{-2}$] \citep{2005ApJ...632L..79W,2010ApJ...714.1256C}.
The global SFRs we obtain using H$\alpha$ + 24\,$\mu$m and only  24\,$\mu$m are in agreement, 19.7\,M$_\odot$\,yr$^{-1}$  and 21.2\,M$_\odot$\,yr$^{-1}$ for NGC\,3110, and 19.3\,M$_\odot$\,yr$^{-1}$ and  27.7\,M$_\odot$\,yr$^{-1}$ for NGC\,232. The choice of either one of the SFR recipes does not have any remarkable impact on our main results. Also, these global SFR measurements agree with those obtained from the IR luminosities in Table\,\ref{tbl-1}.}

\begin{figure*}[htbp]
\begin{center}
\includegraphics[width=8.5cm]{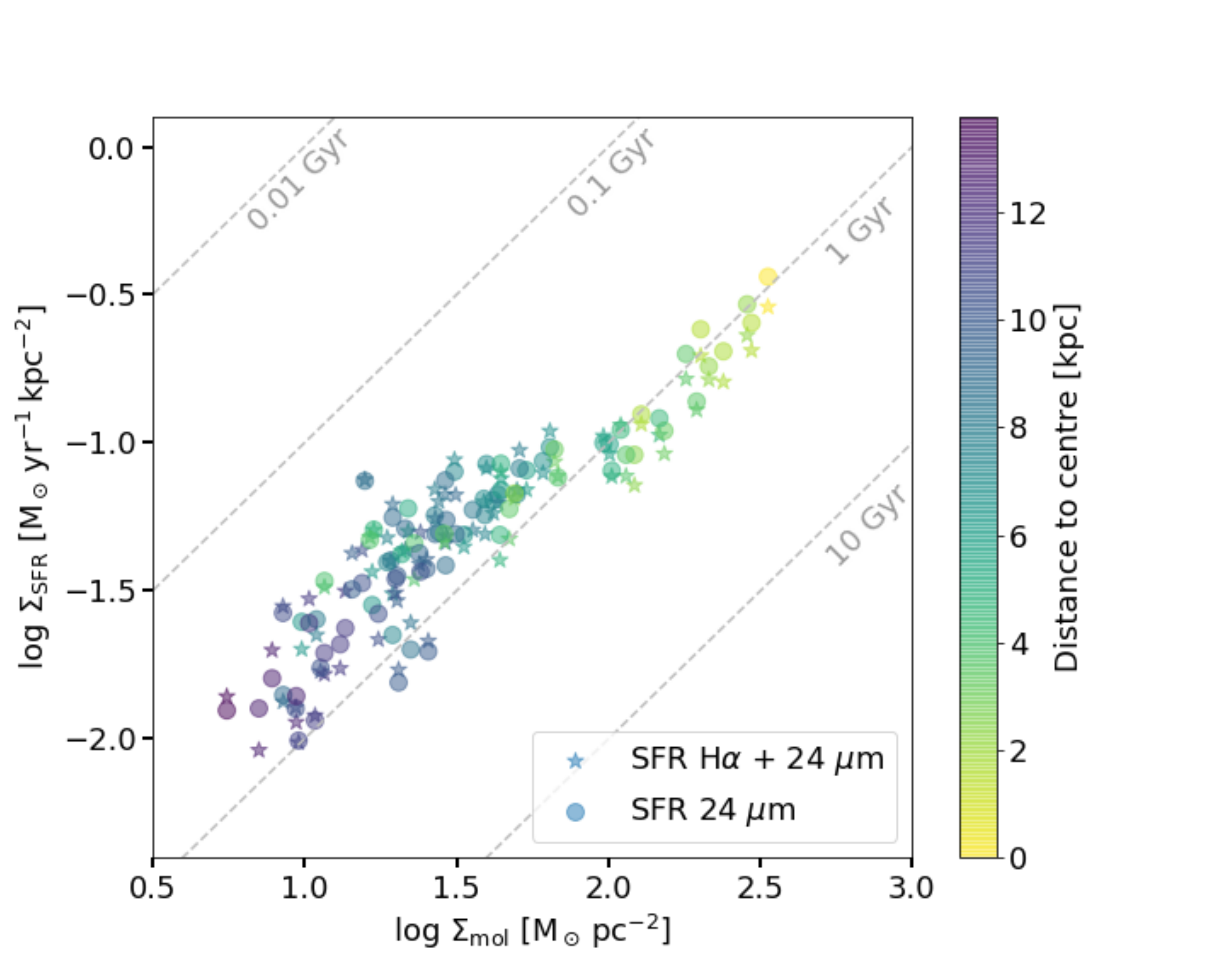}
\includegraphics[width=8.5cm]{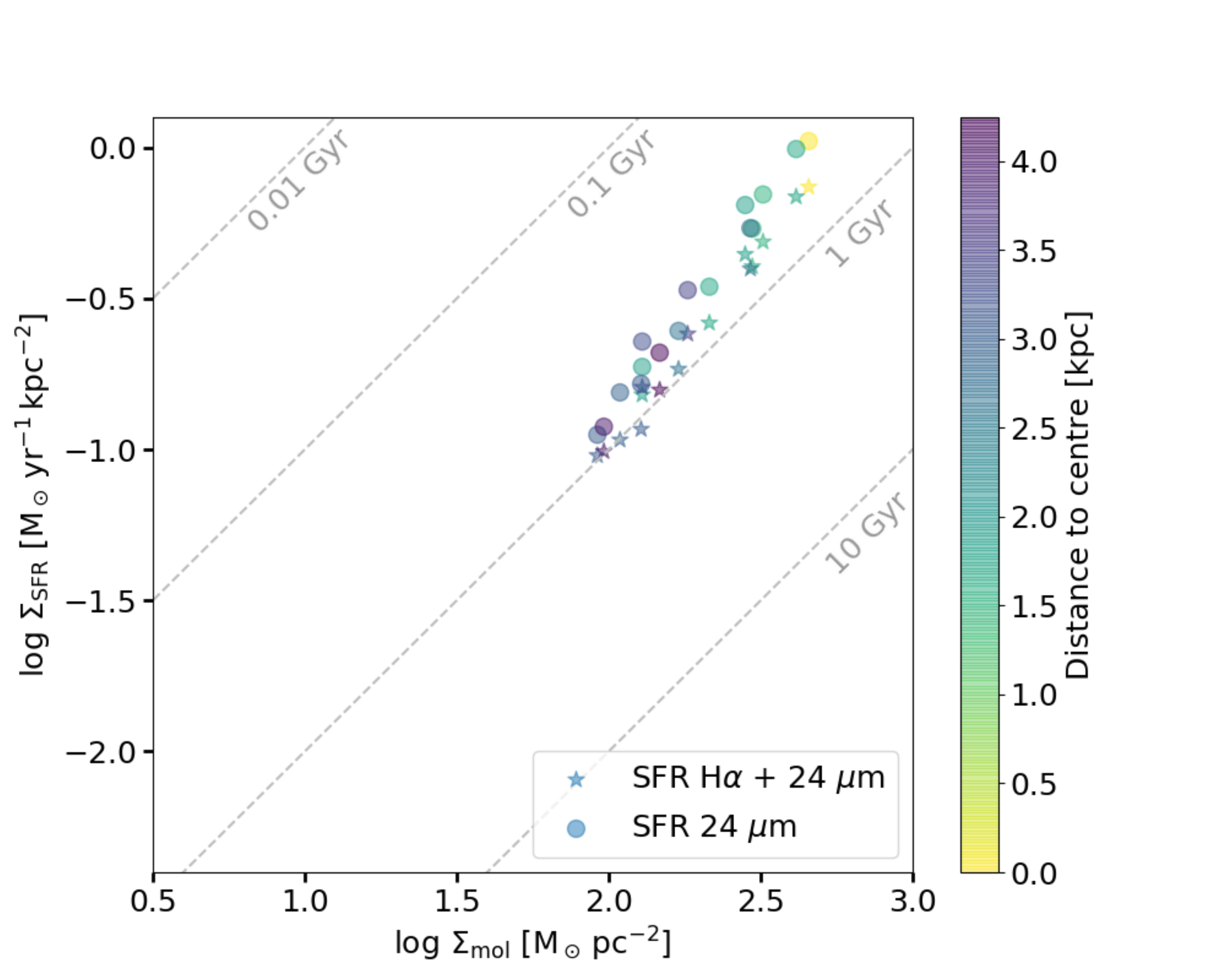}

\caption{ Star formation rate surface density ($\Sigma_{\rm SFR}$) estimated from H$\alpha$ and 24\,$\mu$m emission as a function of molecular gas surface density ($\Sigma_{\rm mol}$) from the CO(2--1) emission obtained for the two early stage merger galaxies NGC\,3110 (left) and NGC\,232 (right). Each data point correspond to a 3.1\arcsec\ pixel (resolution of the CO(2--1) data). Note that the resolution element of the 24\,$\mu$m data is 6\arcsec. Dashed lines indicate constant depletion time lines expressed in Gyr. 
The color scale of the data points represents the distance from the center in kpc. Star symbols show data points with $\Sigma_{\rm SFR}$ estimated from H$\alpha$ and 24\,$\mu$m data, while circles show data points with $\Sigma_{\rm SFR}$ estimated from 24\,$\mu$m only (see \S\,\ref{subsec:resolvedsflaw} for details).}
\label{figsfr}
\end{center}
\end{figure*}

We present in Fig.\,\ref{figsfr} the $\Sigma_{\rm SFR}$ versus $\Sigma_{\rm mol}$ plots for both galaxies.
The data points  corresponding to the circumnuclear regions show $\Sigma_{\rm mol}$ up to $\sim$ 10$^{2.5}$\,M$_\odot$~pc$^{-2}$, and depletion times of 1\,Gyr for NGC\,3110 and 0.8\,Gyr NGC\,232.
In the plot for NGC\,232 (Fig.\,\ref{figsfr}, right) we only show data points for the inner 12\arcsec\ because the 24\,$\mu$m map have artifacts caused by the point spread function (PSF) and the $\Sigma_{\rm SFR}$ values beyond could be uncertain. 

In the case of NGC\,3110, it is possible to compare the SF law for the circumnuclear and arm regions. The depletion times are { on} average 0.5\,dex smaller at large radii than in the circumnuclear region. This may mean that the circumnuclear regions are not as efficient as the arms in transforming gas into stars. However, a smaller $X_{\rm CO}$ factor towards the center may contribute to the observed trend.  Usually the $X_{\rm CO}$ factor is smaller in the nuclei of galaxies, especially in starbursts \citep{2013ARA&A..51..207B}. A factor of 4 could explain this, i.e. from $X_{\rm CO}$ = 2 $\times$ 10$^{20}$\,cm$^{-2}$\,(K\,\kms)$^{-1}$, as we assumed, down to a $X_{\rm CO}$ = 0.5 $\times$ 10$^{20}$\,cm$^{-2}$\,(K\,\kms)$^{-1}$ in the circumnuclear regions, which might be a reasonable value in the central regions of a starburst galaxy. 
{ Assuming that $X_{\rm CO}$ factor does not vary, integrated fluxes about three times larger would be needed along the spiral arms to account for the 0.5\,dex difference between the circumnuclear regions and spiral arms. It is unlikely that the data suffer from such a large amount of missing flux along the arms. Most of the missing flux should originate from a more extended component because the CO emission corresponding to the spiral arms (and in each channel) is expected to be relatively compact.}

\begin{figure*}[htbp]
\begin{center}
\includegraphics[width=8.5cm]{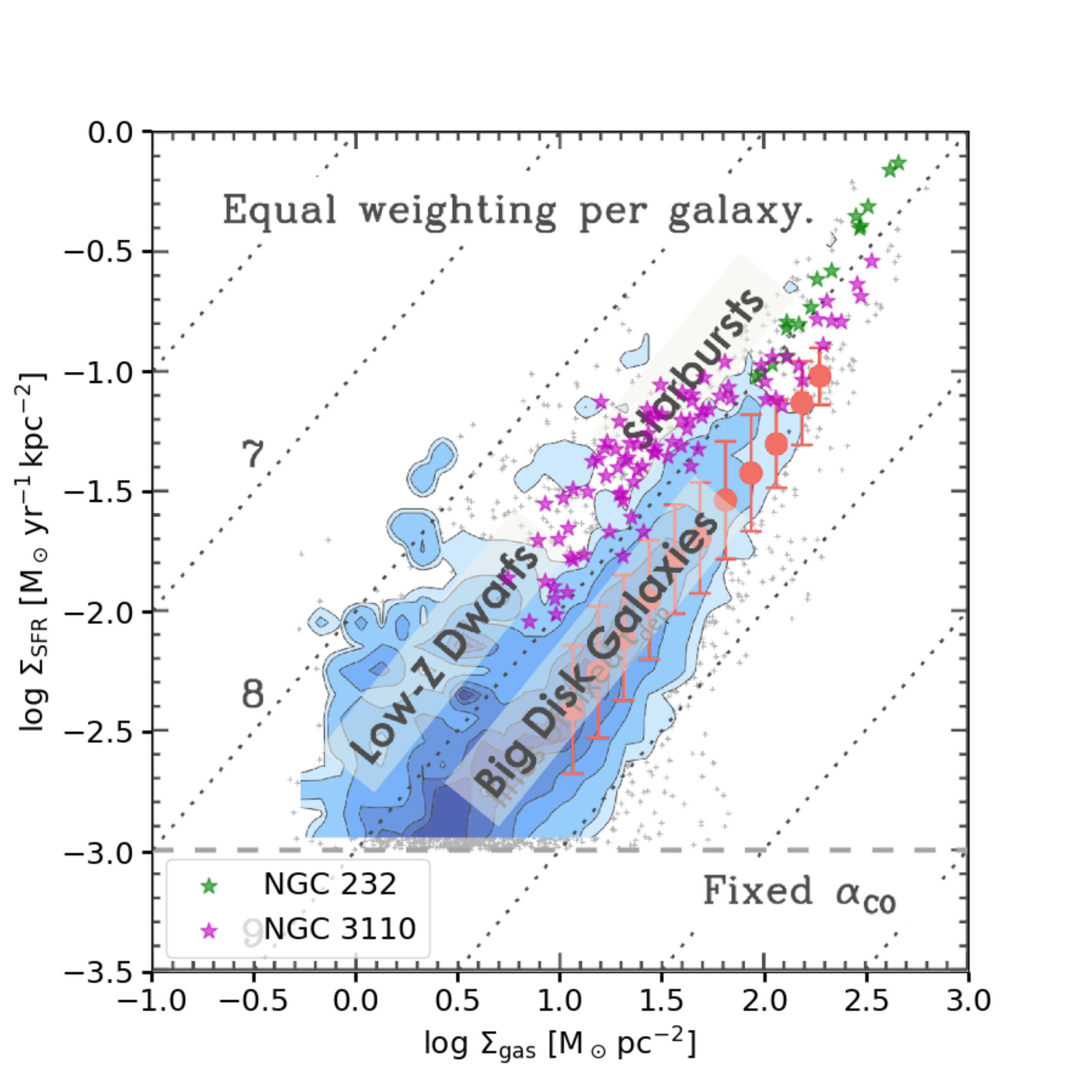}
\caption{ Star formation rate surface density ($\Sigma_{\rm SFR}$) estimated from H$\alpha$ and 24\,$\mu$m emission as a function of molecular gas surface density ($\Sigma_{\rm mol}$), extracted from Fig.\,1 of \citet{2013AJ....146...19L}, together with the data points for NGC\,3110 (magenta) and NGC\,232 (green) in Fig.\,\ref{figsfr} to be able to compare.}
\label{figsfr2}
\end{center}
\end{figure*}

Fig.\,\ref{figsfr2} shows the $\Sigma_{\rm SFR}$ versus $\Sigma_{\rm mol}$ data points of NGC\,3110 and NGC\,232 over the plot of Fig.\,1 of \citet{2013AJ....146...19L} for all the data points in the HERACLES survey composed of 30 nearby disk galaxies.  The adopted recipes to obtain $\Sigma_{\rm mol}$ and $\Sigma_{\rm SFR}$ { are comparable}.
First, we can see that the data points in this study represent relatively high $\Sigma_{\rm mol}$ and $\Sigma_{\rm SFR}$ compared to 1\,kpc regions in other "normal" galaxies. 
The depletion time of "normal" large disk galaxies as in \citet{2008AJ....136.2846B} and \citet{2013AJ....146...19L} is $\sim$ 2\,Gyr. 
Especially in the spiral arms of NGC\,3110 the depletion times due to SFRs are { on} average $\sim$0.5\,dex larger than that in more quiescent galaxies. Regions at the arms far away from the center lead to the "low metallicity dwarf galaxy region" in \citet{2013AJ....146...19L}'s plot. Although SFEs are higher in the spiral arms, the circumunclear regions seem to follow similar SFEs as in "normal" big disk galaxies if our assumed $X_{\rm CO}$ factor holds.

\subsection{Distribution of Super Stellar Clusters in NGC\,3110}
\label{subsec:sscs}

Super star clusters (SSCs) are massive ($>$10$^5$ M$_\odot$) and compact (a few parsec size) star clusters. 
SSCs and other complexes of clusters  are usually found in merging systems close to dense molecular clouds \citep[e.g.][]{1995AJ....109..960W,2007AJ....133.1067W,2012ApJ...760L..25E,2017A&A...600A.139H}, and starburst dwarf galaxies \citep[e.g.][]{2015PASJ...67L...1M,2015Natur.519..331T,2018arXiv180810089M}. 
The properties of SSCs seem to be different depending on the stage of the merger: SSCs in post-mergers are usually more luminous, larger and redder than in interacting systems in the early stages \citep{2011AJ....142...79M}.

\begin{figure*}[htbp]
\begin{center}
\includegraphics[width=7.5cm]{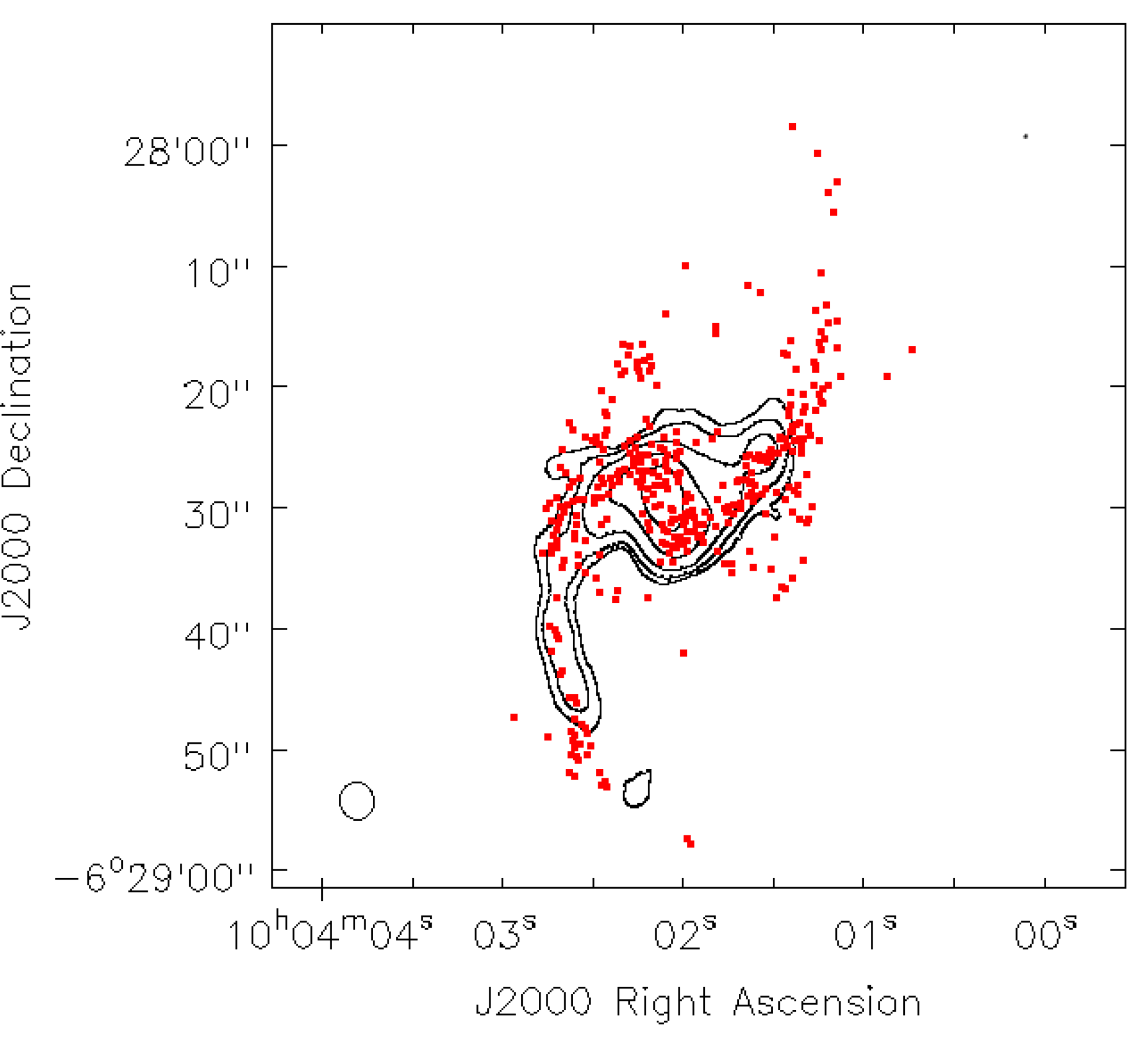}
\includegraphics[width=10cm]{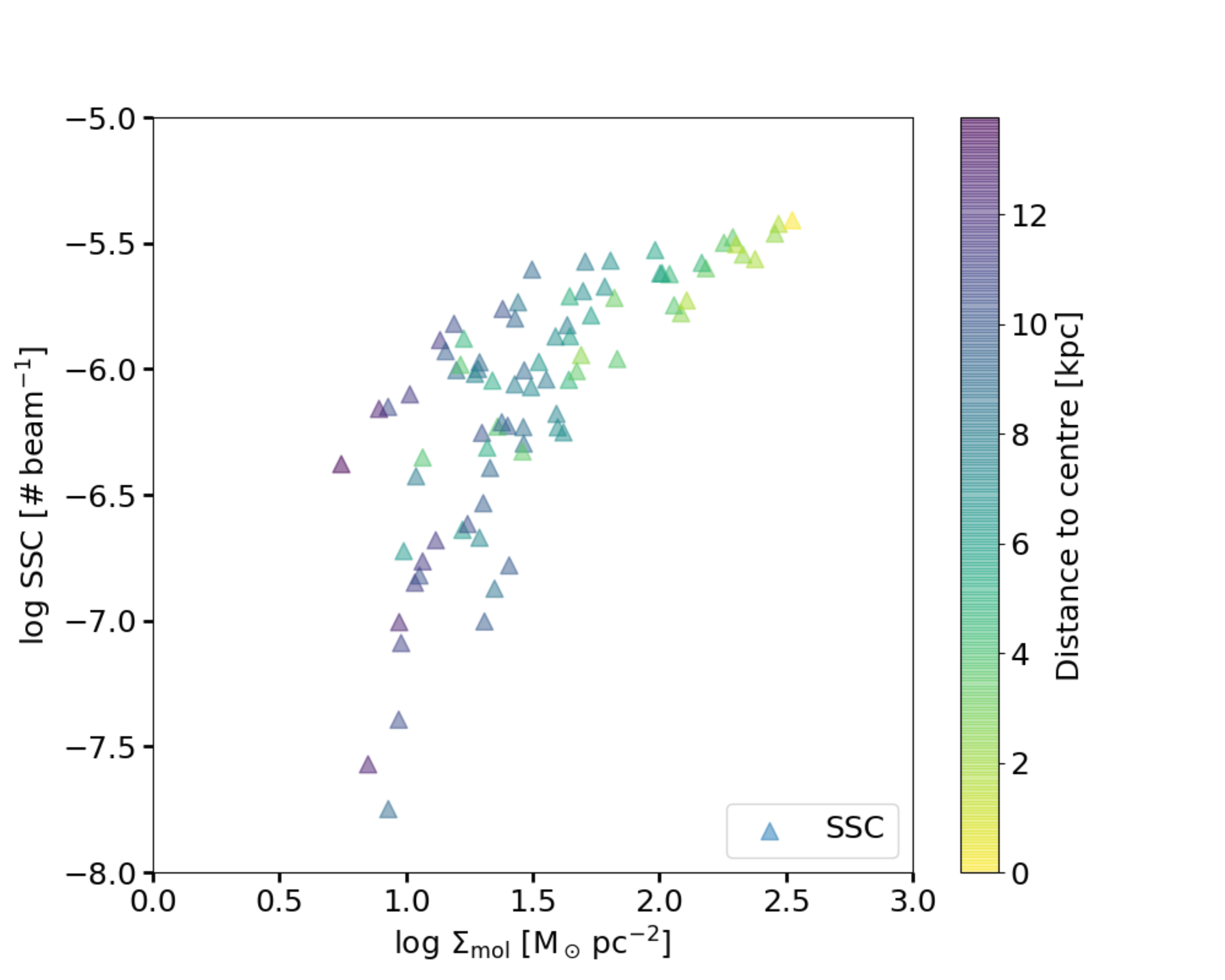}
\caption{  Left) Distribution of identified SSCs based on Ks-band image of NGC\,3110 obtained with VLT/NACO, with the CO(2--1) contours on top as in Fig.\,6. Right) Pixel to pixel correlation between SSC number surface density and molecular gas surface density ($\Sigma_{\rm mol}$). The color scale of the data points represents the distance from the center in kpc.}
\label{figsscs}
\end{center}
\end{figure*}

NGC\,3110 was observed with VLT/NACO as part of the SUNBIRD (SUperNovae and starBursts in the InfraReD) survey \citep{2017IAUS..316...70R}. 
Fig.\,\ref{figsscs} (left panel) shows the location of the Ks-band candidate SSCs as compared with our CO map.
The resolution of the Ks-band map is  $\sim$0\farcs1.
It is remarkable that at least $\sim$ 350 SSCs were found in this object and they are located mostly along the spiral arms and its circumnuclear component \citep{2015PhDT.......214R}. Although SSCs are commonly seen in (intermediate/late stage) mergers \citep[e.g.][]{1995AJ....109..960W,2007AJ....133.1067W}, this proves that even in an initial phase of the merger, and even when the interacting galaxy mass ratio is so large, a substantial population of SSCs can be formed. 
We see that the distribution of SSCs in the northern and southern arms seem more symmetric to each other than in CO emission. Even though the number of SSCs formed along the two arms is similar, the molecular gas surface densities are at least a factor of five smaller or even mostly depleted in the northern arm. 

{ Fig.\,\ref{figsscs} (right panel) shows the SSC number surface density (to the resolution of the CO(2--1) map) versus $\Sigma_{\rm gas}$ plot.  There is a correlation in the sense that regions with higher gas surface densities host a larger number of SSCs. The Spearman's correlation coefficient (log scale) is $\rho$ = 0.8.}

Furthermore, the luminosity functions of SSCs separating the circumnuclear regions and outer regions (arms) have been studied by \citet{2015PhDT.......214R}, 
and it has been found that in the circumnuclear regions the luminosity function is flatter than in the external parts ($\alpha$ = 1.96 $\pm$ 0.08 versus $\alpha$ = 2.35 $\pm$ 0.09). 
This may mean that the formation of the most massive SSCs depends on their environment and occur preferentially in regions with high molecular gas surface densities (and SFR surface densities). The difference in the LF power-law slopes could also be partly caused by blending effects and the difficulty to detect faint NIR clusters in the circumnuclear regions. However, the blending effect is unlikely to be important for targets less distant than 100 Mpc \citep{2013MNRAS.431..554R}.

\section{Discussion}
 \label{discussion}

We discuss the properties of the molecular gas derived from our CO(2--1) observations of NGC\,3110 and NGC\,232, as well as the resulting star formation activities, in the context of published numerical simulations of interacting systems and other observational work. { We also present our own numerical simulations trying to reproduce the properties of the NGC\,3110 system.}

\subsection{Comparison of distribution and SF activities with numerical simulations}
\label{subsec:simulations}
It is known from numerical simulations that early stage merger interactions may generate asymmetric two- armed spirals, warps, bars, and tidal tails that extend well beyond the main body of the galaxy (see \S\,1). A number of numerical studies have explored the parameter space of minor encounter/merger scenarios \citep[e.g.][]{2011MNRAS.414.2498S,2015ApJ...807...73O,2015MNRAS.449.3911P}, including how changes in mass and impact parameters affect the stellar and gaseous components for a galaxy interacting with a companion that is at least an order of magnitude less massive than the host, which matches reasonably well the situation of NGC\,3110. With these kind of low mass companions a $m$=2 response (i.e. spiral arms) is easily produced (within reasonable approach distances and velocity differences). As in the simulations, our NGC\,3110 maps show the two-armed molecular component which dominates outside a few kpc. Higher mass companions are seen to disrupt the disks to the extent that spirals are not as clear, resulting in rings and less regular features, possibly in agreement with the less prominent spiral arms in NGC\,232. It may also be possible that the interaction in NGC\,232 is simply not oriented in the correct manner (i.e. against rotation) which has been seen to produce only a minor morphological response \citep{2010ApJ...725..353D,2018ApJ...857....6L}. Stronger tidal forces are, however, more efficient at driving spiral features into the inner disc ($<$ 2\,kpc) than weaker interactions. A gaseous disk is detected in NGC\,232 at inner radii, with little CO or H$\alpha$ emission present in the form of spiral arms, which might be in agreement with a strong interaction scenario. Also, these simulations showed that higher mass companion galaxies may reduce the effective size of the main galaxy by a few kpc compared to interactions with lighter companions, causing rapid infall and stripping of outer material which both aid in reducing disc scale lengths.

\begin{figure*}[htbp]
\begin{center}
\includegraphics[width=13cm]{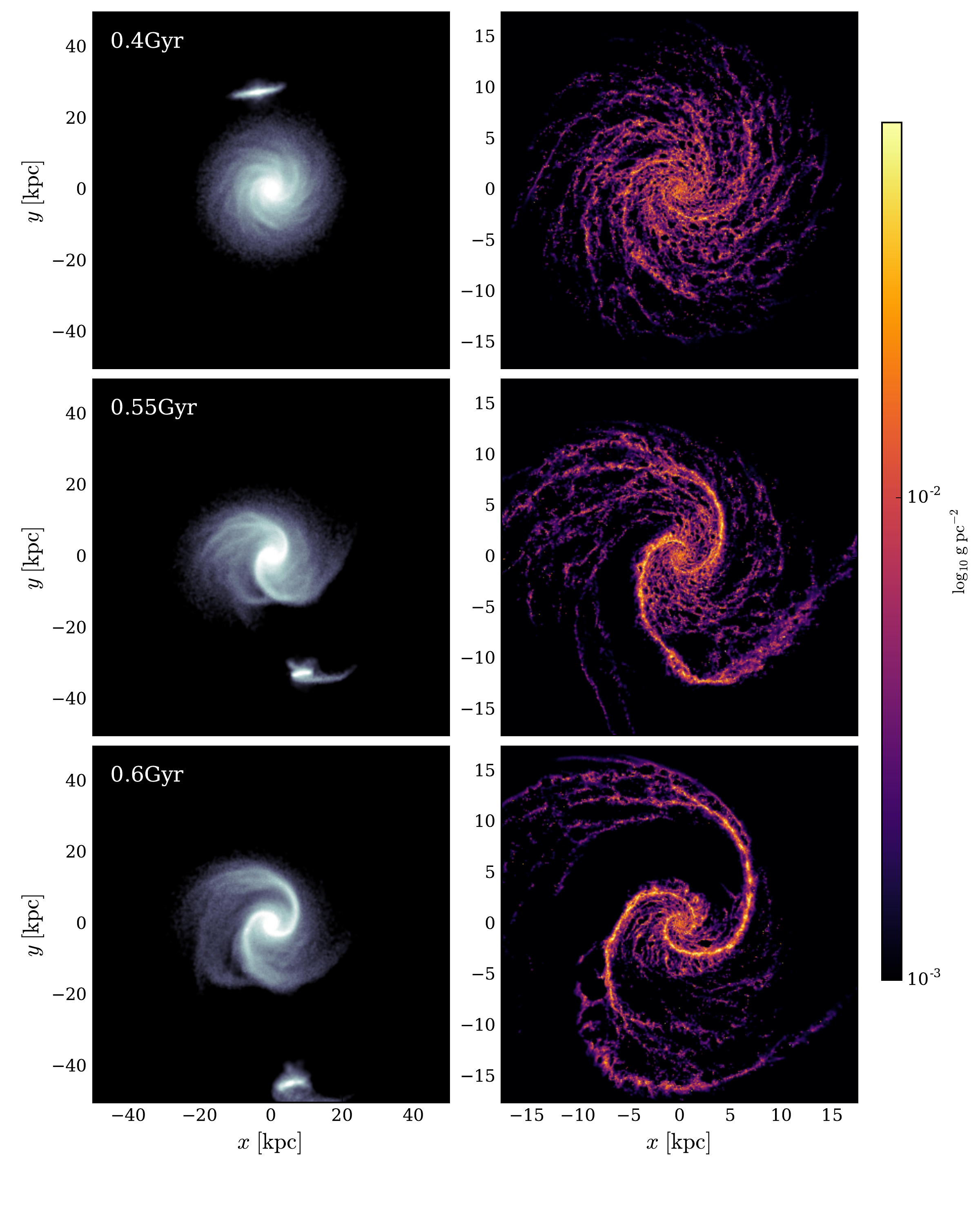}
\caption{ Numerical simulations reproducing the morphological properties of the NGC\,3110 system. The plots show, from top to bottom, different epochs of the early-stage merger, at 0.40, 0.55 (best match), and 0.60\,Gyr. Left panels show the stellar component while right panels the gaseous component. Note the different spatial scales between left and right.}
\label{simulation}
\end{center}
\end{figure*}

{ 
Preliminary numerical simulations including self gravity, cooling, star formation and
feedback of a tidally induced two-armed galaxy with similar properties as NGC 3110 are
briefly presented here, carried out with the \textsc{Gasoline2} SPH code \citep{2017MNRAS.471.2357W}. The evolution of the merger in different epochs around the first approach is
shown in Fig.\,\ref{simulation}. For more details, please refer to \citet{2017MNRAS.468.4189P}, for a similar
experiment of a fly-by interaction between a disk galaxy and smaller companion,
although not optimized to reproduce the properties of this system. The primary galaxy
has a mass model and rotation curve constrained to that observed for NGC\,3110, with
velocity dispersions in the stars set to slow bar formation and create a multi-armed spiral
when evolved in isolation. We assume a total mass ratio of 14:1, a similar velocity
difference and separation as observed in the NGC 3110 system. The tidal forces induce
a two-armed structure in the simulation that matches the distribution seen in our maps of
NGC\,3110 reasonably well at about 150\,Myrs after the closest approach (see Fig.\,\ref{simulation}),
with clear asymmetries seen between the two spiral arms and a kink in orientation at the
end of the southern arm.}

\begin{figure*}[htbp]
\begin{center}
\includegraphics[width=12cm]{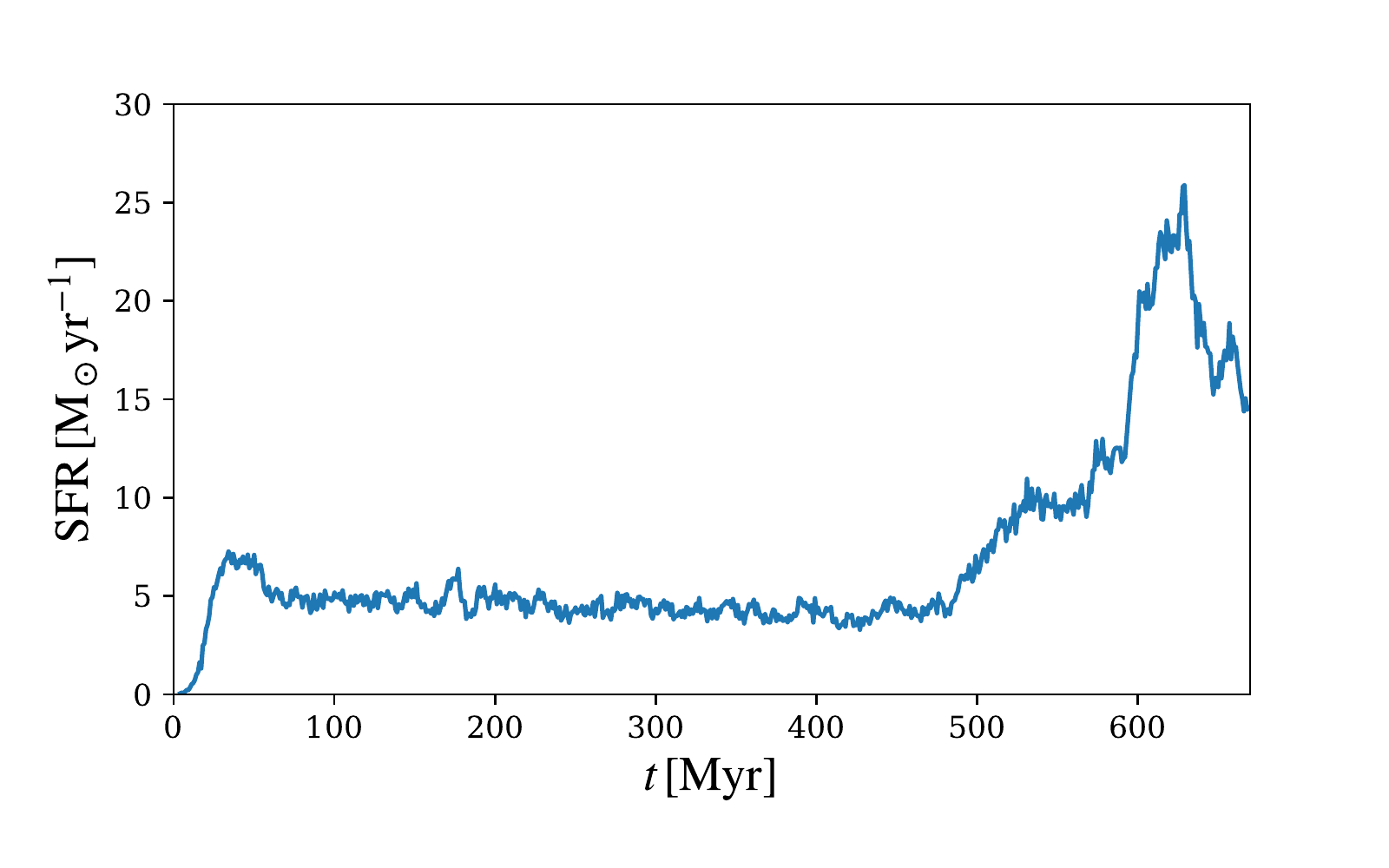}
\caption{ SFR history obtained from our numerical simulations of NGC\,3110 during the first 700\,Myrs since tidal interaction started. Closest
approach occurs at 460\,Myr, and the current epoch based on a morphological match is at about 550\,Myrs. }
\label{simulationsfr}
\end{center}
\end{figure*}

However, the central star forming activities found in the observations  are not as well reproduced by numerical simulations.
The galaxy mass ratio has been identified in other numerical simulations as a main parameter characterizing the resulting merger driven starburst. But the induced SF for mass ratios larger than 10:1 is seen to be almost negligible in other works \citep[e.g.][]{2008MNRAS.384..386C}. This is in contrast with our observational results, where the total SFR (SFRs $\sim$20\,M$_\odot$\,yr$^{-1}$) is as high as that expected in other numerical simulations for some similarly sized galaxy interactions. The starburst efficiency depends on the structure of the primary galaxy, and in particular, the lack of a massive stellar bulge will make the disk unstable and will enhance merger-driven SF even for large mass ratio mergers \citep{2008MNRAS.384..386C}. The SFRs in the simulations of \citet{2017MNRAS.468.4189P} (Fig.\,8) for a system of mass ratio 10:1 are also low, probably as a result of the inclusion of a massive bulge (and halo). Additionally the gas only constitutes 10\% of the baryonic component in the simulation, which is substantially lower than both our target galaxies, which would also result in a lower SFR in comparison. In these simulations the peak is 6\,M$_\odot$/yr and average ~2 M$_\odot$/yr. Other simulations in the literature of stronger bound interactions (clearly resulting in mergers) can reproduce SFRs of up to 100\,M$_\odot$/yr \citep{2015MNRAS.446.2038R,2013MNRAS.430.1901H}, though result in clearly disrupted disks and clear gas bridges between the interacting pairs. { The SFR history of our simulations is presented in Fig.\,\ref{simulationsfr}. Although there is a peak of about 25\,M$_\odot$\,yr$^{-1}$ at about 620\,Myrs since tidal interaction started, { that corresponds to $\sim$70\,Myrs after our best morphological match}.  At any rate, the current status of NGC\,3110 may represent one of the highest SFR episodes as a result of the interaction with the minor companion (before coalescence), likely just after closest approach.} The specific case of the NGC\,3110 interaction and the resulting morphology in dense gas, clusters, and localized SF will be the subject of a future numerical study with higher spatial resolution than presented here.

\subsection{Conversion from \ion{H}{1} into molecular gas and gas inflow}
In the relatively early stages of the interaction of the NGC\,3110 and NGC\,232 systems we found an excess of molecular gas which can be at least partially explained as a result of the conversion of \ion{H}{1}  (\S\,\ref{subsec:def}).
An excess of molecular gas is also seen in the early evolutionary phase of more extreme systems such as Hickson Compact Groups, HCGs, which might also be linked to \ion{H}{1} deficiencies \citep{2012A&A...540A..96M}.
These results are in agreement with simulations, where an increase in the molecular to \ion{H}{1} fraction even in the early stages of a merger can be seen, with one the largest events occurring immediately after the first close approach \citep{2004ApJ...616..199I}. Using single-dish observations, \citet{2017ApJ...844...96Y} found that the molecular gas mass in the central several kpc seems to be constant from early stage to late stage mergers, and they inferred that molecular gas inflows replenish the consumed gas by SF. The excess of molecular gas in the two galaxies studied here can probably be explained partly by molecular gas inflow from the outer regions, but also by the conversion from \ion{H}{1} into molecular gas.
{ \citet{2017PASJ...69...66K} found that there is a high global molecular gas fraction in a sample of four mid-stage mergers and they suggest an efficient transition from atomic to molecular gas due to external pressure as a result of the interaction. The molecular hydrogen gas mass to total (atomic plus molecular hydrogen) gas mass averaged over their sample of galaxies is 0.71 $\pm$ 0.15, in agreement with the 0.66 for NGC\,3110 and 0.78 for NGC\,232 we obtain following their convention}.
Once \ion{H}{1} is consumed or lost, then one would naturally find that the molecular content decreases with decreasing projected separation between the nuclei of the merging galaxies, as SF consumes the available gas and the merger stage advances \citep{1999ApJ...512L..99G}.

In NGC\,3110, the molecular gas concentration might be as large as that produced by a bar (\S\,\ref{discuconcentration}). 
{ We argue that interactions with a minor companion can be a mechanism to build large gas central concentrations in galaxies, triggering starbursts and feeding AGNs, without the need of a bar.} One of our results is that there is a net positive inflow rate to the circumnuclear regions.
The { additional} centrally concentrated mass due to the interaction can be estimated to be a few percent of the total molecular gas mass, by comparing the observed surface density of the interacting system with the average of non-barred galaxies.
In the case of NGC\,3110, { the extra mass that is expected to have been deposited into the inner 500\,pc is $\sim$ 5 $\times$ 10$^8$ M$_\odot$ (as estimated from \S\,\ref{discuconcentration}) in about 150\,Myrs (from the best morphological match with the numerical simulations, \S\,\ref{subsec:simulations})}, and therefore the molecular gas inflow  rate is around 3\,M$_\odot$\,yr$^{-1}$. The molecular gas will accumulate in the inner region because the observed SFR in that area is twice smaller than the inflow rate.

There is a peak in the SF history of mergers occurring mostly in centrally concentrated starbursts triggered during coalescence, when shocks drive large quantities of gas and dust into the nuclear regions (e.g. \citealt{2006asup.book..285L}). Gas can also be channeled into the nuclear regions of the progenitors due to the tidal forces during the first close approach, and this should be a noteworthy epoch in the evolution of these systems. 
 We find in the two objects under study that starbursts with gas surface densities of $\gtrsim$ 10$^{2.5}$\,M$_\odot$ can be found which, although small by an order of magnitude compared to those in ULIRGs,  are considerably larger than in an average non-interacting galaxy. 
We also note that soon after the first approach spiral arms and other filaments are formed, where gas is efficiently compressed and high SFRs produced { (see Fig.\,\ref{simulation})}. This is in agreement with high resolution numerical simulations in \citet{2010ApJ...720L.149T}, who argue that one of the dominant processes in the increase of SFR is actually gas fragmentation into massive and dense clouds along filaments which can be far away from the center and where SF occurs efficiently.
Moreover, from the derived SF history of the numerical simulation of a system like Antennae \citep[Fig.\,2,][]{2010ApJ...720L.149T}, we can see that although during coalescence there are extreme values of SFR (with a peak in SFR one order of magnitude larger than during the approaching phase), the total amount of SF during that shorter period of about 100\,Myr, is comparable to that in the period between the first and second pericenters ($\sim$200\,Myrs). { To our knowledge simulations of minor mergers (whose merging process lasts a longer period of time) such as the case of NGC\,3110 do exist in the literature but do not in general study the SF history, and it would be worth further investigation.} 

\subsection{Distribution of molecular gas, \ion{H}{2} regions, and formation of SSCs in NGC\,3110 }
\label{subsec:off}

\begin{figure*}[htbp]
\begin{center}
\includegraphics[width=8cm]{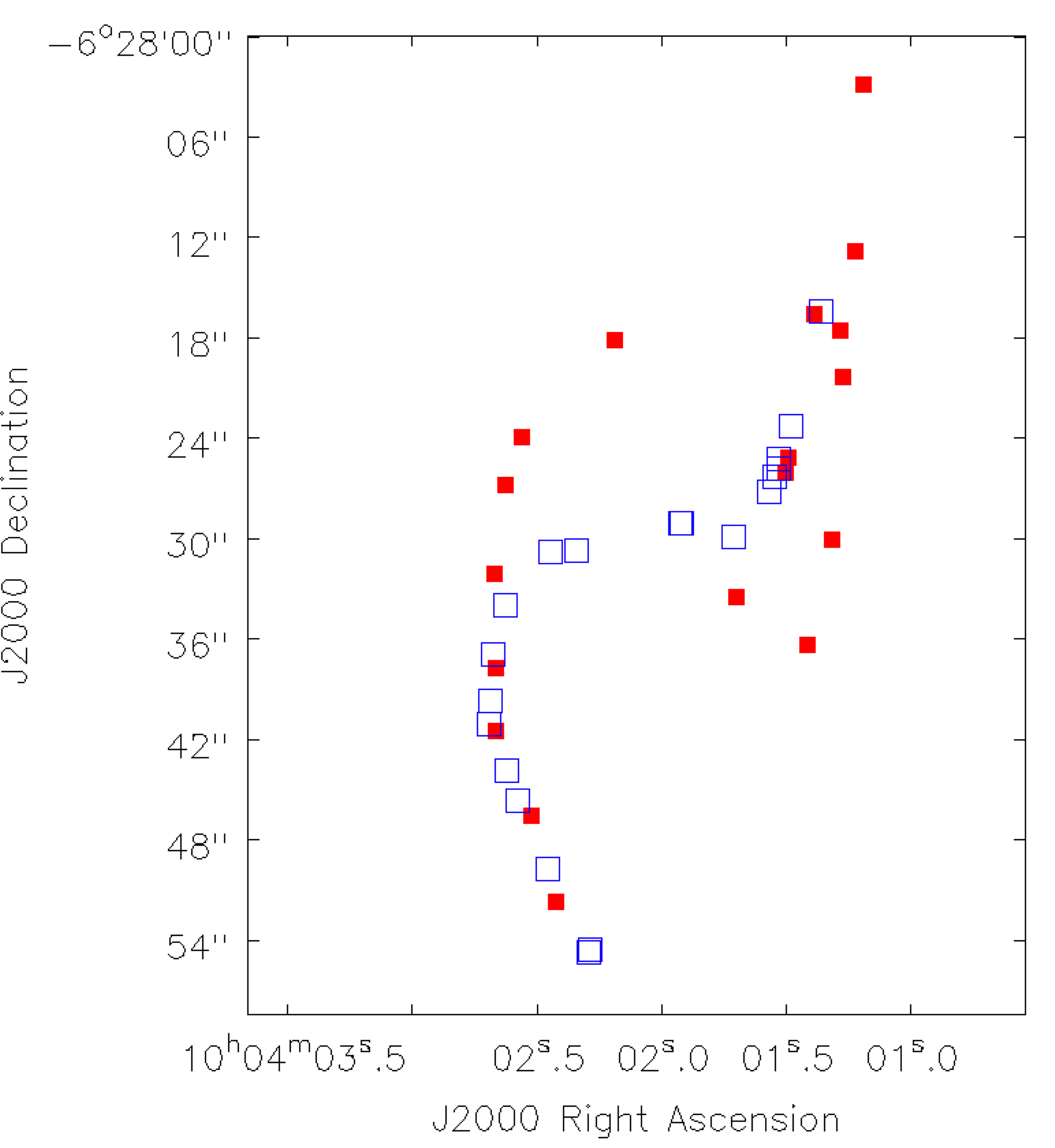}
\caption{Data points representing the location of some of the main molecular clumps (blue open squares) and \ion{H}{2} regions (red filled squares) along the two spiral arms obtained from two dimensional Gaussian fits on the images shown in Fig.\ref{fig2} (left). { See \S.\ref{subsec:off} for details.}}
\label{figoffsets}
\end{center}
\end{figure*}

 In NGC\,3110, H$\alpha$ and the molecular gas emissions are well correlated spatially. 
 In the CO interferometric observations we estimate that the position accuracy is of the order of $\theta$ / (2 $\times$ S/N) \citep{1988ApJ...330..809R}, where $\theta$ is the beam size and S/N is the signal to noise ratio,  so for regions detected with S/N $\sim$ 5 the accuracy is roughly 1/10 of the beam size. If we add other sources of uncertainties such as phase calibration, we estimate that the accuracy is no better than 1/5, or 0\farcs6 (210\,pc). In Fig.\,\ref{figoffsets} 
we present the comparison of the locations of the main molecular clumps and \ion{H}{2} regions along the two spiral arms. 
{ The identification of the molecular clumps was performed on the CO(2--1) channel map. For each channel we selected the main clumps above a threshold of 0.09\,Jy\,beam$^{-1}$ along the arms (excluding the circumnuclear regions) and we fitted Gaussian curves around the peaks to obtain their central positions. For the H$\alpha$ data points, we used a threshold of 6 $\times$ 10$^{-16}$\,erg\,s$^{-1}$\,cm$^{-2}$ in the H$\alpha$ map and obtained the central positions in the same manner. We note that although the identification of molecular clumps and HII regions is not complete, it serves to investigate a possible offset between both components.}

The lack of an offset in the southern arm is in agreement with simulations that suggest that shocks in dynamic arms show little to no spatial offsets between stellar and gas arms \citep{2015PASJ...67L...4B,2011ApJ...735....1W}. Simulations of tidal arms tell a more mixed picture, with some simulations showing little evidence of offsets \citep[e.g.][]{2010MNRAS.403..625D,2011MNRAS.414.2498S}, and others showing clear and often very large offsets (up to half a kpc) { that} are not uniform in time or location in the disk \citep{2016MNRAS.458.3990P,2017MNRAS.468.4189P}. It is important to note however that M51, a clearly interacting system, displays some of the clearest evidence for offsets between arm tracers \citep[e.g.][]{2017MNRAS.465..460E}. It may be that offsets in the southern arm have simply not manifested yet, or that they are not detectible with the current spatial resolution. We cannot { discount} that an offset exists in the northern arm as large portions of it remain undetected in CO emission.

Hundreds of SSCs were found along the filamentary structures of NGC\,3110 (\S\,\ref{subsec:sscs}). 
SSCs have been seen to form in mergers, even at several kpc from their center, as in Antennae galaxies \citep{1995AJ....109..960W}. However, 
our comparison proves { that in an initial phase of the merger, and even when the interacting galaxy mass ratio is large, a substantial population of SSCs can be formed}. 
The distribution of SSCs in the northern and southern arms seem to be more symmetric to each other than in the CO emission, and match well the \ion{H}{2} regions. Molecular gas surface densities are at least a factor of five smaller  in the northern arm. There is also a region almost at the end of the southern arm where there is a higher density of SSCs, bright H$\alpha$, but quite weak CO emission.
Although in general the molecular clumps and \ion{H}{2} regions along the southern arm are well { co-located}, we find that there are regions which are weak in CO and are clearly detected in H$\alpha$ emission, specifically the northern side of the opposite arm, which is less confined than the one in the south, as well as about 6--12\arcsec\ to the NE and SW from the nucleus. These could be explained by molecular gas being consumed by SF, and then ionized and destroyed by the radiation and winds of the resulting SSCs. The life-time of  the molecular arms can be quite reduced due to this, in addition to the gas flows to the center. If the arms are not replenished by external gas, the life time will be of the order of 0.5\,Gyr based on the SFEs we find there.

\section{Summary and conclusions}
\label{conclusion}

{
We presented maps of the bulk of molecular gas of the spiral galaxies NGC\,3110 and NGC\,232 as traced by the CO(2--1) line using the Submillimeter Array (SMA), both in an early-stage phase merger and classified as LIRGs. While NGC\,3110 is interacting with a minor companion whose stellar mass is smaller by a factor of 14 at a distance of 38\,kpc and velocity difference of $\Delta V$= 235\,\kms , NGC\,232 is interacting with a similarly sized object at a distance of 50\,kpc and $\Delta V$=120\,\kms  (projected along the line of sight). The angular resolution achieved in our CO(2--1) observations with SMA is $\sim$ 3\farcs2, or 1.1 and 1.3\,kpc, respectively. Our results illustrate the properties of molecular gas and star formation in the early stages of an interaction { well before coalescence}. Our main findings are summarized as follows:

\begin{itemize}
\item The tidal interactions are likely the primary cause to induce in the main galaxies the formation of the spiral arms and the starburst events.
The molecular gas properties of the two objects are remarkably different. While NGC\,3110 presents molecular gas along the (stellar) spiral arms, no detection is found to our sensitivity limit along the spiral arms of NGC\,232. The distribution and kinematics of the molecular gas and H$\alpha$ emission of NGC\,232 are found to be more centrally concentrated and chaotic than in the case of NGC\,3110. These different properties are likely due to the larger tidal forces exerted on NGC\,232.

\item There is an excess of molecular gas mass ($DEF_{mol}$ = --0.3 -- --0.6) compared with non-interacting galaxies with similar stellar luminosities and morphological types. At the same time we find \ion{H}{1} deficiency of about $DEF_{HI}$ = 0.4 -- 0.5 in the two objects, which lead us to argue that the molecular gas excess might be partly due to the conversion of atomic gas content into molecular gas.

\item The molecular gas concentration, calculated as the ratio of surface densities corresponding to the circumnuclear region and the whole optical disk, is at the high end of non-strongly interacting and non-barred objects, and comparable to the low end of barred galaxies. This suggests that in the early stages of the interaction, the main responsible mechanism for the gas to be driven to galactic centers is the interaction itself, at least in the case of NGC\,3110, and its amplitude can be as important as in a barred system. By comparing its morphology to numerical simulations with similar parameters and the observed concentrations of non-barred non-interacting galaxies from the literature, we estimated that the gas flow rate is $\sim$3 M$_\odot$ yr$^{-1}$ in the case of NGC\,3110. With this gas flow rate the current SFR will be sustained and may probably even increase.

\item The molecular gas and SFR surface densities in the circumnuclear regions of these two objects ($\Sigma_{\rm mol}~\gtrsim10^{2.5}$\,M$_\sun$\, pc$^{-2}$, $\Sigma_{\rm SFR}~\simeq10^{-0.5}$\,M$_\sun$\, pc$^{-2}$, for 1 kpc) are one order of magnitude higher than in non-interacting objects, but smaller than in ULIRGs also by an order of magnitude. In terms of SFE all the molecular clouds are in an intermediate position between the disk galaxies and the starbursts. In NGC\,3110 the gas depletion time is 0.5\,dex smaller in the spiral arms compared to the circumnuclear region. A possible explanation is that SFE is higher along the spiral arms, although we note that a different $X_{\rm CO}$ conversion factor { and missing flux} could affect this conclusion.

\item Numerical simulations agree well with our observations in terms of morphological appearance, and in particular for the case of interactions with small companions such as NGC\,3110.  The distribution of the two-arms found in the light companion simulations matches reasonably well the distribution seen in our maps of NGC\,3110 at $\sim$ 150\,Myrs after the closest approach. { The current status of NGC\,3110 may represent one of the highest SFR episodes as a result of the interaction with the minor companion before coalescence}. We also note differences with the numerical simulations, in the sense that resulting SFR and molecular gas densities are not as large { or at the same epoch} as in our observational results in the minor companion case. The more chaotic distribution in NGC\,232 and the weaker gaseous spiral arms (if any) are in general well reproduced by simulations of more massive companions.

\item A large population of $>$350 super stellar clusters (SSCs) were found in NGC\,3110 (to our knowledge no such observations exist for NGC\,232 yet). They are mostly confined within the circumnuclear regions and along the spiral arms. { There is a correlation between molecular gas and the existence of SSCs, in the sense that regions with higher molecular gas surface densities host a larger number of SSCs.}

\item In general  we find that molecular clumps, \ion{H}{2} regions, and SSCs are well collocated. We do not find large offsets (to about 200\,pc) between the stellar, ionized, and molecular components  along most of the southern arm. However, we note that there are some deviations. { Some regions that present weak or no CO emission are clearly detected in H$\alpha$ emission, and have a large population of SSCs}. This is most obvious in the northern side of the northern arm, which is less confined than the one in the S, as well as about 6--12\arcsec\ to the NE and SW from the nucleus, and the far end of the southern arm. { These could be explained by differences in the gas densities on the different sides of the galaxy as seen in simulations, radial gas inflows, and/or molecular gas being consumed by SF, and then ionized and destroyed by the radiation and winds of the resulting star formation, and in particular SSCs.}

\end{itemize}
}

\acknowledgements{We thank the referee for the careful reading and valuable suggestions which helped to improve this paper substantially. This research made use of the NASA/IPAC Extragalactic
Database (NED), which is operated by the Jet Propulsion Laboratory,
California Institute of Technology, under contract with the National Aeronautics
and Space Administration. We acknowledge the usage of the HyperLeda database (http://leda.univ-lyon1.fr). 
We thank Dr. T. Hattori and Dr. H. Schmitt for kindly providing H$\alpha$ maps used in this paper to derive the SF laws. 
DE was supported by a Marie Curie International Fellowship within the Sixth European Community Framework Programme (MOIF-CT-2006-40298).
SV acknowledges support by the research projects AYA2014-53506-P and AYA2017-84897-P from the Spanish Ministerio de Economia y Competitividad, from the European Regional Development Funds (FEDER) and the Junta de Andalucia (Spain) grants FQM108.
TS acknowledges funding from the European Research Council (ERC) under the European Union's Horizon 2020 research and innovation programme (grant agreement No. 694343).
LVM acknowledges support from the grant AYA2015-65973-C3-1-R (MINECO/FEDER, UE).
SaM is supported by the Ministry of Science and Technology (MOST) of Taiwan, MOST 106-2112-M-001-011 and 107-2119-M-001-020.
MAF is grateful for financial support from the CONICYT Astronomy Program CAS-CONICYT project No.\,CAS17002, sponsored by the Chinese Academy of Sciences (CAS), through a grant to the CAS South America Center for Astronomy (CASSACA) in Santiago, Chile.
This research made use of Astropy, a community-developed core Python (http://www.python.org) package for Astronomy \citep{2013A&A...558A..33A, 2018arXiv180102634T}; ipython \citep{PER-GRA:2007}; matplotlib \citep{Hunter:2007}; APLpy, an open-source plotting package for Python \citep{2012ascl.soft08017R} and NumPy \citep{2011arXiv1102.1523V}.
{We utilized the pynbody python package \citep{2013ascl.soft05002P} for post-processing and
analyzing of the tipsy files created by gasoline2. Special thanks to the authors of gasoline2
\citep{2017MNRAS.471.2357W} for providing the numerical code used to perform the simulations.}
}

\facilities{SMA,Spitzer}

\bibliography{FlyBy}

\begin{thebibliography}{}
\expandafter\ifx\csname natexlab\endcsname\relax\def\natexlab#1{#1}\fi

\bibitem[{{Albrecht} {et~al.}(2007){Albrecht}, {Kr{\"u}gel}, \&
  {Chini}}]{2007A&A...462..575A}
{Albrecht}, M., {Kr{\"u}gel}, E., \& {Chini}, R. 2007, \aap, 462, 575

\bibitem[{{Argudo-Fern{\'a}ndez} {et~al.}(2013){Argudo-Fern{\'a}ndez},
  {Verley}, {Bergond}, {Sulentic}, {Sabater}, {Fern{\'a}ndez Lorenzo}, {Leon},
  {Espada}, {Verdes-Montenegro}, {Santander-Vela}, {Ruiz}, \& {S{\'a}nchez-
  Exp{\'o}sito}}]{2013A&A...560A...9A}
{Argudo-Fern{\'a}ndez}, M., {Verley}, S., {Bergond}, G., {et~al.} 2013, \aap,
  560, A9

\bibitem[{{Argudo-Fern{\'a}ndez} {et~al.}(2014){Argudo-Fern{\'a}ndez},
  {Verley}, {Bergond}, {Sulentic}, {Sabater}, {Fern{\'a}ndez Lorenzo},
  {Espada}, {Leon}, {S{\'a}nchez-Exp{\'o}sito}, {Santander-Vela}, \&
  {Verdes-Montenegro}}]{2014A&A...564A..94A}
---. 2014, \aap, 564, A94

\bibitem[{{Argudo-Fern{\'a}ndez} {et~al.}(2015){Argudo-Fern{\'a}ndez},
  {Verley}, {Bergond}, {Duarte Puertas}, {Ramos Carmona}, {Sabater},
  {Fern{\'a}ndez Lorenzo}, {Espada}, {Sulentic}, {Ruiz}, \&
  {Leon}}]{2015A&A...578A.110A}
---. 2015, \aap, 578, A110

\bibitem[{{Armus} {et~al.}(2009){Armus}, {Mazzarella}, {Evans}, {Surace},
  {Sanders}, {Iwasawa}, {Frayer}, {Howell}, {Chan}, {Petric}, {Vavilkin},
  {Kim}, {Haan}, {Inami}, {Murphy}, {Appleton}, {Barnes}, {Bothun}, {Bridge},
  {Charmandaris}, {Jensen}, {Kewley}, {Lord}, {Madore}, {Marshall},
  {Melbourne}, {Rich}, {Satyapal}, {Schulz}, {Spoon}, {Sturm}, {U}, {Veilleux},
  \& {Xu}}]{2009PASP..121..559A}
{Armus}, L., {Mazzarella}, J.~M., {Evans}, A.~S., {et~al.} 2009, \pasp, 121,
  559

\bibitem[{{Astropy Collaboration} {et~al.}(2013){Astropy Collaboration},
  {Robitaille}, {Tollerud}, {Greenfield}, {Droettboom}, {Bray}, {Aldcroft},
  {Davis}, {Ginsburg}, {Price-Whelan}, {Kerzendorf}, {Conley}, {Crighton},
  {Barbary}, {Muna}, {Ferguson}, {Grollier}, {Parikh}, {Nair}, {Unther},
  {Deil}, {Woillez}, {Conseil}, {Kramer}, {Turner}, {Singer}, {Fox}, {Weaver},
  {Zabalza}, {Edwards}, {Azalee Bostroem}, {Burke}, {Casey}, {Crawford},
  {Dencheva}, {Ely}, {Jenness}, {Labrie}, {Lim}, {Pierfederici}, {Pontzen},
  {Ptak}, {Refsdal}, {Servillat}, \& {Streicher}}]{2013A&A...558A..33A}
{Astropy Collaboration}, {Robitaille}, T.~P., {Tollerud}, E.~J., {et~al.} 2013,
  \aap, 558, doi:10.1051/0004-6361/201322068

\bibitem[{{Baba} {et~al.}(2015){Baba}, {Morokuma-Matsui}, \&
  {Egusa}}]{2015PASJ...67L...4B}
{Baba}, J., {Morokuma-Matsui}, K., \& {Egusa}, F. 2015, \pasj, 67, L4

\bibitem[{{Barnes} {et~al.}(2001){Barnes}, {Staveley-Smith}, {de Blok},
  {Oosterloo}, {Stewart}, {Wright}, {Banks}, {Bhathal}, {Boyce}, {Calabretta},
  {Disney}, {Drinkwater}, {Ekers}, {Freeman}, {Gibson}, {Green}, {Haynes}, {te
  Lintel Hekkert}, {Henning}, {Jerjen}, {Juraszek}, {Kesteven}, {Kilborn},
  {Knezek}, {Koribalski}, {Kraan-Korteweg}, {Malin}, {Marquarding}, {Minchin},
  {Mould}, {Price}, {Putman}, {Ryder}, {Sadler}, {Schr{\"o}der}, {Stootman},
  {Webster}, {Wilson}, \& {Ye}}]{2001MNRAS.322..486B}
{Barnes}, D.~G., {Staveley-Smith}, L., {de Blok}, W.~J.~G., {et~al.} 2001,
  \mnras, 322, 486

\bibitem[{{Barnes} \& {Hernquist}(1992)}]{1992ARAA..30..705B}
{Barnes}, J.~E., \& {Hernquist}, L. 1992, \araa, 30, 705

\bibitem[{{Barnes} \& {Hernquist}(1996)}]{1996ApJ...471..115B}
---. 1996, \apj, 471, 115

\bibitem[{{Bigiel} {et~al.}(2008){Bigiel}, {Leroy}, {Walter}, {Brinks}, {de
  Blok}, {Madore}, \& {Thornley}}]{2008AJ....136.2846B}
{Bigiel}, F., {Leroy}, A., {Walter}, F., {et~al.} 2008, \aj, 136, 2846

\bibitem[{{Bolatto} {et~al.}(2013){Bolatto}, {Wolfire}, \&
  {Leroy}}]{2013ARA&A..51..207B}
{Bolatto}, A.~D., {Wolfire}, M., \& {Leroy}, A.~K. 2013, \araa, 51, 207

\bibitem[{{Bridge} {et~al.}(2010){Bridge}, {Carlberg}, \&
  {Sullivan}}]{2010ApJ...709.1067B}
{Bridge}, C.~R., {Carlberg}, R.~G., \& {Sullivan}, M. 2010, \apj, 709, 1067

\bibitem[{{Bryant} \& {Scoville}(1999)}]{1999AJ....117.2632B}
{Bryant}, P.~M., \& {Scoville}, N.~Z. 1999, \aj, 117, 2632

\bibitem[{{Bushouse}(1987)}]{1987ApJ...320...49B}
{Bushouse}, H.~A. 1987, \apj, 320, 49

\bibitem[{{Calzetti} {et~al.}(2007){Calzetti}, {Kennicutt}, {Engelbracht},
  {Leitherer}, {Draine}, {Kewley}, {Moustakas}, {Sosey}, {Dale}, {Gordon},
  {Helou}, {Hollenbach}, {Armus}, {Bendo}, {Bot}, {Buckalew}, {Jarrett}, {Li},
  {Meyer}, {Murphy}, {Prescott}, {Regan}, {Rieke}, {Roussel}, {Sheth}, {Smith},
  {Thornley}, \& {Walter}}]{2007ApJ...666..870C}
{Calzetti}, D., {Kennicutt}, R.~C., {Engelbracht}, C.~W., {et~al.} 2007, \apj,
  666, 870

\bibitem[{{Calzetti} {et~al.}(2010){Calzetti}, {Wu}, {Hong}, {Kennicutt},
  {Lee}, {Dale}, {Engelbracht}, {van Zee}, {Draine}, {Hao}, {Gordon},
  {Moustakas}, {Murphy}, {Regan}, {Begum}, {Block}, {Dalcanton}, {Funes}, {Gil
  de Paz}, {Johnson}, {Sakai}, {Skillman}, {Walter}, {Weisz}, {Williams}, \&
  {Wu}}]{2010ApJ...714.1256C}
{Calzetti}, D., {Wu}, S.-Y., {Hong}, S., {et~al.} 2010, \apj, 714, 1256

\bibitem[{{Chini} {et~al.}(1996){Chini}, {Kruegel}, \&
  {Lemke}}]{1996A&AS..118...47C}
{Chini}, R., {Kruegel}, E., \& {Lemke}, R. 1996, \aaps, 118, 47

\bibitem[{{Colina} {et~al.}(2015){Colina}, {Piqueras L{\'o}pez}, {Arribas},
  {Riffel}, {Riffel}, {Rodriguez-Ardila}, {Pastoriza}, {Storchi-Bergmann},
  {Alonso-Herrero}, \& {Sales}}]{2015A&A...578A..48C}
{Colina}, L., {Piqueras L{\'o}pez}, J., {Arribas}, S., {et~al.} 2015, \aap,
  578, A48

\bibitem[{{Corbett} {et~al.}(2002){Corbett}, {Norris}, {Heisler}, {Dopita},
  {Appleton}, {Struck}, {Murphy}, {Marston}, {Charmandaris}, {Kewley}, \&
  {Zezas}}]{2002ApJ...564..650C}
{Corbett}, E.~A., {Norris}, R.~P., {Heisler}, C.~A., {et~al.} 2002, \apj, 564,
  650

\bibitem[{{Corbett} {et~al.}(2003){Corbett}, {Kewley}, {Appleton},
  {Charmandaris}, {Dopita}, {Heisler}, {Norris}, {Zezas}, \&
  {Marston}}]{2003ApJ...583..670C}
{Corbett}, E.~A., {Kewley}, L., {Appleton}, P.~N., {et~al.} 2003, \apj, 583,
  670

\bibitem[{{Cox}(2000)}]{2000asqu.book.....C}
{Cox}, A.~N. 2000, {Allen's astrophysical quantities}

\bibitem[{{Cox} {et~al.}(2008){Cox}, {Jonsson}, {Somerville}, {Primack}, \&
  {Dekel}}]{2008MNRAS.384..386C}
{Cox}, T.~J., {Jonsson}, P., {Somerville}, R.~S., {Primack}, J.~R., \& {Dekel},
  A. 2008, \mnras, 384, 386

\bibitem[{{Coziol} {et~al.}(2000){Coziol}, {Iovino}, \& {de
  Carvalho}}]{2000AJ....120...47C}
{Coziol}, R., {Iovino}, A., \& {de Carvalho}, R.~R. 2000, \aj, 120, 47

\bibitem[{{Daddi} {et~al.}(2010){Daddi}, {Elbaz}, {Walter}, {Bournaud},
  {Salmi}, {Carilli}, {Dannerbauer}, {Dickinson}, {Monaco}, \&
  {Riechers}}]{2010ApJ...714L.118D}
{Daddi}, E., {Elbaz}, D., {Walter}, F., {et~al.} 2010, \apjl, 714, L118

\bibitem[{{Dame} {et~al.}(2001){Dame}, {Hartmann}, \&
  {Thaddeus}}]{2001ApJ...547..792D}
{Dame}, T.~M., {Hartmann}, D., \& {Thaddeus}, P. 2001, \apj, 547, 792

\bibitem[{{de Vaucouleurs} {et~al.}(1991){de Vaucouleurs}, {de Vaucouleurs},
  {Corwin}, {Buta}, {Paturel}, \& {Fouqu{\'e}}}]{1991rc3..book.....D}
{de Vaucouleurs}, G., {de Vaucouleurs}, A., {Corwin}, Jr., H.~G., {et~al.}
  1991, {Third Reference Catalogue of Bright Galaxies. Volume I: Explanations
  and references. Volume II: Data for galaxies between 0$^{h}$ and 12$^{h}$.
  Volume III: Data for galaxies between 12$^{h}$ and 24$^{h}$.}

\bibitem[{{Dobbs} {et~al.}(2010){Dobbs}, {Theis}, {Pringle}, \&
  {Bate}}]{2010MNRAS.403..625D}
{Dobbs}, C.~L., {Theis}, C., {Pringle}, J.~E., \& {Bate}, M.~R. 2010, \mnras,
  403, 625

\bibitem[{{D'Onghia} {et~al.}(2010){D'Onghia}, {Vogelsberger},
  {Faucher-Giguere}, \& {Hernquist}}]{2010ApJ...725..353D}
{D'Onghia}, E., {Vogelsberger}, M., {Faucher-Giguere}, C.-A., \& {Hernquist},
  L. 2010, \apj, 725, 353

\bibitem[{{Dopita} {et~al.}(2002){Dopita}, {Pereira}, {Kewley}, \&
  {Capaccioli}}]{2002ApJS..143...47D}
{Dopita}, M.~A., {Pereira}, M., {Kewley}, L.~J., \& {Capaccioli}, M. 2002,
  \apjs, 143, 47

\bibitem[{{Downes} \& {Solomon}(1998)}]{1998ApJ...507..615D}
{Downes}, D., \& {Solomon}, P.~M. 1998, \apj, 507, 615

\bibitem[{{Dubinski} \& {Chakrabarty}(2009)}]{2009ApJ...703.2068D}
{Dubinski}, J., \& {Chakrabarty}, D. 2009, \apj, 703, 2068

\bibitem[{{Egusa} {et~al.}(2017){Egusa}, {Mentuch Cooper}, {Koda}, \&
  {Baba}}]{2017MNRAS.465..460E}
{Egusa}, F., {Mentuch Cooper}, E., {Koda}, J., \& {Baba}, J. 2017, \mnras, 465,
  460

\bibitem[{{Ellison} {et~al.}(2008){Ellison}, {Patton}, {Simard}, \&
  {McConnachie}}]{2008AJ....135.1877E}
{Ellison}, S.~L., {Patton}, D.~R., {Simard}, L., \& {McConnachie}, A.~W. 2008,
  \aj, 135, 1877

\bibitem[{{Espada} {et~al.}(2011){Espada}, {Verdes-Montenegro}, {Huchtmeier},
  {Sulentic}, {Verley}, {Leon}, \& {Sabater}}]{2011A&A...532A.117E}
{Espada}, D., {Verdes-Montenegro}, L., {Huchtmeier}, W.~K., {et~al.} 2011,
  \aap, 532, A117

\bibitem[{{Espada} {et~al.}(2010){Espada}, {Martin}, {Hsieh}, {Ho},
  {Matsushita}, {Verdes-Montenegro}, {Sabater}, {Verley}, {Krips}, \&
  {Espigares}}]{2010gama.conf...97E}
{Espada}, D., {Martin}, S., {Hsieh}, P.-Y., {et~al.} 2010, in Galaxies and
  their Masks, ed. D.~L. {Block}, K.~C. {Freeman}, \& I.~{Puerari}, 97

\bibitem[{{Espada} {et~al.}(2012){Espada}, {Komugi}, {Muller}, {Nakanishi},
  {Saito}, {Tatematsu}, {Iguchi}, {Hasegawa}, {Mizuno}, {Iono}, {Matsushita},
  {Trejo}, {Chapillon}, {Takahashi}, {Su}, {Kawamura}, {Akiyama}, {Hiramatsu},
  {Nagai}, {Miura}, {Kurono}, {Sawada}, {Higuchi}, {Tachihara}, {Saigo}, \&
  {Kamazaki}}]{2012ApJ...760L..25E}
{Espada}, D., {Komugi}, S., {Muller}, E., {et~al.} 2012, \apjl, 760, L25

\bibitem[{{Gao} \& {Solomon}(1999)}]{1999ApJ...512L..99G}
{Gao}, Y., \& {Solomon}, P.~M. 1999, \apjl, 512, L99

\bibitem[{{Gordon} {et~al.}(2005){Gordon}, {Rieke}, {Engelbracht}, {Muzerolle},
  {Stansberry}, {Misselt}, {Morrison}, {Cadien}, {Young}, {Dole}, {Kelly},
  {Alonso-Herrero}, {Egami}, {Su}, {Papovich}, {Smith}, {Hines}, {Rieke},
  {Blaylock}, {P{\'e}rez-Gonz{\'a}lez}, {Le Floc'h}, {Hinz}, {Latter},
  {Hesselroth}, {Frayer}, {Noriega-Crespo}, {Masci}, {Padgett}, {Smylie}, \&
  {Haegel}}]{2005PASP..117..503G}
{Gordon}, K.~D., {Rieke}, G.~H., {Engelbracht}, C.~W., {et~al.} 2005, \pasp,
  117, 503

\bibitem[{{Greve} {et~al.}(2009){Greve}, {Papadopoulos}, {Gao}, \&
  {Radford}}]{2009ApJ...692.1432G}
{Greve}, T.~R., {Papadopoulos}, P.~P., {Gao}, Y., \& {Radford}, S.~J.~E. 2009,
  \apj, 692, 1432

\bibitem[{{Hattori} {et~al.}(2004){Hattori}, {Yoshida}, {Ohtani}, {Sugai},
  {Ishigaki}, {Sasaki}, {Hayashi}, {Ozaki}, {Ishii}, \&
  {Kawai}}]{2004AJ....127..736H}
{Hattori}, T., {Yoshida}, M., {Ohtani}, H., {et~al.} 2004, \aj, 127, 736

\bibitem[{{Herrera} \& {Boulanger}(2017)}]{2017A&A...600A.139H}
{Herrera}, C.~N., \& {Boulanger}, F. 2017, \aap, 600, A139

\bibitem[{{Ho} {et~al.}(2004){Ho}, {Moran}, \& {Lo}}]{2004ApJ...616L...1H}
{Ho}, P.~T.~P., {Moran}, J.~M., \& {Lo}, K.~Y. 2004, \apjl, 616, L1

\bibitem[{{Hopkins} {et~al.}(2013){Hopkins}, {Cox}, {Hernquist}, {Narayanan},
  {Hayward}, \& {Murray}}]{2013MNRAS.430.1901H}
{Hopkins}, P.~F., {Cox}, T.~J., {Hernquist}, L., {et~al.} 2013, \mnras, 430,
  1901

\bibitem[{Hunter(2007)}]{Hunter:2007}
Hunter, J.~D. 2007, Computing In Science \& Engineering, 9, 90

\bibitem[{{Imanishi} \& {Nakanishi}(2013)}]{2013AJ....146...47I}
{Imanishi}, M., \& {Nakanishi}, K. 2013, \aj, 146, 47

\bibitem[{{Imanishi} {et~al.}(2009){Imanishi}, {Nakanishi}, {Tamura}, \&
  {Peng}}]{2009AJ....137.3581I}
{Imanishi}, M., {Nakanishi}, K., {Tamura}, Y., \& {Peng}, C.-H. 2009, \aj, 137,
  3581

\bibitem[{{Iono} {et~al.}(2005){Iono}, {Yun}, \& {Ho}}]{2005ApJS..158....1I}
{Iono}, D., {Yun}, M.~S., \& {Ho}, P.~T.~P. 2005, \apjs, 158, 1

\bibitem[{{Iono} {et~al.}(2004){Iono}, {Yun}, \& {Mihos}}]{2004ApJ...616..199I}
{Iono}, D., {Yun}, M.~S., \& {Mihos}, J.~C. 2004, \apj, 616, 199

\bibitem[{{Iono} {et~al.}(2009){Iono}, {Wilson}, {Yun}, {Baker}, {Petitpas},
  {Peck}, {Krips}, {Cox}, {Matsushita}, {Mihos}, \&
  {Pihlstrom}}]{2009ApJ...695.1537I}
{Iono}, D., {Wilson}, C.~D., {Yun}, M.~S., {et~al.} 2009, \apj, 695, 1537

\bibitem[{{Iono} {et~al.}(2013){Iono}, {Saito}, {Yun}, {Kawabe}, {Espada},
  {Hagiwara}, {Imanishi}, {Izumi}, {Kohno}, {Motohara}, {Nakanishi}, {Sugai},
  {Tateuchi}, {Tamura}, {Ueda}, \& {Yoshii}}]{2013PASJ...65L...7I}
{Iono}, D., {Saito}, T., {Yun}, M.~S., {et~al.} 2013, \pasj, 65, L7

\bibitem[{{Iovino}(2002)}]{2002AJ....124.2471I}
{Iovino}, A. 2002, \aj, 124, 2471

\bibitem[{{Jones} {et~al.}(2018){Jones}, {Espada}, {Verdes-Montenegro},
  {Huchtmeier}, {Lisenfeld}, {Leon}, {Sulentic}, {Sabater}, {Jones}, {Sanchez},
  \& {Garrido}}]{2017arXiv171003034J}
{Jones}, M.~G., {Espada}, D., {Verdes-Montenegro}, L., {et~al.} 2018, \aap,
  609, A17

\bibitem[{{Kaneko} {et~al.}(2017){Kaneko}, {Kuno}, {Iono}, {Tamura}, {Tosaki},
  {Nakanishi}, \& {Sawada}}]{2017PASJ...69...66K}
{Kaneko}, H., {Kuno}, N., {Iono}, D., {et~al.} 2017, Publications of the
  Astronomical Society of Japan, 69, 66

\bibitem[{{Kennicutt}(1998{\natexlab{a}})}]{1998ARA&A..36..189K}
{Kennicutt}, Robert~C., J. 1998{\natexlab{a}}, Annual Review of Astronomy and
  Astrophysics, 36, 189

\bibitem[{{Kennicutt}(1998{\natexlab{b}})}]{1998ApJ...498..541K}
{Kennicutt}, Jr., R.~C. 1998{\natexlab{b}}, \apj, 498, 541

\bibitem[{{Kim} {et~al.}(2014){Kim}, {Peirani}, {Kim}, {Ann}, {An}, \&
  {Yoon}}]{2014ApJ...789...90K}
{Kim}, J.~H., {Peirani}, S., {Kim}, S., {et~al.} 2014, \apj, 789, 90

\bibitem[{{Kroupa}(2001)}]{2001MNRAS.322..231K}
{Kroupa}, P. 2001, \mnras, 322, 231

\bibitem[{{Lang} {et~al.}(2014){Lang}, {Holley-Bockelmann}, \&
  {Sinha}}]{2014ApJ...790L..33L}
{Lang}, M., {Holley-Bockelmann}, K., \& {Sinha}, M. 2014, \apjl, 790, L33

\bibitem[{{Larson} {et~al.}(2016){Larson}, {Sanders}, {Barnes}, {Ishida},
  {Evans}, {U}, {Mazzarella}, {Kim}, {Privon}, {Mirabel}, \&
  {Flewelling}}]{2016ApJ...825..128L}
{Larson}, K.~L., {Sanders}, D.~B., {Barnes}, J.~E., {et~al.} 2016, \apj, 825,
  128

\bibitem[{{Leitherer} {et~al.}(1999){Leitherer}, {Schaerer}, {Goldader},
  {Delgado}, {Robert}, {Kune}, {de Mello}, {Devost}, \&
  {Heckman}}]{1999ApJS..123....3L}
{Leitherer}, C., {Schaerer}, D., {Goldader}, J.~D., {et~al.} 1999, \apjs, 123,
  3

\bibitem[{{Leroy} {et~al.}(2013){Leroy}, {Walter}, {Sandstrom}, {Schruba},
  {Munoz-Mateos}, {Bigiel}, {Bolatto}, {Brinks}, {de Blok}, {Meidt}, {Rix},
  {Rosolowsky}, {Schinnerer}, {Schuster}, \& {Usero}}]{2013AJ....146...19L}
{Leroy}, A.~K., {Walter}, F., {Sandstrom}, K., {et~al.} 2013, \aj, 146, 19

\bibitem[{{Lisenfeld} {et~al.}(2011){Lisenfeld}, {Espada}, {Verdes-Montenegro},
  {Kuno}, {Leon}, {Sabater}, {Sato}, {Sulentic}, {Verley}, \&
  {Yun}}]{2011A&A...534A.102L}
{Lisenfeld}, U., {Espada}, D., {Verdes-Montenegro}, L., {et~al.} 2011, \aap,
  534, A102

\bibitem[{{{\L}okas}(2018)}]{2018ApJ...857....6L}
{{\L}okas}, E.~L. 2018, \apj, 857, 6

\bibitem[{{Lonsdale} {et~al.}(2006){Lonsdale}, {Farrah}, \&
  {Smith}}]{2006asup.book..285L}
{Lonsdale}, C.~J., {Farrah}, D., \& {Smith}, H.~E. 2006, {Ultraluminous
  Infrared Galaxies}, ed. J.~W. {Mason}, 285

\bibitem[{{L{\'o}pez-Cob{\'a}} {et~al.}(2017){L{\'o}pez-Cob{\'a}},
  {S{\'a}nchez}, {Cruz-Gonz{\'a}lez}, {Binette}, {Galbany}, {Kr{\"u}hler},
  {Rodr{\'{\i}}guez}, {Barrera-Ballesteros}, {S{\'a}nchez-Menguiano},
  {Walcher}, {Aquino-Ort{\'{\i}}z}, \& {Anderson}}]{2017ApJ...850L..17L}
{L{\'o}pez-Cob{\'a}}, C., {S{\'a}nchez}, S.~F., {Cruz-Gonz{\'a}lez}, I.,
  {et~al.} 2017, \apjl, 850, L17

\bibitem[{{L{\'o}pez-Sanjuan} {et~al.}(2015){L{\'o}pez-Sanjuan}, {Cenarro},
  {Varela}, {Viironen}, {Molino}, {Ben{\'{\i}}tez}, {Arnalte-Mur}, {Ascaso},
  {D{\'{\i}}az-Garc{\'{\i}}a}, {Fern{\'a}ndez-Soto}, {Jim{\'e}nez-Teja},
  {M{\'a}rquez}, {Masegosa}, {Moles}, {Povi{\'c}}, {Aguerri}, {Alfaro},
  {Aparicio-Villegas}, {Broadhurst}, {Cabrera-Ca{\~n}o}, {Castander}, {Cepa},
  {Cervi{\~n}o}, {Crist{\'o}bal-Hornillos}, {Del Olmo}, {Gonz{\'a}lez Delgado},
  {Husillos}, {Infante}, {Mart{\'{\i}}nez}, {Perea}, {Prada}, \&
  {Quintana}}]{2015A&A...576A..53L}
{L{\'o}pez-Sanjuan}, C., {Cenarro}, A.~J., {Varela}, J., {et~al.} 2015, \aap,
  576, A53

\bibitem[{{Makarov} {et~al.}(2014){Makarov}, {Prugniel}, {Terekhova},
  {Courtois}, \& {Vauglin}}]{2014A&A...570A..13M}
{Makarov}, D., {Prugniel}, P., {Terekhova}, N., {Courtois}, H., \& {Vauglin},
  I. 2014, \aap, 570, A13

\bibitem[{{Malkan} {et~al.}(1998){Malkan}, {Gorjian}, \&
  {Tam}}]{1998ApJS..117...25M}
{Malkan}, M.~A., {Gorjian}, V., \& {Tam}, R. 1998, \apjs, 117, 25

\bibitem[{{Martin} {et~al.}(1991){Martin}, {Bottinelli}, {Gouguenheim}, \&
  {Dennefeld}}]{1991A&A...245..393M}
{Martin}, J.~M., {Bottinelli}, L., {Gouguenheim}, L., \& {Dennefeld}, M. 1991,
  \aap, 245, 393

\bibitem[{{Mart{\'{\i}}n} {et~al.}(2010){Mart{\'{\i}}n}, {George}, {Wilner}, \&
  {Espada}}]{2010AJ....139.2241M}
{Mart{\'{\i}}n}, S., {George}, M.~R., {Wilner}, D.~J., \& {Espada}, D. 2010,
  \aj, 139, 2241

\bibitem[{{Martinez-Badenes} {et~al.}(2012){Martinez-Badenes}, {Lisenfeld},
  {Espada}, {Verdes-Montenegro}, {Garc{\'{\i}}a-Burillo}, {Leon}, {Sulentic},
  \& {Yun}}]{2012A&A...540A..96M}
{Martinez-Badenes}, V., {Lisenfeld}, U., {Espada}, D., {et~al.} 2012, \aap,
  540, A96

\bibitem[{{McMullin} {et~al.}(2007){McMullin}, {Waters}, {Schiebel}, {Young},
  \& {Golap}}]{2007ASPC..376..127M}
{McMullin}, J.~P., {Waters}, B., {Schiebel}, D., {Young}, W., \& {Golap}, K.
  2007, in Astronomical Society of the Pacific Conference Series, Vol. 376,
  Astronomical Data Analysis Software and Systems XVI, ed. R.~A. {Shaw},
  F.~{Hill}, \& D.~J. {Bell}, 127

\bibitem[{{Mihos} \& {Hernquist}(1996)}]{1996ApJ...464..641M}
{Mihos}, J.~C., \& {Hernquist}, L. 1996, \apj, 464, 641

\bibitem[{{Mihos} {et~al.}(1992){Mihos}, {Richstone}, \&
  {Bothun}}]{1992ApJ...400..153M}
{Mihos}, J.~C., {Richstone}, D.~O., \& {Bothun}, G.~D. 1992, \apj, 400, 153

\bibitem[{{Mirabel} {et~al.}(1990){Mirabel}, {Booth}, {Johansson}, {Garay}, \&
  {Sanders}}]{1990A&A...236..327M}
{Mirabel}, I.~F., {Booth}, R.~S., {Johansson}, L.~E.~B., {Garay}, G., \&
  {Sanders}, D.~B. 1990, \aap, 236, 327

\bibitem[{{Miralles-Caballero} {et~al.}(2011){Miralles-Caballero}, {Colina},
  {Arribas}, \& {Duc}}]{2011AJ....142...79M}
{Miralles-Caballero}, D., {Colina}, L., {Arribas}, S., \& {Duc}, P.-A. 2011,
  \aj, 142, 79

\bibitem[{{Miura} {et~al.}(2018){Miura}, {Espada}, {Hirota}, {Nakanishi},
  {Bendo}, \& {Sugai}}]{2018arXiv180810089M}
{Miura}, R.~E., {Espada}, D., {Hirota}, A., {et~al.} 2018, ArXiv e-prints,
  arXiv:1808.10089

\bibitem[{{Miura} {et~al.}(2015){Miura}, {Espada}, {Sugai}, {Nakanishi}, \&
  {Hirota}}]{2015PASJ...67L...1M}
{Miura}, R.~E., {Espada}, D., {Sugai}, H., {Nakanishi}, K., \& {Hirota}, A.
  2015, \pasj, 67, L1

\bibitem[{{Moore} {et~al.}(1996){Moore}, {Katz}, {Lake}, {Dressler}, \&
  {Oemler}}]{1996Natur.379..613M}
{Moore}, B., {Katz}, N., {Lake}, G., {Dressler}, A., \& {Oemler}, A. 1996,
  \nat, 379, 613

\bibitem[{{Oh} {et~al.}(2015){Oh}, {Kim}, \& {Lee}}]{2015ApJ...807...73O}
{Oh}, S.~H., {Kim}, W.-T., \& {Lee}, H.~M. 2015, \apj, 807, 73

\bibitem[{{Papadopoulos} {et~al.}(2012){Papadopoulos}, {van der Werf},
  {Xilouris}, {Isaak}, \& {Gao}}]{2012ApJ...751...10P}
{Papadopoulos}, P.~P., {van der Werf}, P., {Xilouris}, E., {Isaak}, K.~G., \&
  {Gao}, Y. 2012, \apj, 751, 10

\bibitem[{{Patton} {et~al.}(2013){Patton}, {Torrey}, {Ellison}, {Mendel}, \&
  {Scudder}}]{2013MNRAS.433L..59P}
{Patton}, D.~R., {Torrey}, P., {Ellison}, S.~L., {Mendel}, J.~T., \& {Scudder},
  J.~M. 2013, \mnras, 433, L59

\bibitem[{Perez \& Granger(2007)}]{PER-GRA:2007}
Perez, F., \& Granger, B.~E. 2007, Computing in Science and Engineering, 9, 21

\bibitem[{{Pettitt} {et~al.}(2015){Pettitt}, {Dobbs}, {Acreman}, \&
  {Bate}}]{2015MNRAS.449.3911P}
{Pettitt}, A.~R., {Dobbs}, C.~L., {Acreman}, D.~M., \& {Bate}, M.~R. 2015,
  \mnras, 449, 3911

\bibitem[{{Pettitt} {et~al.}(2016){Pettitt}, {Tasker}, \&
  {Wadsley}}]{2016MNRAS.458.3990P}
{Pettitt}, A.~R., {Tasker}, E.~J., \& {Wadsley}, J.~W. 2016, \mnras, 458, 3990

\bibitem[{{Pettitt} {et~al.}(2017){Pettitt}, {Tasker}, {Wadsley}, {Keller}, \&
  {Benincasa}}]{2017MNRAS.468.4189P}
{Pettitt}, A.~R., {Tasker}, E.~J., {Wadsley}, J.~W., {Keller}, B.~W., \&
  {Benincasa}, S.~M. 2017, \mnras, 468, 4189

\bibitem[{{Pettitt} \& {Wadsley}(2018)}]{2018MNRAS.474.5645P}
{Pettitt}, A.~R., \& {Wadsley}, J.~W. 2018, \mnras, 474, 5645

\bibitem[{{Pontzen} {et~al.}(2013){Pontzen}, {Ro{\v{s}}kar}, {Stinson}, \&
  {Woods}}]{2013ascl.soft05002P}
{Pontzen}, A., {Ro{\v{s}}kar}, R., {Stinson}, G., \& {Woods}, R. 2013,
  {pynbody: N-Body/SPH analysis for python}, Astrophysics Source Code Library,
  , , ascl:1305.002

\bibitem[{{Prandoni} {et~al.}(1994){Prandoni}, {Iovino}, \&
  {MacGillivray}}]{1994AJ....107.1235P}
{Prandoni}, I., {Iovino}, A., \& {MacGillivray}, H.~T. 1994, \aj, 107, 1235

\bibitem[{{Randriamanakoto}(2015)}]{2015PhDT.......214R}
{Randriamanakoto}, Z. 2015, PhD thesis, Department of Astronomy, University of
  Cape Town, South Africa

\bibitem[{{Randriamanakoto} \& {V{\"a}is{\"a}nen}(2017)}]{2017IAUS..316...70R}
{Randriamanakoto}, Z., \& {V{\"a}is{\"a}nen}, P. 2017, in IAU Symposium, Vol.
  316, Formation, Evolution, and Survival of Massive Star Clusters, ed.
  C.~{Charbonnel} \& A.~{Nota}, 70--76

\bibitem[{{Randriamanakoto} {et~al.}(2013){Randriamanakoto},
  {V{\"a}is{\"a}nen}, {Ryder}, {Kankare}, {Kotilainen}, \&
  {Mattila}}]{2013MNRAS.431..554R}
{Randriamanakoto}, Z., {V{\"a}is{\"a}nen}, P., {Ryder}, S., {et~al.} 2013,
  \mnras, 431, 554

\bibitem[{{Reid} {et~al.}(1988){Reid}, {Schneps}, {Moran}, {Gwinn}, {Genzel},
  {Downes}, \& {Roennaeng}}]{1988ApJ...330..809R}
{Reid}, M.~J., {Schneps}, M.~H., {Moran}, J.~M., {et~al.} 1988, \apj, 330, 809

\bibitem[{{Renaud} {et~al.}(2015){Renaud}, {Bournaud}, \&
  {Duc}}]{2015MNRAS.446.2038R}
{Renaud}, F., {Bournaud}, F., \& {Duc}, P.-A. 2015, \mnras, 446, 2038

\bibitem[{{Richter} {et~al.}(1994){Richter}, {Sackett}, \&
  {Sparke}}]{1994AJ....107...99R}
{Richter}, O.-G., {Sackett}, P.~D., \& {Sparke}, L.~S. 1994, \aj, 107, 99

\bibitem[{{Rieke} {et~al.}(2004){Rieke}, {Young}, {Cadien}, {Engelbracht},
  {Gordon}, {Kelly}, {Low}, {Misselt}, {Morrison}, {Muzerolle}, {Rivlis},
  {Stansberry}, {Beeman}, {Haller}, {Frayer}, {Latter}, {Noriega-Crespo},
  {Padgett}, {Hines}, {Bean}, {Burmester}, {Heim}, {Glenn}, {Ordonez},
  {Schwenker}, {Siewert}, {Strecker}, {Tennant}, {Troeltzsch}, {Unruh},
  {Warden}, {Ade}, {Alonso-Herrero}, {Blaylock}, {Dole}, {Egami}, {Hinz}, {Le
  Floc'h}, {Papovich}, {Perez-Gonzalez}, {Rieke}, {Smith}, {Su}, {Bennett},
  {Henderson}, {Lu}, {Masci}, {Pesenson}, {Rebull}, {Rho}, {Keene}, {Stolovy},
  {Wachter}, {Wheaton}, {Richards}, {Garner}, {Hegge}, {Henderson}, {MacFeely},
  {Michika}, {Miller}, {Neitenbach}, {Winghart}, {Woodruff}, {Arens},
  {Beichman}, {Gaalema}, {Gautier}, {Lada}, {Mould}, {Neugebauer}, \&
  {Stapelfeldt}}]{2004SPIE.5487...50R}
{Rieke}, G.~H., {Young}, E.~T., {Cadien}, J., {et~al.} 2004, in \procspie, Vol.
  5487, Optical, Infrared, and Millimeter Space Telescopes, ed. J.~C. {Mather},
  50--61

\bibitem[{{Robitaille} \& {Bressert}(2012)}]{2012ascl.soft08017R}
{Robitaille}, T., \& {Bressert}, E. 2012, {APLpy: Astronomical Plotting Library
  in Python}, Astrophysics Source Code Library, , , ascl:1208.017

\bibitem[{{Sabater} {et~al.}(2013){Sabater}, {Best}, \&
  {Argudo-Fern{\'a}ndez}}]{2013MNRAS.430..638S}
{Sabater}, J., {Best}, P.~N., \& {Argudo-Fern{\'a}ndez}, M. 2013, \mnras, 430,
  638

\bibitem[{{Saito} {et~al.}(2015){Saito}, {Iono}, {Yun}, {Ueda}, {Nakanishi},
  {Sugai}, {Espada}, {Imanishi}, {Motohara}, {Hagiwara}, {Tateuchi}, {Lee}, \&
  {Kawabe}}]{2015ApJ...803...60S}
{Saito}, T., {Iono}, D., {Yun}, M.~S., {et~al.} 2015, \apj, 803, 60

\bibitem[{{Saito} {et~al.}(2016){Saito}, {Iono}, {Xu}, {Ueda}, {Nakanishi},
  {Yun}, {Kaneko}, {Yamashita}, {Lee}, {Espada}, {Motohara}, \&
  {Kawabe}}]{2016PASJ...68...20S}
{Saito}, T., {Iono}, D., {Xu}, C.~K., {et~al.} 2016, \pasj, 68, 20

\bibitem[{{Saito} {et~al.}(2017{\natexlab{a}}){Saito}, {Iono}, {Espada},
  {Nakanishi}, {Ueda}, {Sugai}, {Takano}, {Yun}, {Imanishi}, {Ohashi}, {Lee},
  {Hagiwara}, {Motohara}, \& {Kawabe}}]{2017ApJ...834....6S}
{Saito}, T., {Iono}, D., {Espada}, D., {et~al.} 2017{\natexlab{a}}, \apj, 834,
  6

\bibitem[{{Saito} {et~al.}(2017{\natexlab{b}}){Saito}, {Iono}, {Xu}, {Sliwa},
  {Ueda}, {Espada}, {Kaneko}, {K{\"o}nig}, {Nakanishi}, {Lee}, {Yun}, {Aalto},
  {Hibbard}, {Yamashita}, {Motohara}, \& {Kawabe}}]{2017ApJ...835..174S}
{Saito}, T., {Iono}, D., {Xu}, C.~K., {et~al.} 2017{\natexlab{b}}, \apj, 835,
  174

\bibitem[{{Sakamoto} {et~al.}(1999){Sakamoto}, {Okumura}, {Ishizuki}, \&
  {Scoville}}]{1999ApJ...525..691S}
{Sakamoto}, K., {Okumura}, S.~K., {Ishizuki}, S., \& {Scoville}, N.~Z. 1999,
  \apj, 525, 691

\bibitem[{{Sanders} {et~al.}(2003){Sanders}, {Mazzarella}, {Kim}, {Surace}, \&
  {Soifer}}]{2003AJ....126.1607S}
{Sanders}, D.~B., {Mazzarella}, J.~M., {Kim}, D.-C., {Surace}, J.~A., \&
  {Soifer}, B.~T. 2003, \aj, 126, 1607

\bibitem[{{Sanders} \& {Mirabel}(1996)}]{1996ARA&A..34..749S}
{Sanders}, D.~B., \& {Mirabel}, I.~F. 1996, \araa, 34, 749

\bibitem[{{Sanders} {et~al.}(1991){Sanders}, {Scoville}, \&
  {Soifer}}]{1991ApJ...370..158S}
{Sanders}, D.~B., {Scoville}, N.~Z., \& {Soifer}, B.~T. 1991, \apj, 370, 158

\bibitem[{{Sault} {et~al.}(1995){Sault}, {Teuben}, \&
  {Wright}}]{1995ASPC...77..433S}
{Sault}, R.~J., {Teuben}, P.~J., \& {Wright}, M.~C.~H. 1995, in Astronomical
  Society of the Pacific Conference Series, Vol.~77, Astronomical Data Analysis
  Software and Systems IV, ed. R.~A. {Shaw}, H.~E. {Payne}, \& J.~J.~E.
  {Hayes}, 433

\bibitem[{{Schmitt} {et~al.}(2006{\natexlab{a}}){Schmitt}, {Calzetti}, {Armus},
  {Giavalisco}, {Heckman}, {Kennicutt}, {Leitherer}, \&
  {Meurer}}]{2006ApJS..164...52S}
{Schmitt}, H.~R., {Calzetti}, D., {Armus}, L., {et~al.} 2006{\natexlab{a}},
  \apjs, 164, 52

\bibitem[{{Schmitt} {et~al.}(2006{\natexlab{b}}){Schmitt}, {Calzetti}, {Armus},
  {Giavalisco}, {Heckman}, {Kennicutt}, {Leitherer}, \&
  {Meurer}}]{2006ApJ...643..173S}
---. 2006{\natexlab{b}}, \apj, 643, 173

\bibitem[{{Scudder} {et~al.}(2012){Scudder}, {Ellison}, {Torrey}, {Patton}, \&
  {Mendel}}]{2012MNRAS.426..549S}
{Scudder}, J.~M., {Ellison}, S.~L., {Torrey}, P., {Patton}, D.~R., \& {Mendel},
  J.~T. 2012, \mnras, 426, 549

\bibitem[{{Sliwa} {et~al.}(2017){Sliwa}, {Wilson}, {Matsushita}, {Peck},
  {Petitpas}, {Saito}, \& {Yun}}]{2017ApJ...840....8S}
{Sliwa}, K., {Wilson}, C.~D., {Matsushita}, S., {et~al.} 2017, \apj, 840, 8

\bibitem[{{Sliwa} {et~al.}(2012){Sliwa}, {Wilson}, {Petitpas}, {Armus},
  {Juvela}, {Matsushita}, {Peck}, \& {Yun}}]{2012ApJ...753...46S}
{Sliwa}, K., {Wilson}, C.~D., {Petitpas}, G.~R., {et~al.} 2012, \apj, 753, 46

\bibitem[{{Smith} {et~al.}(2007){Smith}, {Struck}, {Hancock}, {Appleton},
  {Charmandaris}, \& {Reach}}]{2007AJ....133..791S}
{Smith}, B.~J., {Struck}, C., {Hancock}, M., {et~al.} 2007, \aj, 133, 791

\bibitem[{{Solomon} \& {Vanden Bout}(2005)}]{2005ARA&A..43..677S}
{Solomon}, P.~M., \& {Vanden Bout}, P.~A. 2005, \araa, 43, 677

\bibitem[{{Springel} {et~al.}(2005){Springel}, {White}, {Jenkins}, {Frenk},
  {Yoshida}, {Gao}, {Navarro}, {Thacker}, {Croton}, {Helly}, {Peacock}, {Cole},
  {Thomas}, {Couchman}, {Evrard}, {Colberg}, \& {Pearce}}]{2005Natur.435..629S}
{Springel}, V., {White}, S.~D.~M., {Jenkins}, A., {et~al.} 2005, \nat, 435, 629

\bibitem[{{Struck} {et~al.}(2011){Struck}, {Dobbs}, \&
  {Hwang}}]{2011MNRAS.414.2498S}
{Struck}, C., {Dobbs}, C.~L., \& {Hwang}, J.-S. 2011, \mnras, 414, 2498

\bibitem[{{Tateuchi} {et~al.}(2015){Tateuchi}, {Konishi}, {Motohara},
  {Takahashi}, {Mitani Kato}, {Kitagawa}, {Todo}, {Toshikawa}, {Sako},
  {Uchimoto}, {Ohsawa}, {Asano}, {Ita}, {Kamizuka}, {Komugi}, {Koshida},
  {Manabe}, {Nakamura}, {Nakashima}, {Okada}, {Takagi}, {Tanab{\'e}},
  {Uchiyama}, {Aoki}, {Doi}, {Handa}, {Kawara}, {Kohno}, {Minezaki}, {Miyata},
  {Morokuma}, {Soyano}, {Tamura}, {Tanaka}, {Tarusawa}, \&
  {Yoshii}}]{2015ApJS..217....1T}
{Tateuchi}, K., {Konishi}, M., {Motohara}, K., {et~al.} 2015, \apjs, 217, 1

\bibitem[{{Teyssier} {et~al.}(2010){Teyssier}, {Chapon}, \&
  {Bournaud}}]{2010ApJ...720L.149T}
{Teyssier}, R., {Chapon}, D., \& {Bournaud}, F. 2010, \apjl, 720, L149

\bibitem[{{The Astropy Collaboration} {et~al.}(2018){The Astropy
  Collaboration}, {Price-Whelan}, {Sip{\'{o}}cz}, {G{\"u}nther}, {Lim},
  {Crawford}, {Conseil}, {Shupe}, {Craig}, {Dencheva}, {Ginsburg},
  {VanderPlas}, {Bradley}, {P{\'e}rez- Su{\'a}rez}, {de Val-Borro}, {Aldcroft},
  {Cruz}, {Robitaille}, {Tollerud}, {Ardelean}, {Babej}, {Bachetti}, {Bakanov},
  {Bamford}, {Barentsen}, {Barmby}, {Baumbach}, {Berry}, {Biscani}, {Boquien},
  {Bostroem}, {Bouma}, {Brammer}, {Bray}, {Breytenbach}, {Buddelmeijer},
  {Burke}, {Calderone}, {Cano Rodr{\'\i}guez}, {Cara}, {Cardoso}, {Cheedella},
  {Copin}, {Crichton}, {D{\'A}vella}, {Deil}, {Depagne}, {Dietrich}, {Donath},
  {Droettboom}, {Earl}, {Erben}, {Fabbro}, {Ferreira}, {Finethy}, {Fox},
  {Garrison}, {Gibbons}, {Goldstein}, {Gommers}, {Greco}, {Greenfield},
  {Groener}, {Grollier}, {Hagen}, {Hirst}, {Homeier}, {Horton}, {Hosseinzadeh},
  {Hu}, {Hunkeler}, {Ivezi{\'c}}, {Jain}, {Jenness}, {Kanarek}, {Kendrew},
  {Kern}, {Kerzendorf}, {Khvalko}, {King}, {Kirkby}, {Kulkarni}, {Kumar},
  {Lee}, {Lenz}, {Littlefair}, {Ma}, {Macleod}, {Mastropietro}, {McCully},
  {Montagnac}, {Morris}, {Mueller}, {Mumford}, {Muna}, {Murphy}, {Nelson},
  {Nguyen}, {Ninan}, {N{\"o}the}, {Ogaz}, {Oh}, {Parejko}, {Parley}, {Pascual},
  {Patil}, {Patil}, {Plunkett}, {Prochaska}, {Rastogi}, {Reddy Janga},
  {Sabater}, {Sakurikar}, {Seifert}, {Sherbert}, {Sherwood-Taylor}, {Shih},
  {Sick}, {Silbiger}, {Singanamalla}, {Singer}, {Sladen}, {Sooley},
  {Sornarajah}, {Streicher}, {Teuben}, {Thomas}, {Tremblay}, {Turner},
  {Terr{\'o}n}, {van Kerkwijk}, {de la Vega}, {Watkins}, {Weaver}, {Whitmore},
  {Woillez}, \& {Zabalza}}]{2018arXiv180102634T}
{The Astropy Collaboration}, {Price-Whelan}, A.~M., {Sip{\'{o}}cz}, B.~M.,
  {et~al.} 2018, ArXiv e-prints, arXiv:1801.02634

\bibitem[{{Thomas} {et~al.}(2002){Thomas}, {Dunne}, {Clemens}, {Alexander},
  {Eales}, \& {Green}}]{2002MNRAS.329..747T}
{Thomas}, H.~C., {Dunne}, L., {Clemens}, M.~S., {et~al.} 2002, \mnras, 329, 747

\bibitem[{{Toomre} \& {Toomre}(1972)}]{1972ApJ...178..623T}
{Toomre}, A., \& {Toomre}, J. 1972, \apj, 178, 623

\bibitem[{{Turner} {et~al.}(2015){Turner}, {Beck}, {Benford}, {Consiglio},
  {Ho}, {Kov{\'a}cs}, {Meier}, \& {Zhao}}]{2015Natur.519..331T}
{Turner}, J.~L., {Beck}, S.~C., {Benford}, D.~J., {et~al.} 2015, \nat, 519, 331

\bibitem[{{Tutukov} \& {Fedorova}(2006)}]{2006ARep...50..785T}
{Tutukov}, A.~V., \& {Fedorova}, A.~V. 2006, Astronomy Reports, 50, 785

\bibitem[{{U} {et~al.}(2012){U}, {Sanders}, {Mazzarella}, {Evans}, {Howell},
  {Surace}, {Armus}, {Iwasawa}, {Kim}, {Casey}, {Vavilkin}, {Dufault},
  {Larson}, {Barnes}, {Chan}, {Frayer}, {Haan}, {Inami}, {Ishida},
  {Kartaltepe}, {Melbourne}, \& {Petric}}]{2012ApJS..203....9U}
{U}, V., {Sanders}, D.~B., {Mazzarella}, J.~M., {et~al.} 2012, \apjs, 203, 9

\bibitem[{{Ueda} {et~al.}(2014){Ueda}, {Iono}, {Yun}, {Crocker}, {Narayanan},
  {Komugi}, {Espada}, {Hatsukade}, {Kaneko}, {Matsuda}, {Tamura}, {Wilner},
  {Kawabe}, \& {Pan}}]{2014ApJS..214....1U}
{Ueda}, J., {Iono}, D., {Yun}, M.~S., {et~al.} 2014, \apjs, 214, 1

\bibitem[{{Van Der Walt} {et~al.}(2011){Van Der Walt}, {Colbert}, \&
  {Varoquaux}}]{2011arXiv1102.1523V}
{Van Der Walt}, S., {Colbert}, S.~C., \& {Varoquaux}, G. 2011, ArXiv e-prints,
  arXiv:1102.1523

\bibitem[{{Veilleux} {et~al.}(1995){Veilleux}, {Kim}, {Sanders}, {Mazzarella},
  \& {Soifer}}]{1995ApJS...98..171V}
{Veilleux}, S., {Kim}, D.-C., {Sanders}, D.~B., {Mazzarella}, J.~M., \&
  {Soifer}, B.~T. 1995, \apjs, 98, 171

\bibitem[{{Verdes-Montenegro} {et~al.}(2005){Verdes-Montenegro}, {Sulentic},
  {Lisenfeld}, {Leon}, {Espada}, {Garcia}, {Sabater}, \&
  {Verley}}]{2005A&A...436..443V}
{Verdes-Montenegro}, L., {Sulentic}, J., {Lisenfeld}, U., {et~al.} 2005, \aap,
  436, 443

\bibitem[{{Verley} {et~al.}(2007){Verley}, {Leon}, {Verdes-Montenegro},
  {Combes}, {Sabater}, {Sulentic}, {Bergond}, {Espada}, {Garc{\'{\i}}a},
  {Lisenfeld}, \& {Odewahn}}]{2007A&A...472..121V}
{Verley}, S., {Leon}, S., {Verdes-Montenegro}, L., {et~al.} 2007, \aap, 472,
  121

\bibitem[{{Wada} {et~al.}(2011){Wada}, {Baba}, \&
  {Saitoh}}]{2011ApJ...735....1W}
{Wada}, K., {Baba}, J., \& {Saitoh}, T.~R. 2011, \apj, 735, 1

\bibitem[{{Wadsley} {et~al.}(2017){Wadsley}, {Keller}, \&
  {Quinn}}]{2017MNRAS.471.2357W}
{Wadsley}, J.~W., {Keller}, B.~W., \& {Quinn}, T.~R. 2017, \mnras, 471, 2357

\bibitem[{{Werner} {et~al.}(2004){Werner}, {Roellig}, {Low}, {Rieke}, {Rieke},
  {Hoffmann}, {Young}, {Houck}, {Brandl}, {Fazio}, {Hora}, {Gehrz}, {Helou},
  {Soifer}, {Stauffer}, {Keene}, {Eisenhardt}, {Gallagher}, {Gautier}, {Irace},
  {Lawrence}, {Simmons}, {Van Cleve}, {Jura}, {Wright}, \&
  {Cruikshank}}]{2004ApJS..154....1W}
{Werner}, M.~W., {Roellig}, T.~L., {Low}, F.~J., {et~al.} 2004, \apjs, 154, 1

\bibitem[{{Whitmore} {et~al.}(2007){Whitmore}, {Chandar}, \&
  {Fall}}]{2007AJ....133.1067W}
{Whitmore}, B.~C., {Chandar}, R., \& {Fall}, S.~M. 2007, \aj, 133, 1067

\bibitem[{{Whitmore} \& {Schweizer}(1995)}]{1995AJ....109..960W}
{Whitmore}, B.~C., \& {Schweizer}, F. 1995, \aj, 109, 960

\bibitem[{{Wilson} {et~al.}(2008){Wilson}, {Petitpas}, {Iono}, {Baker}, {Peck},
  {Krips}, {Warren}, {Golding}, {Atkinson}, {Armus}, {Cox}, {Ho}, {Juvela},
  {Matsushita}, {Mihos}, {Pihlstrom}, \& {Yun}}]{2008ApJS..178..189W}
{Wilson}, C.~D., {Petitpas}, G.~R., {Iono}, D., {et~al.} 2008, \apjs, 178, 189

\bibitem[{{Woods} \& {Geller}(2007)}]{2007AJ....134..527W}
{Woods}, D.~F., \& {Geller}, M.~J. 2007, \aj, 134, 527

\bibitem[{{Wu} {et~al.}(2005){Wu}, {Cao}, {Hao}, {Liu}, {Wang}, {Xia}, {Deng},
  \& {Young}}]{2005ApJ...632L..79W}
{Wu}, H., {Cao}, C., {Hao}, C.-N., {et~al.} 2005, \apj, 632, L79

\bibitem[{{Xu} {et~al.}(2014){Xu}, {Cao}, {Lu}, {Gao}, {van der Werf}, {Evans},
  {Mazzarella}, {Chu}, {Haan}, {Diaz-Santos}, {Meijerink}, {Zhao}, {Appleton},
  {Armus}, {Charmandaris}, {Lord}, {Murphy}, {Sanders}, {Schulz}, \&
  {Stierwalt}}]{2014ApJ...787...48X}
{Xu}, C.~K., {Cao}, C., {Lu}, N., {et~al.} 2014, \apj, 787, 48

\bibitem[{{Yamashita} {et~al.}(2017){Yamashita}, {Komugi}, {Matsuhara},
  {Armus}, {Inami}, {Ueda}, {Iono}, {Kohno}, {Evans}, \&
  {Arimatsu}}]{2017ApJ...844...96Y}
{Yamashita}, T., {Komugi}, S., {Matsuhara}, H., {et~al.} 2017, \apj, 844, 96

\bibitem[{{Yao} {et~al.}(2003){Yao}, {Seaquist}, {Kuno}, \&
  {Dunne}}]{2003ApJ...588..771Y}
{Yao}, L., {Seaquist}, E.~R., {Kuno}, N., \& {Dunne}, L. 2003, \apj, 588, 771

\bibitem[{{Yun} \& {Carilli}(2002)}]{2002ApJ...568...88Y}
{Yun}, M.~S., \& {Carilli}, C.~L. 2002, \apj, 568, 88

\bibitem[{{Zink} {et~al.}(2000){Zink}, {Lester}, {Doppmann}, \&
  {Harvey}}]{2000ApJS..131..413Z}
{Zink}, E.~C., {Lester}, D.~F., {Doppmann}, G., \& {Harvey}, P.~M. 2000, \apjs,
  131, 413

\end{thebibliography}

\begin{table*}
\begin{center}
\caption{General properties of NGC\,3110 and NGC\,232 \label{tbl-1}}
\begin{tabular}{lll}
\tableline
   &   NGC\,3110 &  NGC\,232   \\\tableline
RA, Dec [J2000]  \tablenotemark{a}                                     & { 10$^{\rm h}$04$^{\rm m}$02\farcs1, -06\arcdeg28\arcmin29\arcsec } & { 00$^{\rm h}$42$^{\rm m}$45\farcs8, -23\arcdeg33\arcmin41\arcsec} \\    
$T$ \tablenotemark{b}                                & SB(rs)b pec  &  SBa(r) pec \\
Mass ratio with companion & { 14  } & { 0.8 }\\
Separation -- Velocity difference to companion & { 38\,kpc -- 235\,\kms } & { 50\,kpc -- 120\,\kms }\\
Nuclear Activity \tablenotemark{c}            & HII, LIRG     & HII, LIRG, LINER \\
$D_{25}$, Incl., PA \tablenotemark{d}      & { 1\farcm86, 65$\arcdeg$, 171$\arcdeg$ }& { 0\farcm97,   47$\arcdeg$,   17$\arcdeg$}   \\                                         
$D$ [Mpc] \tablenotemark{e}                          & { 75.2 }  -- { 352\,pc }  & { 90.5 } -- { 420\,pc} \\
log($L_{\rm B}$[L$_\odot$]) \tablenotemark{f} & { 11.01}   & { 10.44 }\\ 
log(\mhi\ [M$_\odot$]) \tablenotemark{g} & { 10.0} &  { 9.50 } \\
log($M_{\rm mol}$ [M$_\odot$]) \tablenotemark{h} & { 10.37} &   { 10.16} \\
$M_{\rm mol}$ / \mhi \tablenotemark{i} & { 2.34} &   { 4.57} \\
log($L_{\rm IR}$ [L$_\odot$]) \tablenotemark{j} & { 11.23 | 11.33}  & { 11.30 | 11.41}\\
$SFR$ [M$_\sun$\,yr$^{-1}$] \tablenotemark{k} &  { 23.2} & {27.9} \\
log($SFE$  [yr$^{-1}$]) \tablenotemark{l} & { -9.00} & { -8.71} \\
log($M_\star$ [M$_\odot$]) \tablenotemark{m} & { 10.78}   & { 10.80 }\\

\tableline
\end{tabular}
\end{center}
\tablenotetext{a}{Coordinates from NED.}
\tablenotetext{b}{Morphological type following the RC3 system \citep{2001MNRAS.322..486B}.}
\tablenotetext{c}{Nuclear activity type from NED.}
\tablenotetext{d}{Major optical axis, inclination and position angle (N to E) from Hyperleda \citep{2014A&A...570A..13M}.}
\tablenotetext{e}{Luminosity distance and linear scale from NED using 3K Microwave Background Radiation as reference frame and assuming  Ho =  73\,\kms\,Mpc$^{-1}$, $\Omega_{matter}$ =   0.27, $\Omega_{vacuum}$ =   0.73.}
\tablenotetext{f}{log($L_{\rm B}$), base 10 logarithm of the blue luminosity obtained as log($L_{\rm B}$) =12.192 + 2 log($D$) -0.4 $\times$ $B_{T,c}$, where $B_{T,c}$ is the apparent B-magnitude corrected by galactic, internal extinction and k-correction, as obtained from Hyperleda.}
\tablenotetext{g}{log($M_{\rm HI}$), base 10 logarithm of the \ion{H}{1} masses from \citet{1991A&A...245..393M}, corrected to our choice of distance.}
\tablenotetext{h}{log($M_{\rm mol}$),  base 10 logarithm of the molecular gas content accounting for elements other than hydrogen. 
These were calculated from the CO(1--0) integrated intensity measurements in \citet{1991ApJ...370..158S} for NGC\,3110 (12m NRAO telescope, characterized by a 55\arcsec\ beam size at 115\,GHz) and in \citet{1990A&A...236..327M} for NGC\,232 (SEST telescope, 44\arcsec\ at 115\,GHz). We used Jy-to-K conversion factors of 30.4 Jy/K for the 12m NRAO, and 19 Jy/K for SEST. The $X$ factor is $X$ = 2 $\times$ 10$^{20}$\,cm$^{-2}$\,(K\,km\,s$^{-1}$)$^{-1}$.}
\tablenotetext{i}{$M_{\rm mol}$/$M_{\rm HI}$,  molecular to atomic gas fraction.}
\tablenotetext{j}{log($L_{\rm IR}$), base 10 logarithm of the 40--400\,$\mu$m and 8--1000\,$\mu$m luminosities (in this order) obtained from the four IRAS bands 12, 25, 60 and 100\,$\mu$m, as provided in the Revised Bright Galaxy Catalog (RBGS, \citealt{2003AJ....126.1607S}). The value was corrected to our choice of distance.}
\tablenotetext{k}{SFR as obtained from the 8--1000\,$\mu$m $L_{\rm IR}$ in the previous row ($h$), following SFR = $L_{\rm IR}$/ (5.8  $\times$ 10$^9$ $L_\odot$)\,M$_\odot$\,yr$^{-1}$ (as in Eq. 3 of \citealt{1998ApJ...498..541K}, see also \citealt{1998ARA&A..36..189K}), and correcting to a Kroupa IMF by a factor 1.59 following \citet{2008AJ....136.2846B}.}
\tablenotetext{l}{log($SFE$ = $SFR$ / $M_{\rm mol}$), base 10 logarithm of the ratio between SFR and $M_{\rm mol}$.}
\tablenotetext{m}{log($M_\star$), base 10 logarithm of the total stellar mass computed by fitting stellar population synthesis models to the observed near-infrared (NIR) through ultraviolet (UV) SEDs \citep{2012ApJS..203....9U,2016ApJ...825..128L}, corrected to our choice of distance.}

\end{table*}

\begin{table*}
\begin{center}
\caption{SMA CO(2--1) observations \label{tbl-2}}
\begin{tabular}{lll}
\tableline
                       & NGC\,3110                               & NGC\,232\\
       \tableline
           Date &    2008/10/11         & 2008/05/08 - 2008/05/21 \\
           Time on source & 162\,min & 89\,min \\
           Antennas/Maximum baseline length & 7 / 53.795\,m & 7 / 75.573 - 90.950\,m \\
Field of view     & 52\arcsec\ $\times$ 52\arcsec\  &  52\arcsec\ $\times$ 52\arcsec\  \\                                
Beam size        & { 3\farcs1 $\times$ 2\farcs8 }       &  {  3\farcs2 $\times$ 2\farcs4 }\\       
P.A. (N to E)       &  { 10.7\arcdeg }                       &   { -86.2\arcdeg }   \\                                                       
rms noise channel 20\,\kms   [mJy/beam]                     &  14   & { 15}  \\                                                          
Calibrators (ampl. / bandpass / phase (separation)) & 3C84 / 3C84 / 1058+015 (15.8\arcdeg)   & Uranus / 3C454.3 / 2258-279 (23.9\arcdeg) \\

\tableline
\end{tabular}
\end{center}
\end{table*}

\begin{table}
\caption{Derived parameters of the different regions detected in CO(2--1)   \label{table2}}
\begin{center}
\begin{tabular}{l r r r}
\tableline
\small Component    &\small Velocity range &\small ${S_{\rm CO(2-1)}}$\tablenotemark{b}  & $M_{\rm mol}$ \tablenotemark{c} \\
                                 &\small [km\,s$^{-1}$]  &\small  [Jy\,km\,s$^{-1}$]  & [M$_\odot$]  \\
\tableline
\\
{ NGC\,3110}\\
Spiral arm NW                                   &   4800 -- 5140 &   116        &  2.1 -- 3.9 $\times$ 10$^{9}$  \\
Spiral arm SE                                    &   4860 -- 5200 &     165      &  3.1 -- 5.7  $\times$ 10$^{9}$ \\
Circumnuclear region \tablenotemark{a} &   4780 -- 5240 &   358      & 6.6 -- 12.2 $\times$ 10$^{9}$    \\
All                                                    &   4760 -- 5240 &   679     &  1.3 -- 2.4 $\times$ 10$^{10}$ \\
{ NGC\,232}\\
All                                                    &  6370 -- 6970   &   295    &   1.4 --2.0 $\times$ 10$^{10}$ \\

\\

\tableline
\end{tabular}
\end{center}
\tablenotetext{a}{The circumnuclear region of NGC\,3110 is defined as the region within an ellipse with major axes 12\farcs4 $\times$ 8\farcs0 (4.6\,kpc $\times$ 2.9\,kpc), centered at its nucleus and with P.A. = 0\arcdeg. }
\tablenotetext{b}{Integrated CO(2--1) flux, corrected by primary beam but not by flux loss.}
\tablenotetext{c}{$M_{\rm mol}$, the molecular gas mass, is calculated as $M_{\rm mol} [{\rm M}_\odot] = 4.3 \times X_{\rm 2}~(R_{2-1/1-0})^{-1}   L_{\rm CO(2-1)}$, where the CO(2--1) luminosity is derived as $L_{\rm CO(2-1)} = 611 \times S_{\rm CO(2-1)}\,D_{\rm L}^2$ [K\,\kms\,pc$^2$].
The factor 1.36  for elements other than hydrogen \citep{2000asqu.book.....C} is taken into account, 
$X_{\rm 2}$ (i.e. $X$ factor normalized to 2 $\times$ $10^{20}$\,cm$^{-2}$\,(K\,\kms)$^{-1}$) is unity, 
$R_{2-1/1-0}$ is the CO(1--0) to CO(2--1) integrated intensity ratio,
$S_{\rm CO(2-1)}$ is the integrated CO(2--1) line flux in Jy\,\kms, and 
$D_{\rm L}$ is the luminosity distance to the source in Mpc.
 $R_{2-1/1-0}$  $\simeq$ 0.8 and 0.46 for NGC\,3110 and NGC\,232, respectively (see \S\,\ref{sub:moleculargasmass}).
 In the range, the second value correspond to a fixed correction factor to account for flux loss (see \S\,\ref{subsect:co2-1emissionline}).}

\end{table}

\begin{table}[htp]
\caption{Molecular gas concentrations ($f_{\rm con}$) and gas-to-dynamical mass ratios ($f_{\rm dyn}$) }
\begin{center}
\begin{tabular}{cccccccc}
\hline
Galaxy        &  $R_{25}$ &  $\Sigma_{\rm mol}^{\rm disk}$& $\Sigma_{\rm mol}^{\rm 1kpc}$   & $f_{\rm con}$ & $M_{\rm mol}^{\rm 1kpc}$ &  $M_{\rm dyn}^{\rm 1kpc}$    &  $f_{\rm dyn}$  \\
                    &  [kpc]        &  [M$_\odot$\,pc$^{-2}$]   & [M$_\odot$\,pc$^{-2}$]                          & & [10$^9$ M$_\odot$]           &   [10$^9$ M$_\odot$]            &                                              \\
                  \hline
NGC\,3110 &      19.6      &    8  &    306    &     38      &  0.9              &       14.9                       &     0.06                   \\                         
NGC\,232  &       12.2      &  21  &    455    &     22    & 1.1               &     39.1                       &      0.03                  \\         \hline
\end{tabular}
\end{center}
\label{tab4}

\end{table}

\end{document}